\documentclass[twocolumn,secnumarabic,amssymb, nobibnotes, aps,prx,showpacs]{revtex4}
\usepackage{comment}
\usepackage{amsmath,amsthm,amssymb}
\usepackage[pdftex]{graphicx}
\usepackage{color}
\usepackage[abs]{overpic}
\def\vector#1{\mbox{\boldmath $#1$}} 
\newcommand{\argmin}{\mathop{\rm arg~min}\limits}

\begin{document}
	\title{Empirical observations of the ultraslow diffusion driven by the fractional dynamics in languages: Dynamical statistical properties of word counts of already popular words} 
	\author{Hayafumi Watanabe$^{1,2}$}\email[E-mail: ]{hayafumi.watanabe@gmail.com}
	\affiliation{$^1$Risk Analysis Research Center, The Institute of Statistical Mathematics, 10-3 Midori-cho, Tachikawa, Tokyo 190-8562, Japan}
	\affiliation{$^2$Joint Support-Center for Data Science Research, The Research Organization of Information and Systems, 10-3 Midori-cho, Tachikawa, Tokyo 190-8562, Japan}
	\affiliation{$^3$Hottolink,Inc., 6 Yonbancho Chiyoda-ku, Tokyo 102-0081, Japan}
	\begin{abstract}
		Ultraslow diffusion (i.e. logarithmic diffusion) has been extensively studied theoretically, but has hardly been observed empirically. 
		In this paper, firstly, we find the ultraslow-like diffusion of the time-series of word counts of already popular words by analysing three different nationwide language databases: (i) newspaper articles (Japanese), (ii) blog articles (Japanese), and (iii) page views of Wikipedia  (English, French, Chinese, and Japanese).
		Secondly, we use theoretical analysis to show that this diffusion is basically explained by the random walk model with the power-law forgetting with the exponent $\beta \approx 0.5$, which is related to the fractional Langevin equation.
		The exponent $\beta$ characterises the speed of forgetting and $\beta \approx 0.5$ corresponds to (i) the border (or thresholds) between the stationary and the nonstationary and 
		(ii) the right-in-the-middle dynamics between the IID noise for $\beta=1$ and the normal random walk for $\beta=0$.
		Thirdly, 
		the generative model of the time-series of word counts of already popular words, which is a kind of Poisson process with the Poisson parameter sampled by the above-mentioned random walk model, can almost reproduce not only the empirical mean-squared displacement but also the power spectrum density 
		and the probability density function. 
	\end{abstract}

\begin{abstract} 
Ultraslow diffusion (i.e. logarithmic diffusion) has been extensively studied theoretically, but has hardly been observed empirically.  In this paper, firstly, we find the ultraslow-like diffusion of the time-series of word counts of already popular words by analysing three different nationwide language databases: (i) newspaper articles (Japanese), (ii) blog articles (Japanese), and (iii) page views of Wikipedia  (English, French, Chinese, and Japanese). 
Secondly, we use theoretical analysis to show that this diffusion is basically explained by the random walk model with the power-law forgetting with the exponent $\beta \approx 0.5$, which is related to the fractional Langevin equation.
The exponent $\beta$ characterises the speed of forgetting and $\beta \approx 0.5$ corresponds to (i) the border (or thresholds) between the stationary and the nonstationary and (ii) the right-in-the-middle dynamics between the IID noise for $\beta=1$ and the normal random walk for $\beta=0$.
Thirdly, the generative model of the time-series of word counts of already popular words, which is a kind of Poisson process with the Poisson parameter sampled by the above-mentioned random walk model, can almost reproduce not only the empirical mean-squared displacement but also the power spectrum density and the probability density function. 
\end{abstract}

	\maketitle



\maketitle
\section{Introduction}
	A language is a typical complex system, and is characterised by well-known language-independent statistical laws such as Zipf's law and Heap's law \cite{altmann2016statistical}.   
	In this study, we investigate the dynamical statistical properties in languages by using massive databases related to word usage that has developed in the past 10 years.
	Especially, we focus on the stability or the slowness of change in the usage of already popular words from the viewpoint of diffusion on a complex system 
	and show that common logarithmic diffusion (i.e. very slow diffusion or change) is approximately observed by some languages or media.   \par
	Diffusion on complex systems, which is an attractive research topic in physics or complex system science, has been extensively studied both theoretically and empirically and 
	has been applied to various systems such as biological or social systems. 
	The diffusion on complex systems is basically characterised by the mean-squared displacement (MSD). 
	The vast majority of studies reported that MSD growth occurs asymptotically according to the power law,
	\textcolor{black}{
	\begin{equation}
	\left< x^2(t) \right> \propto t^{\alpha}.
	\end{equation}
	}
	In the case of $\alpha=1$, the diffusion corresponds to normal diffusion, such as the diffusion of particles in water, which is modelled using a random walk. 
	In other cases, it is known as anomalous diffusion, especially, it is termed subdiffusion for $0<\alpha<1$ and superdiffusion for $\alpha>1$. 
	Many complex systems have been shown to exhibit this power-law type of anomalous diffusion in diverse areas, such as physics, chemistry, geophysics, biology, and economy \cite{metzler2000random, da2014ultraslow}. 
	In theoretical studies,
	\textcolor{black}{anomalous diffusion is explained using the correlation of random noise (e.g. a random walk in disordered media) \cite{bouchaud1990anomalous}, a finite-variance (e.g. a Levy flight) \cite{bouchaud1990anomalous, metzler2000random}, a power-law wait time (e.g. a continuous random walk) \cite{bouchaud1990anomalous, metzler2000random, burov2011single}, and a long memory (e.g. a fractional random walk) \cite{lowen2005fractal, burov2011single}.}
	\par
	%
	%
	Another class of anomalous diffusion is predicted by theories, where the MSD growths logarithmically,  
	\begin{equation}
	\left<x^2(t) \right> \propto \log(t)^{\alpha} \label{eq_usd}
	\end{equation}
	This type of diffusion is known as ``ultraslow diffusion''.
	One of the best-known examples that was first discovered is the diffusion in a disordered medium (it is known as Sinai diffusion for $\alpha=4$ ) \cite{sinai1983limiting}. 
	Thereafter, other types of models that explain ultraslow diffusion have also been proposed such as continuous random walk (CTRW) with the waiting time generated by the logarithmic-form probability density function \cite{godec2014localisation}, 
	CTRW with waiting time generated by the power-form probability density function and the excluded volume effect \cite{sanders2014severe}, temporal change of diffusion coefficients \cite{bodrova2015ultraslow}, spatial changes \cite{cherstvy2013population}, and \textcolor{black}{fractional dynamics \cite{eab2011fractional}.} \par
	Although many theoretical studies of ultraslow diffusion have been reported, we were unable to find empirical examples thereof. \textcolor{black}{
		A rare example of diffusion related to the logarithmic function, which is similar but different from the ``ultraslow diffusion'' defined by Eq. \ref{eq_usd} (i.e. $\propto \log(t)^{\alpha}$), is the mobility of humans measured by mobile phone data.
		 In this study, by using both data and models, the authors insisted that the MSD grows according to $\log(\log(t))^{\alpha}$ or becomes saturated (i.e. the MSD grows slower than $\log(t)^\alpha$). This diffusion is mainly explained by the CTRW and preferential return (to home) effects \cite{song2010modelling}.}
	 Diffusion resembling this very slow diffusion was also observed in the mobility of monkeys and the authors maintained that this diffusion may be explained by the heterogeneity of the space such as by Sinai's model \cite{boyer2011non}.
	Note that logarithmic ``relaxation'' phenomena, which are known as ``ageing'', are observed in many systems such as paper crumpling \cite{matan2002crumpling} and grain compalification \cite{richard2005slow}.
	\par
	We investigate the stability or dynamic usage of already popular words.
	In other words, we focus only on the dynamics of the ``mature phase'' in the life trajectory of words (consisting of an ``infant phase'', an ``adolescent phase'', and a ``mature phase'') \cite{petersen2012statistical,gerlach2013stochastic}.
	\textcolor{black}{The pioneer study of a stability and variation of language from the viewpoint on the dynamical statistical property was given by  Erez Lieberman et. al \cite{lieberman2007quantifying}. 
 In this study, the regularization of English verbs (i.e., change from irregular to regular verbs) over the past 1200 years was investigated, and the 0.5th power law of the regularization rate as a function of word frequencies (i.e., higher frequency words involve less changes, or are more stable) was noted.  
 This study quantified the stability of language on a historical timescale (i.e., from 100 to 1000 years). In contrast, our study focuses on stability on a shorter timescale (i.e., from 1 day to 10 years).   }
	Note that some findings relating to the dynamics or properties of words in the ``infant phase'', the ``adolescent phase'' or total life trajectory (i.e. from birth to death) were conducted by using the Google Ngram data corpus (in which word frequencies occurring in printed books from 1520 to 2000 are given) \cite{michel2011quantitative,petersen2012statistical,gerlach2013stochastic}.
	In these studies, the authors found some statistical properties such as the typical time to reach the ``adolescent phase'' is about 20 or 30 years; the MSD is superdiffusion and the dynamics are related to Yule's, Simons, Gibrat's, and preferential attachment processes  \cite{petersen2012statistical,gerlach2013stochastic}. \par
	\textcolor{black}{Note that physicists have studied linguistic phenomena using the concepts of complex systems \cite{link1}, such as competitive dynamics \cite{abrams2003linguistics}, statistical laws \cite{altmann2016statistical}, complex networks \cite{cong2014approaching}, the phase transition and the information theory \cite{i2005zipf}. }
	\par
	In this paper, in order to quantify ``the stability'' or ``the speed of change'' of the usage of already popular words (i.e. the mature phase) precisely, we measure the MSD by using actual data and introduce the time-evolution model of frequencies of words for it. 
	In addition, we clarify the dynamics behind this diffusion. \par
	  \textcolor{black}{Firstly, we investigate the MSD of the time series of word counts of three actual types of data: (i) newspapers for 10 years (Japanese), (ii) blogs for 5 years (Japanese) and (iii) Wikipedia page views for 2 years (English, French, Chinese and Japanese). This approach enabled us to observe an ultraslow-like diffusion for all data (Figs. \ref{single_msd} and \ref{ensemble_msd}).} \par
\textcolor{black}{Secondly, we discuss the relation between empirical results and the random walk model with the power-law forgetting given by Eq. \ref{eq_rw}, which is related to the fractional Langevin equation, and can essentially explain the ultraslow diffusion. } \par 
	\textcolor{black}{Thirdly, we introduce a model of word counts sampled from the Poisson process (Eq. \ref{eq_rd}) of which the rate is generated by the previously mentioned random walk model (Eq. \ref{eq_rw}), in order to connect the ultraslow diffusion explained by Eq. \ref{eq_rw}, with peculiar properties of word count data, such as discreteness(i.e. counts take $0,1,2,3,\cdots$).} 
	\textcolor{black}{
		In addition, we show that the model can consistently reproduce the following empirical dynamical statistical properties (Fig. \ref{fig_sim}):
	} 
	\textcolor{black}{
		\begin{itemize}
		  \item[(i)] Mean squared displacement [MSD], 
		  \begin{equation}
		  	\sigma_j(L) = \sum_{t=1}^{T-L}\frac{(f_j(t+L)-f_j(t))^2}{T-L}
		  	\label{def_msd0}
		\end{equation}
		 \item[(ii)] Power spectral density [PSD] (periodogram),
			\begin{equation}
				P(\nu;f_j) = \frac{1}{T} \left|\sum_{t=1}^{T} \exp(- i 2 \pi \nu t) f_j(t)  \right|^2 
				\label{def_psd0}
			\end{equation}
		\item[(iii)] Probability density function [PDF] (histogram)
		\begin{eqnarray}
				&&q(x;\{f_j(t+L)-f_j(t)\})   \nonumber \\
				&&= \frac{\#\{t|x-\frac{\delta}{2}  \leq f_j(t+L)-f_j(t) < x+\frac{\delta}{2} \}}{\delta \times (T-L)}, \label{def_pdf0}
			\end{eqnarray}			
		\end{itemize}
where $f_j(t)$ \textcolor{black}{$(t=1,2,3 \cdots ,T)$} is the word count scaled by the database size at the date $t$ defined by Section \ref{sec2_2}; $T$ is the last date of observation; $L$ is a time lag (positive integer, $1 \leq L \leq T$); $\nu$ is a (spectral) frequency; $\delta>0$ is the bin size of a histogram; $x$ represents the value of a bin of a histogram; and $\#A$ means the number of elements of the set $A$. 
	} \par
	Finally, we conclude with a discussion. 
	\begin{figure*}
		\begin{minipage}{0.32\hsize}
			\centering
			\begin{overpic}[scale=.35,tics=5]%
				{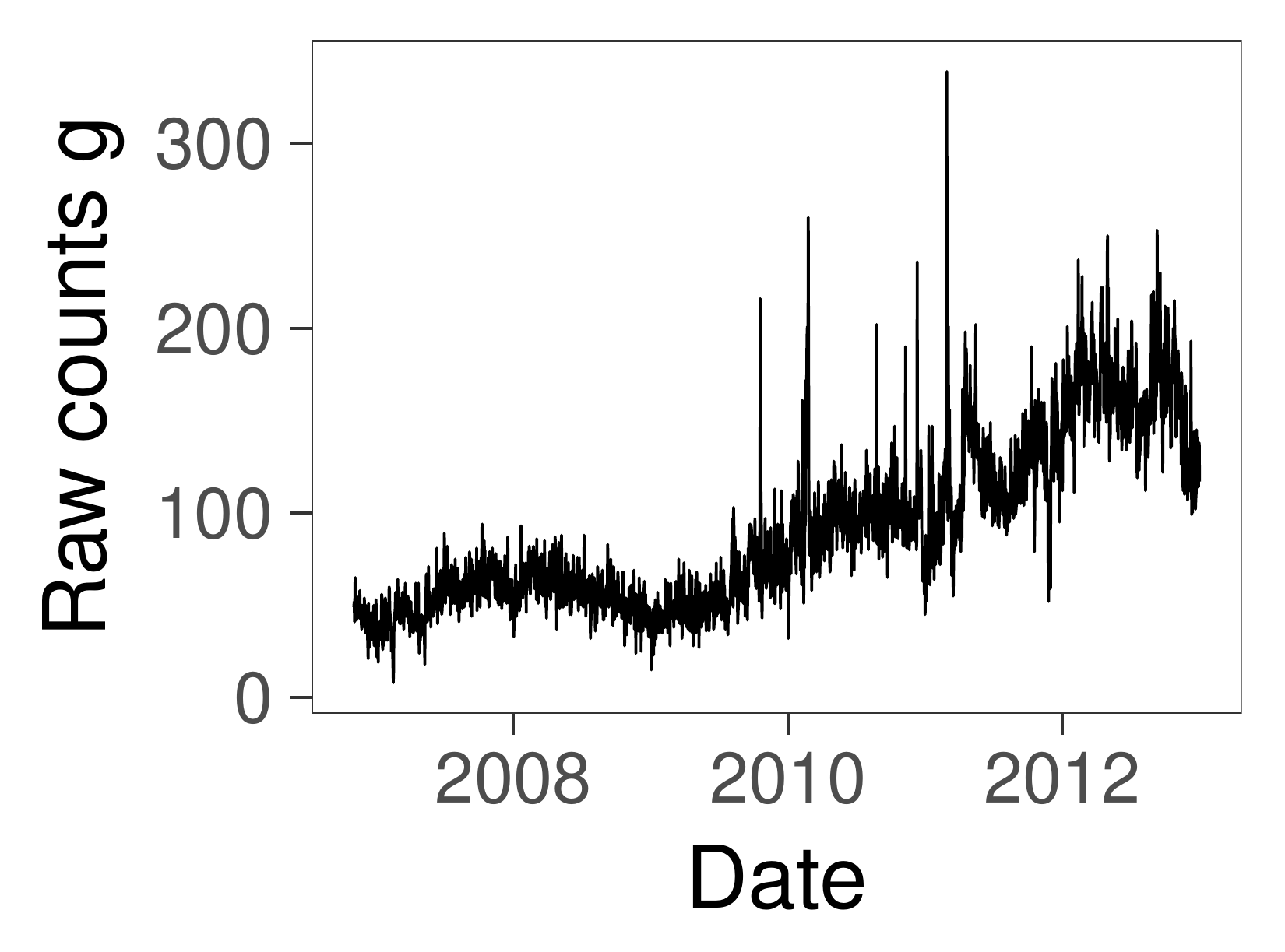}	
				\put(44,105){(a)}	
			\end{overpic}
		\end{minipage}
		\begin{minipage}{0.32\hsize}
			\centering
			\begin{overpic}[scale=.35,tics=5]%
				{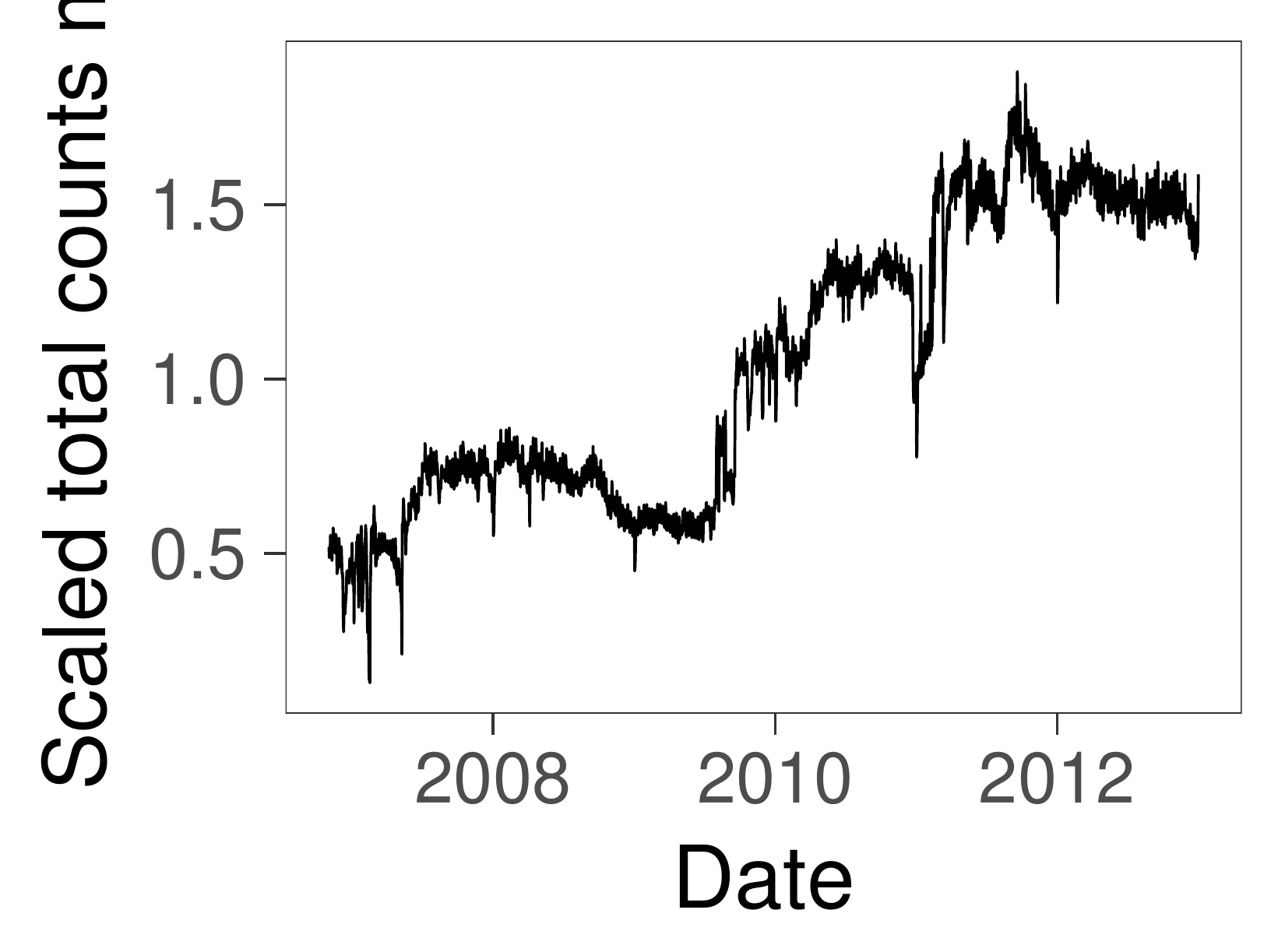}
				\put(40,105){(b)}	
			\end{overpic}
		\end{minipage}
		\begin{minipage}{0.32\hsize}
			\centering
			\begin{overpic}[scale=.35,tics=5]%
				{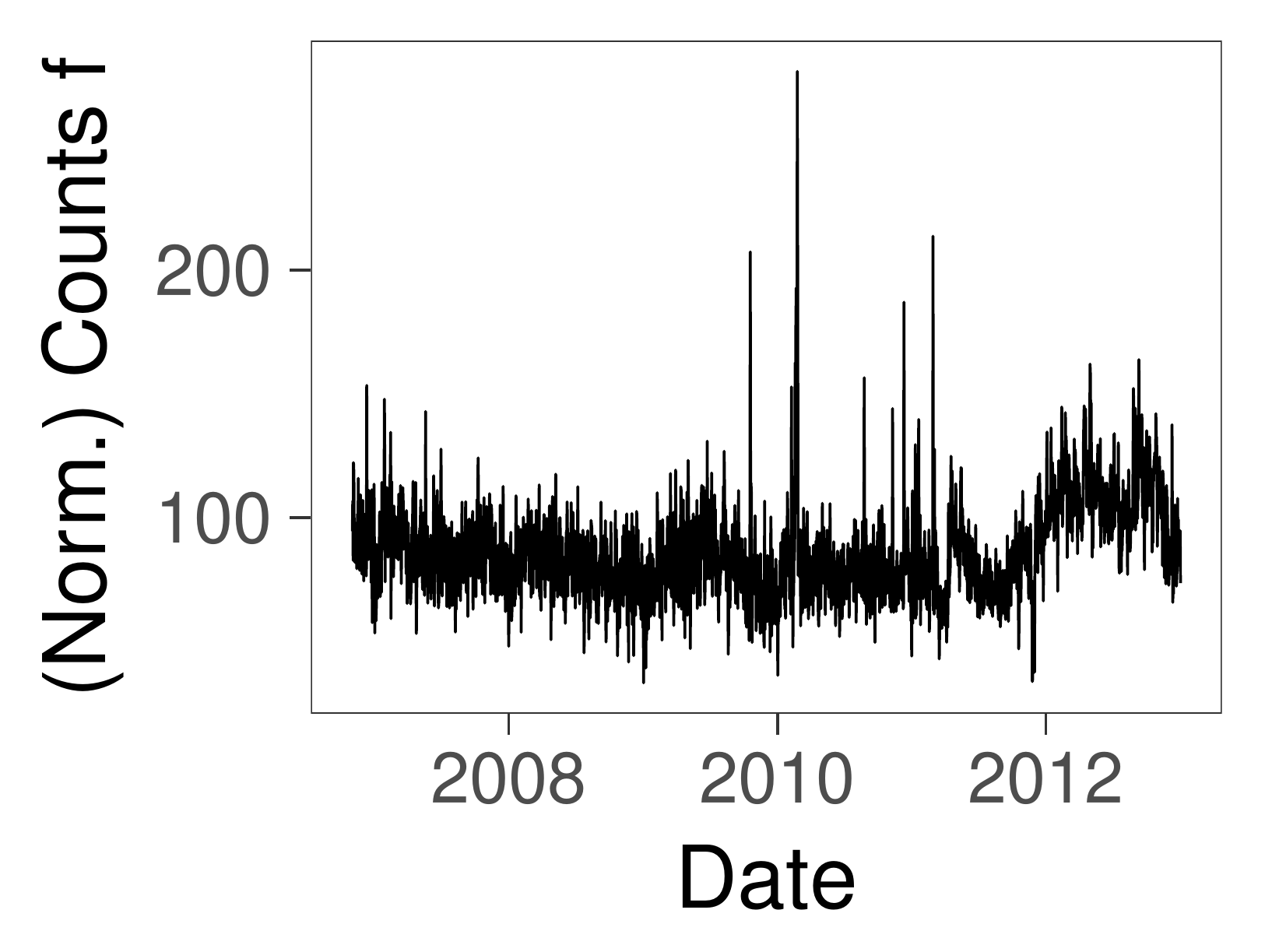}
				\put(42,105){(c)}	
			\end{overpic}
		\end{minipage}
		\caption{
			Time-series of word appearance on Japanese blogs:
			(a) Example of a daily time-series of raw \textcolor{black}{word appearance} for ``Sanada (i.e. well-known family name)'', $g_j(t)$ on Japanese blogs.
			(b) Daily time-series of the normalised total number of blogs (i.e., the normalised scale of database), $m(t)$, which is estimtaed by Appendix \ref{app_m}.
			(c) Daily corresponding time series of word appearances scaled by the normalised total number of blogs  (i.e., the normalised scale of database), $f_j(t)=g_j(t)/m(t)$. 
			We can confirm that the time-variation of raw word appearances $g_j(t)$ shown in panel (a) is almost the same as that of the total number of blogs $m(t)$ shown in panel (b).
		}
		\label{tseries}
	\end{figure*}
	
	\section{Data set}
	Our data analysis involved analysing the daily time-series of the word counts in newspapers (Japanese), blogs (Japanese), and Wikipedia page views (English, French, Chinese, and Japanese). 
	\subsection{Data sources}
	\paragraph{\bf Newspapers} 
	We obtained the time-series of word appearance per day in nation-wide Japanese newspapers by using the ``Shinbun trend in NIKKEI Telecom'' database, which was provided by Nikkei Inc.
	  \textcolor{black}{Using the database, we obtained the daily number of articles containing a keyword from 80 newspapers published in Japan between Jan. 2005 and Sep. 2017 \cite{nikkei}. Note that if an article contains more than two focused keywords (e.g. key word ``dog'': ``There is a dog. The dog is big.''), the system counts it as one article. }
	 We used the top 10,000 ranked words in frequency order as keywords. We referred to the pages entitled \textcolor{black}{"Wiktionary:Frequency lists"} in Wiktionary \cite{wikitionaly} to obtain the rank of the word frequency. \par
	\paragraph{\bf Blogs}
	\textcolor{black}{We obtained the time-series of the daily number of articles containing a keyword in nationwide Japanese blogs using a large database of Japanese blogs ("Kuchikomi@kakaricho") provided by Hottolink, Inc.
	This database contains 3 billion articles of Japanese blogs, which covers 90\% of Japanese blogs from Nov. 2006 to Dec. 2012 \cite{RD_base}. Note that in common with the newspaper data, if one article contains more than two focused keywords, the system counts it as one article.}
	  We used 1,771 basic adjectives and 60,476 nouns as keywords from ipa-dic \cite{idadic}. \par
	\paragraph{\bf Wikipedia page views}
	\textcolor{black}{We obtained daily Wikipedia pageviews using PageviewAPI, which is a public API developed and maintained by the Wikimedia Foundation. This API provides analytical data about article pageviews (or the number of page loads) of Wikipedia. By inputting an article title as a keyword (e.g. ``dog''), a time period (e.g. from 1st Jan. 2017 to 31st Nov. 2017) and a language (e.g. English Wikipedia) to the API, we can obtain time-series of count data on how many times people have visited the focused article (e.g. the number of loads or pageviews of the ``dog'' page in the English Wikipedia) per day during a given time period. Although Wikipedia page views are not the word appearance of a keyword in documents unlike newspaper data and blog data, they are often used to investigate the daily changes of concerns of keywords (or article) in common with newspaper and blog data.
	}
	 We obtained the data of the English, French, Chinese, and Japanese pages from Jul. 2015 to Sep. 2017 \cite{wikipedia_pageview}. We used the top 10,000 ranked words in frequency order as keywords \cite{RD_base} with respect to each language.  To obtain the rank of the word frequency, we referred to the pages entitled \textcolor{black}{"Wiktionary:Frequency lists"} in Wiktionary as is the case with the newspaper data \cite{wikitionaly}. 
	\subsection{Normalised time-series of word appearances}
	\label{sec2_2}
	\textcolor{black}{
	We define herein as follows the notation of the time-series of the word counts $g_j(t)$ and the normalised word counts $f_j(t)$:} 
	\begin{itemize}  
		\item \textcolor{black}{$g_j(t)$ $(t=1,2,3,\cdots,T)$ $(j=1,2,3,\cdots,W)$ is the raw daily counts of the $j$-th word within the nationwide dataset (Fig.\ref{tseries}(a)), where $T$ is the last date of observation, and $W$ is the number of observed keywords.} \par
		\textcolor{black}{Concretely speaking, for the newspaper and blog data, $g_j(t)$ corresponds to the daily number of articles containing the $j$-th keyword in the database. For the Wikipedia page view data, it corresponds to the daily page view of an article entitled the $j$-th keyword (how many times people have visited the focused article).}
		\textcolor{black}{
		\item $f_j(t)=g_j(t)/m(t)$ is the time-series of the daily count normalised by the temporal scale of database $m(t)$ (Fig. \ref{tseries} (c)).} \par 
			\textcolor{black}{$f_j(t)$ corresponds to the original time deviation of the $j$-th word separated from the effects of deviations in the scale of database $m(t)$ (Figs. \ref{tseries} (b) and (c)).}  
			\textcolor{black}{The scale of database $m(t)$ almost corresponds to the (normalised) total number of articles (i.e. temporal database size) for the newspaper and blog data. For the Wikipedia data, it conceivable that $m(t)$ almost corresponds to the (normalised) total temporal number of users of Wikipedia of a focused language ($m(t)$ does not correspond to the size or number of articles of Wikipedia of a focused language itself.).
		} 
		\textcolor{black}{
		 	$m(t)$ is estimated herein by the ensemble median of the number of words at time $t$, as described in the Appendix \ref{app_m}. $m(t)$ assumes that $\sum_{t=1}^{T}m(t)/T=1$ for normalisation (Fig.\ref{tseries}(b)).} \par
		 	
	\end{itemize}

	\begin{figure*}
		\begin{minipage}{0.32\hsize}
\begin{overpic}[scale=0.35,tics=5]
{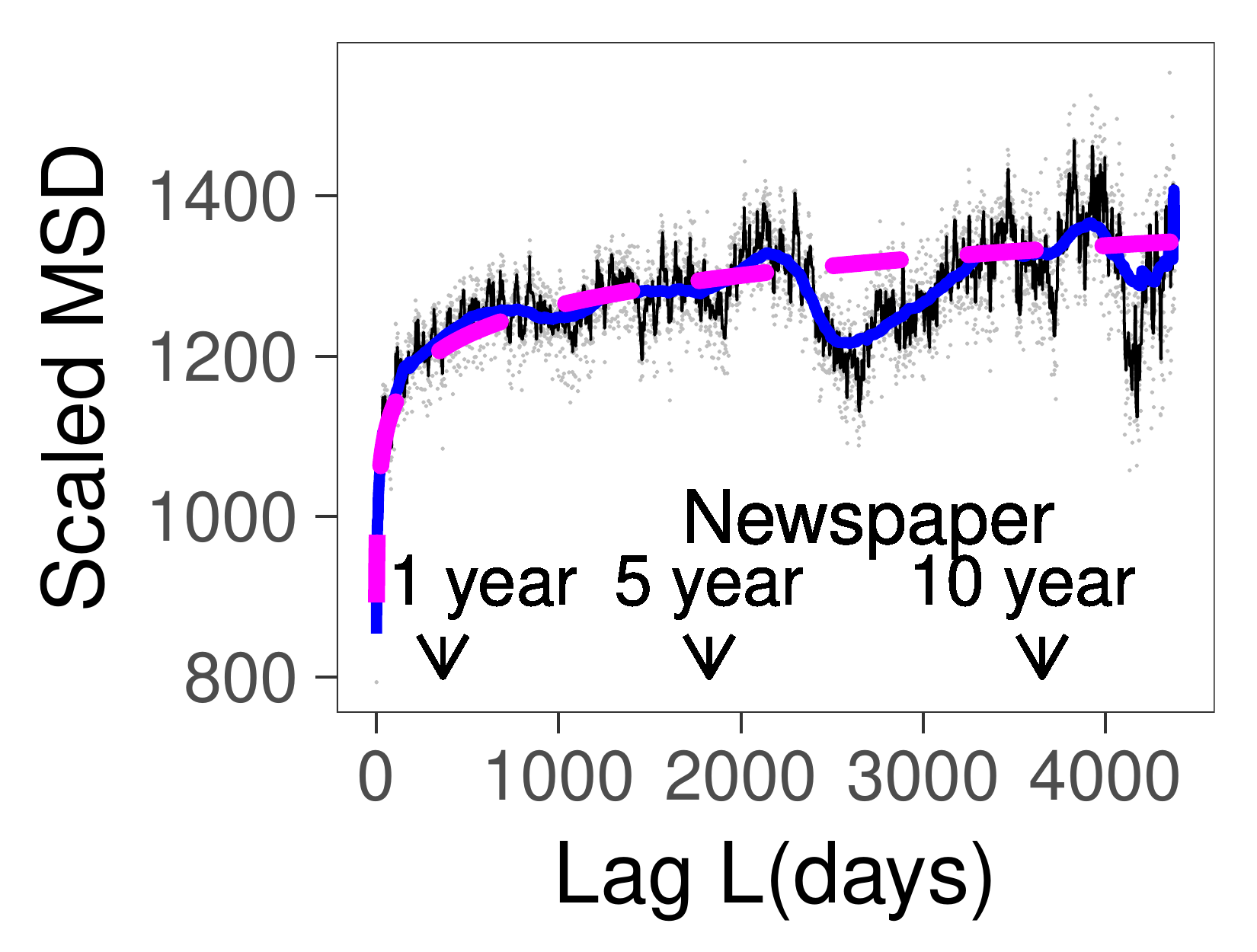}
\put(50,110){(a)}
\end{overpic}
\end{minipage}
\begin{minipage}{0.32\hsize}
\begin{overpic}[scale=0.35,tics=5]
{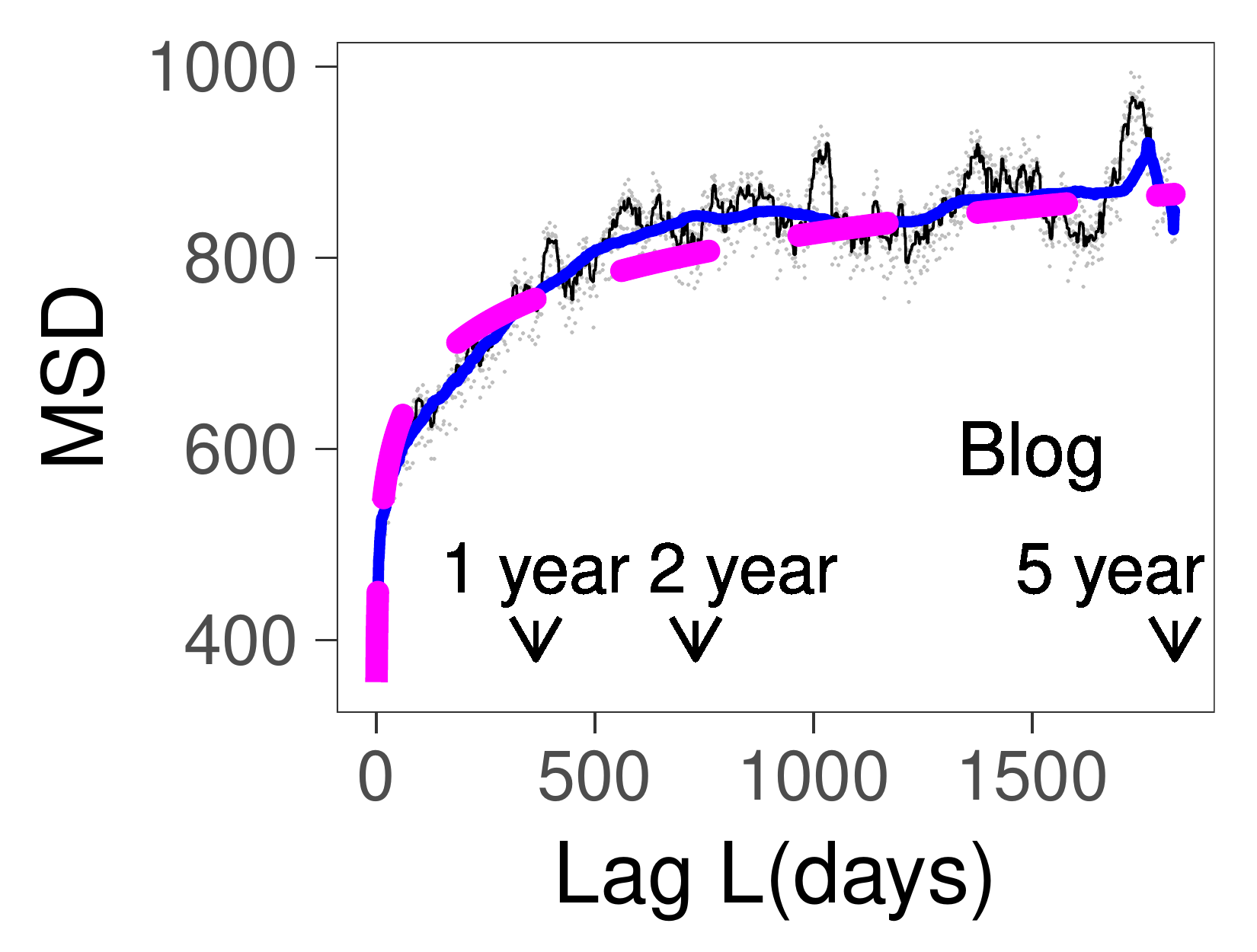}
\put(50,112){(b)}
\end{overpic}
\end{minipage}
\begin{minipage}{0.32\hsize}
\begin{overpic}[scale=0.35,tics=5]
{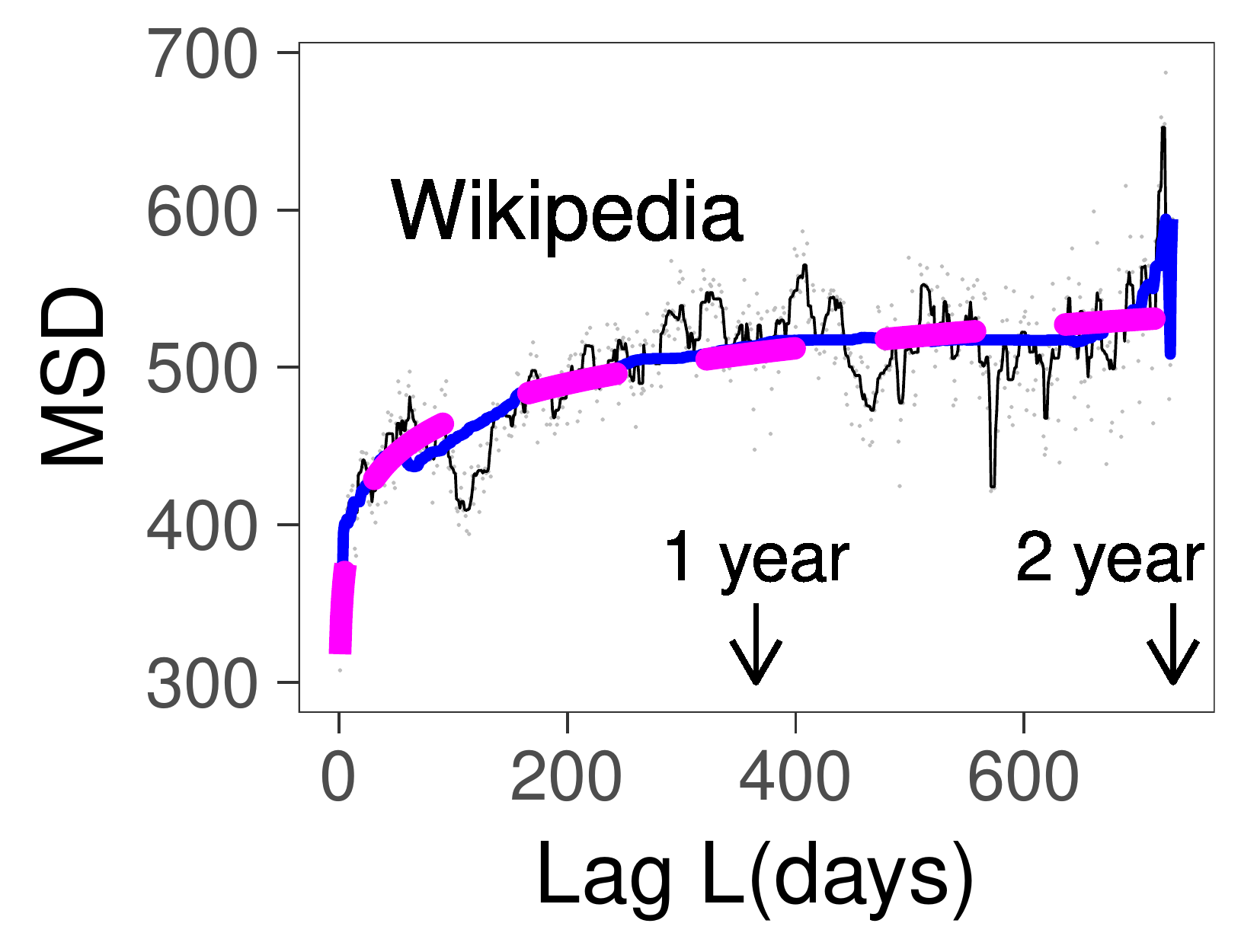}
\put(50,113){(c)}
\end{overpic}
\end{minipage}
\begin{minipage}{0.32\hsize}
\begin{overpic}[scale=0.35,tics=5]
{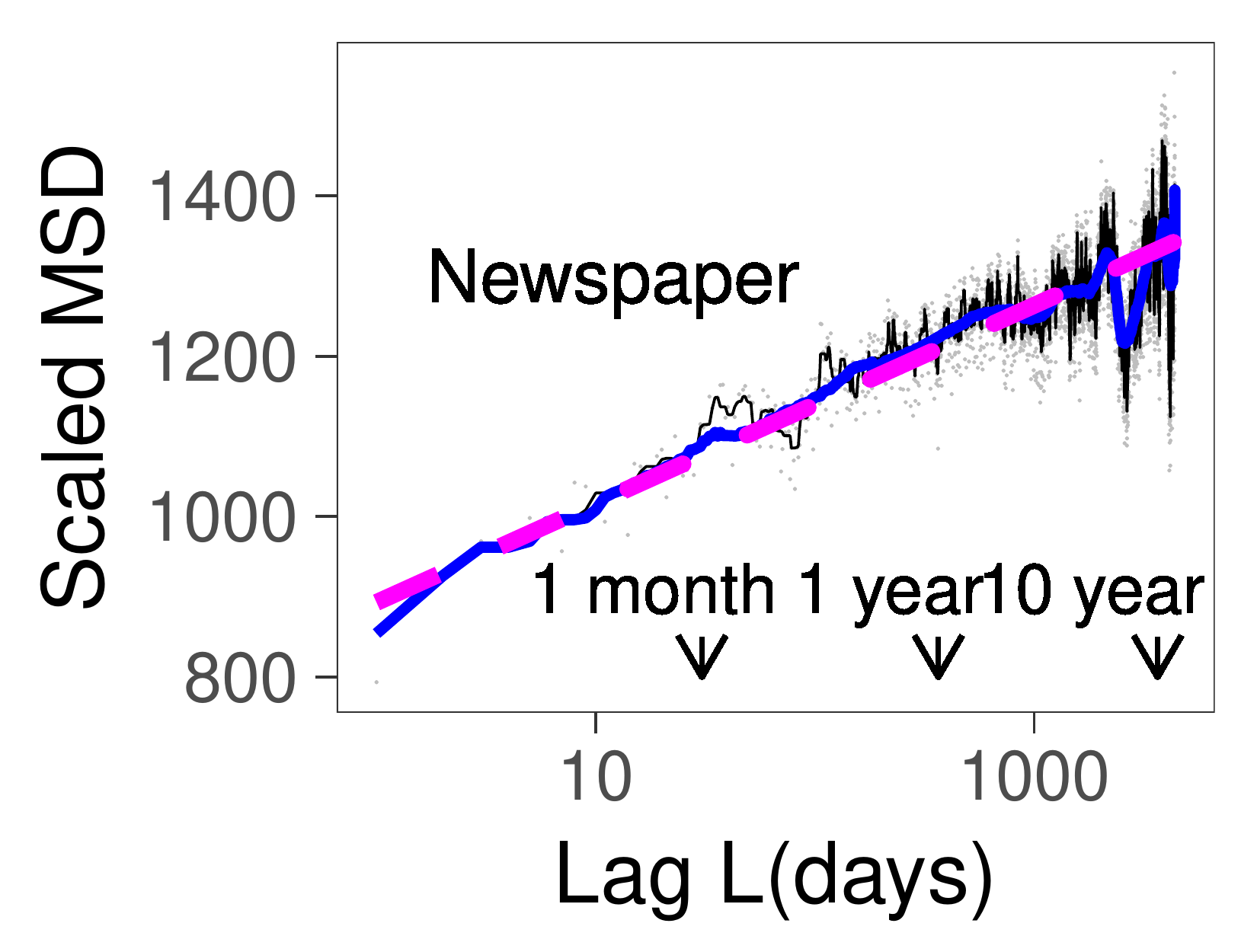}
\put(50,110){(d)}
\end{overpic}
\end{minipage}
\begin{minipage}{0.32\hsize}
\begin{overpic}[scale=0.35,tics=5]
{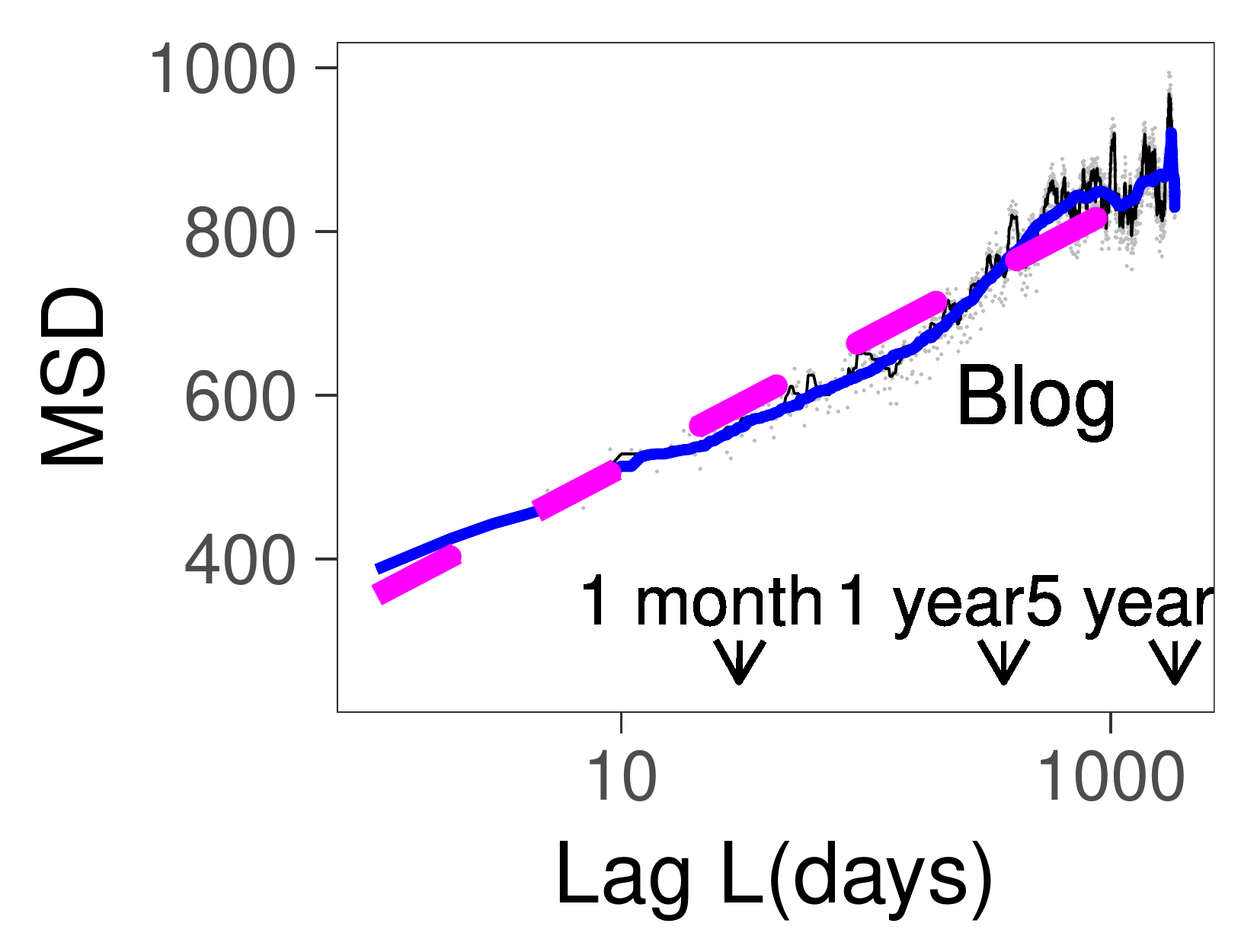}
\put(50,110){(e)}
\end{overpic}
\end{minipage}
\begin{minipage}{0.32\hsize}
\begin{overpic}[scale=0.35,tics=5]
{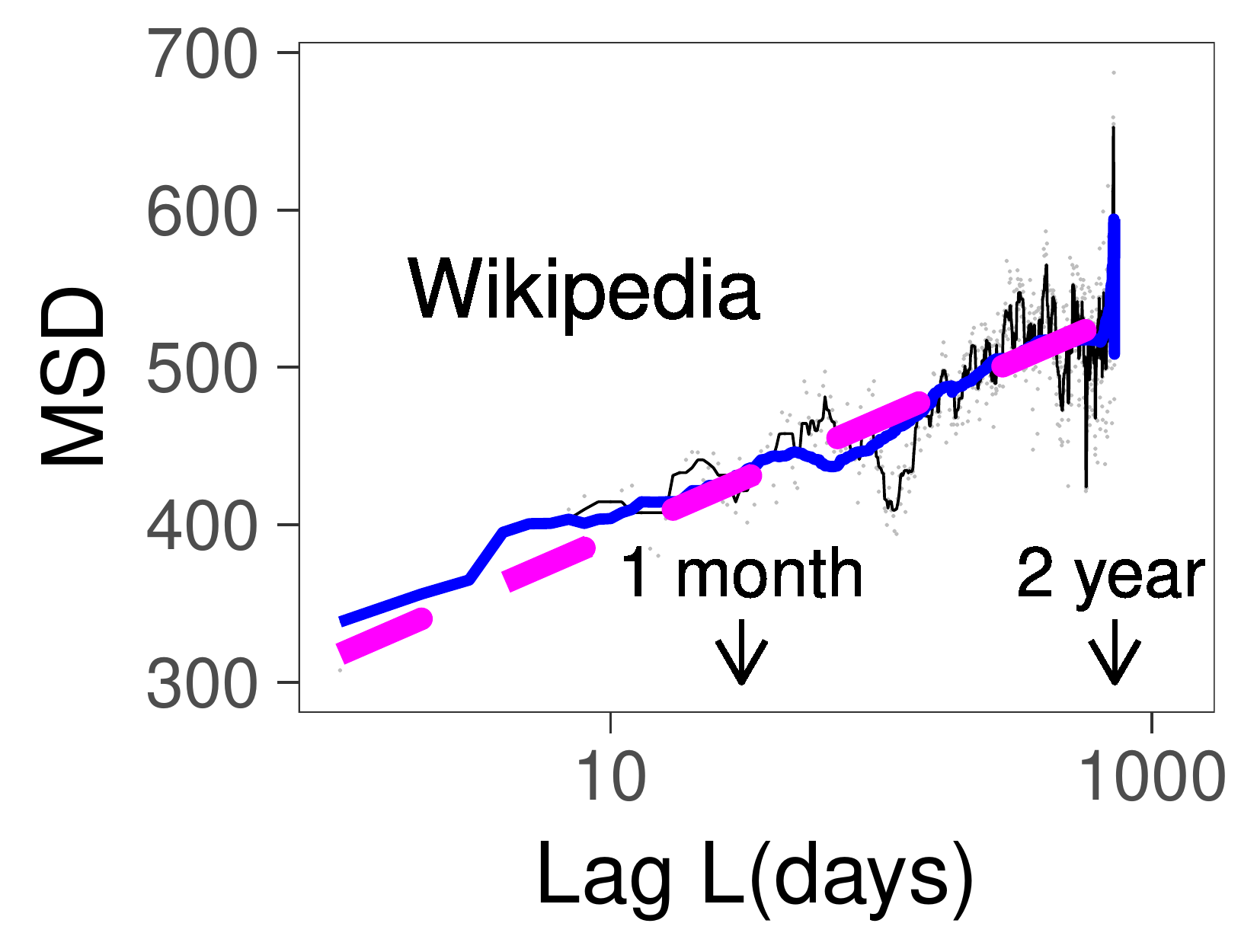}
\put(50,110){(f)}
\end{overpic}
\end{minipage}

		\caption{Empirical MSD $\sigma_j(L)^2$ given by Eq. \ref{eq_msd_base} for typical words. 
			The grey dots indicate the raw MSD, the thin black solid line indicates the corresponding 7-day moving median, and the thick solid  blue line the 365-day moving median.
			The magenta thick dashed line corresponds to the logarithmic curve $\sigma_j(L)=a_j\log(L)+b_j$ given by Eq. \ref{eq_msd_single}. 
			\textcolor{black}{Panel (a) shows the newspaper data for ``Tatiba (position or standpoint in English )''  ($a_j=66.6$ and $b_j=1449$), panel (b) the blog data  for ``Sanada (well-known Japanese family name)'' ($a_j=68.0$ and $b_j=356$), and panel (c) the English Wikipedia page views for ``Handle'' ($a_j=32.5$ and $b_j=317$).} 
			Panels (d), (e), and (f) are the corresponding figures on a semi-logarithmic scale.  
			The results in these figures confirm that the logarithmic curves substantially agree with the empirical data. }
		\label{single_msd}
	\end{figure*}
	
	\begin{figure*}
		\begin{minipage}{0.32\hsize}
			\begin{overpic}[scale=0.35,tics=5]
				{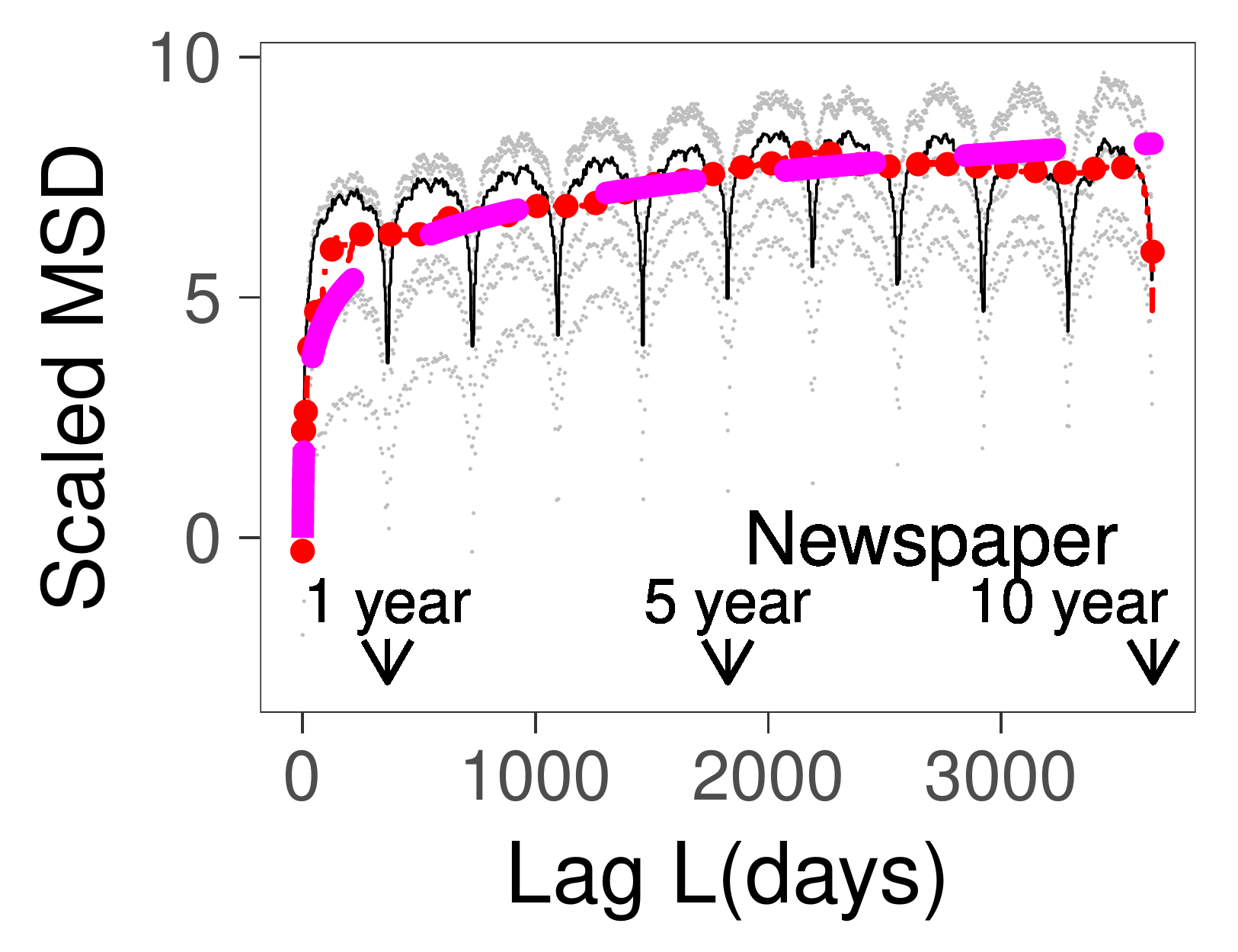}
				\put(42,110){(a)}
			\end{overpic}
		\end{minipage}
		\begin{minipage}{0.32\hsize}
			\begin{overpic}[scale=0.35,tics=5]
				{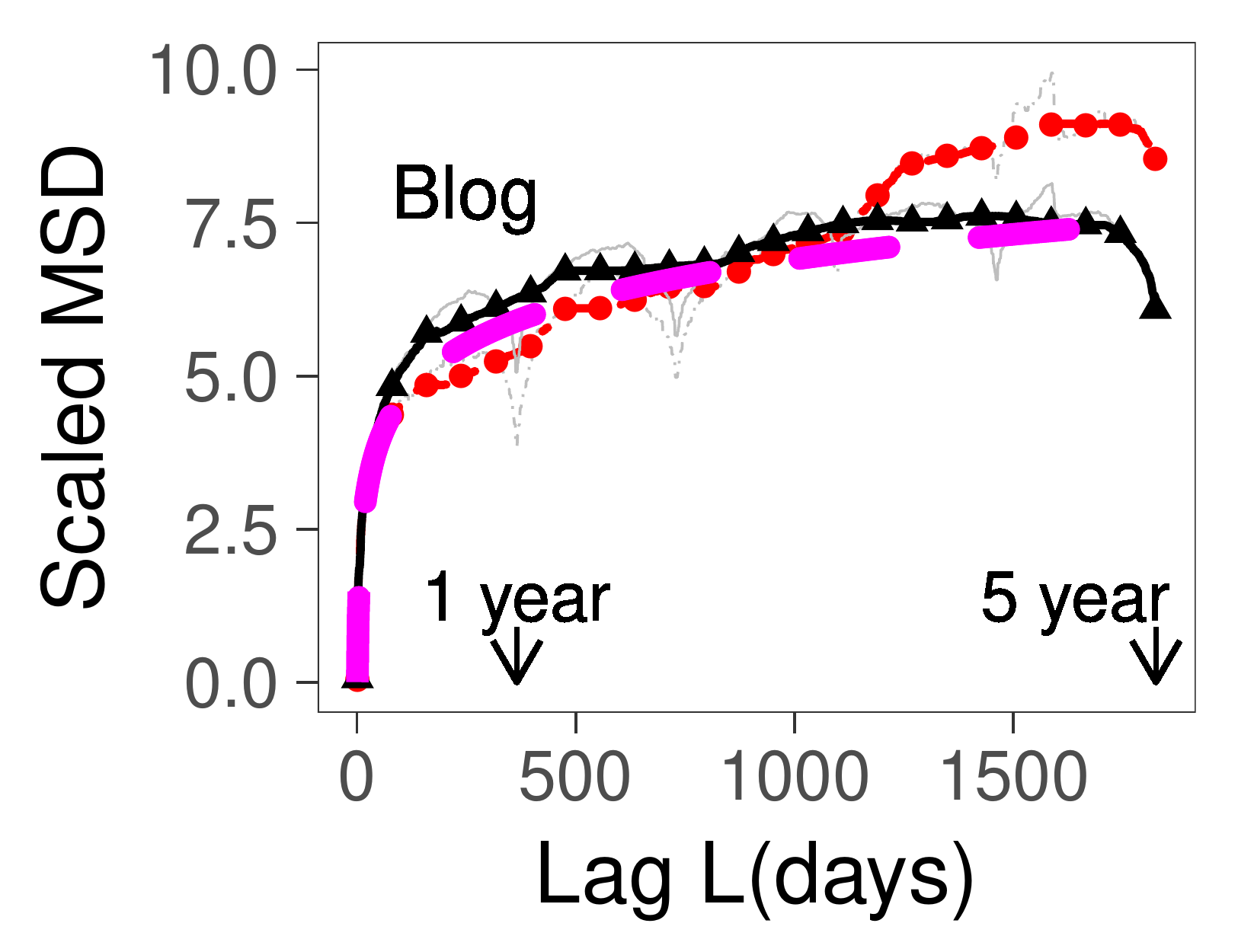}
				\put(46,110){(b)}
			\end{overpic}
		\end{minipage}
		\begin{minipage}{0.32\hsize}
			\begin{overpic}[scale=0.35,tics=5]
				{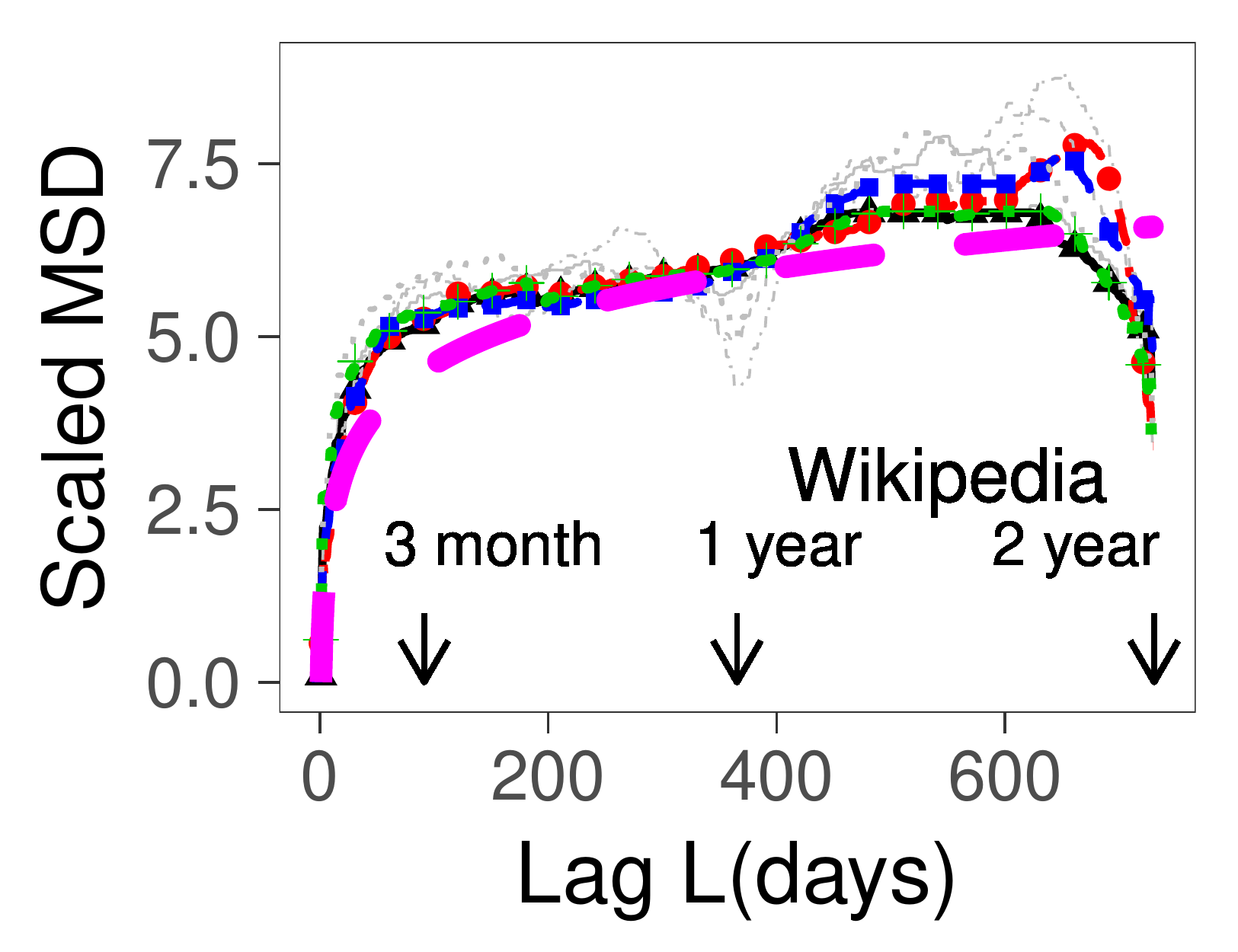}
				\put(42,110){(c)}
			\end{overpic}
		\end{minipage}
		\begin{minipage}{0.32\hsize}
			\begin{overpic}[scale=0.35,tics=5]
				{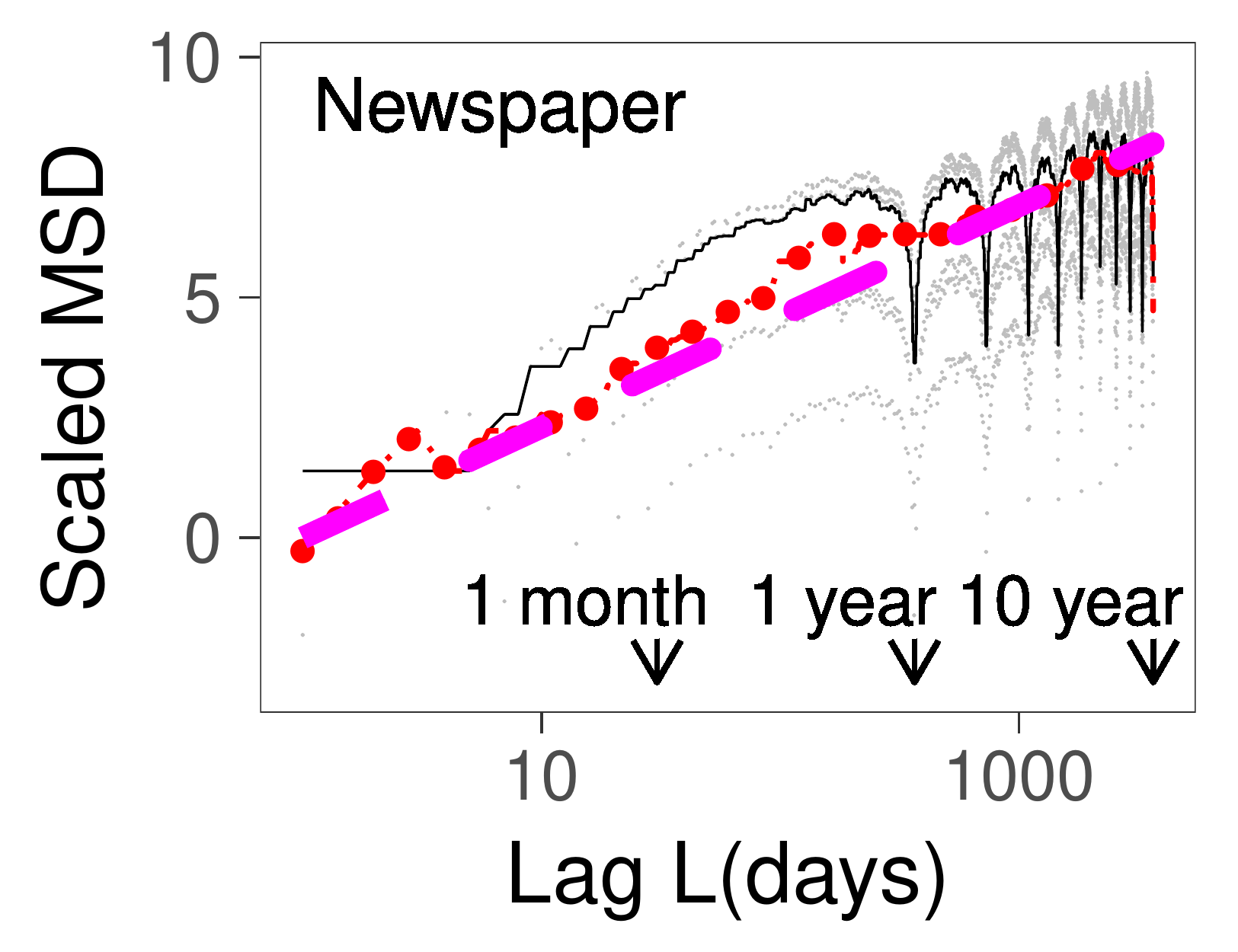}
				\put(42,102){(d)}
			\end{overpic}
		\end{minipage}
		\begin{minipage}{0.32\hsize}
			\begin{overpic}[scale=0.35,tics=5]
				{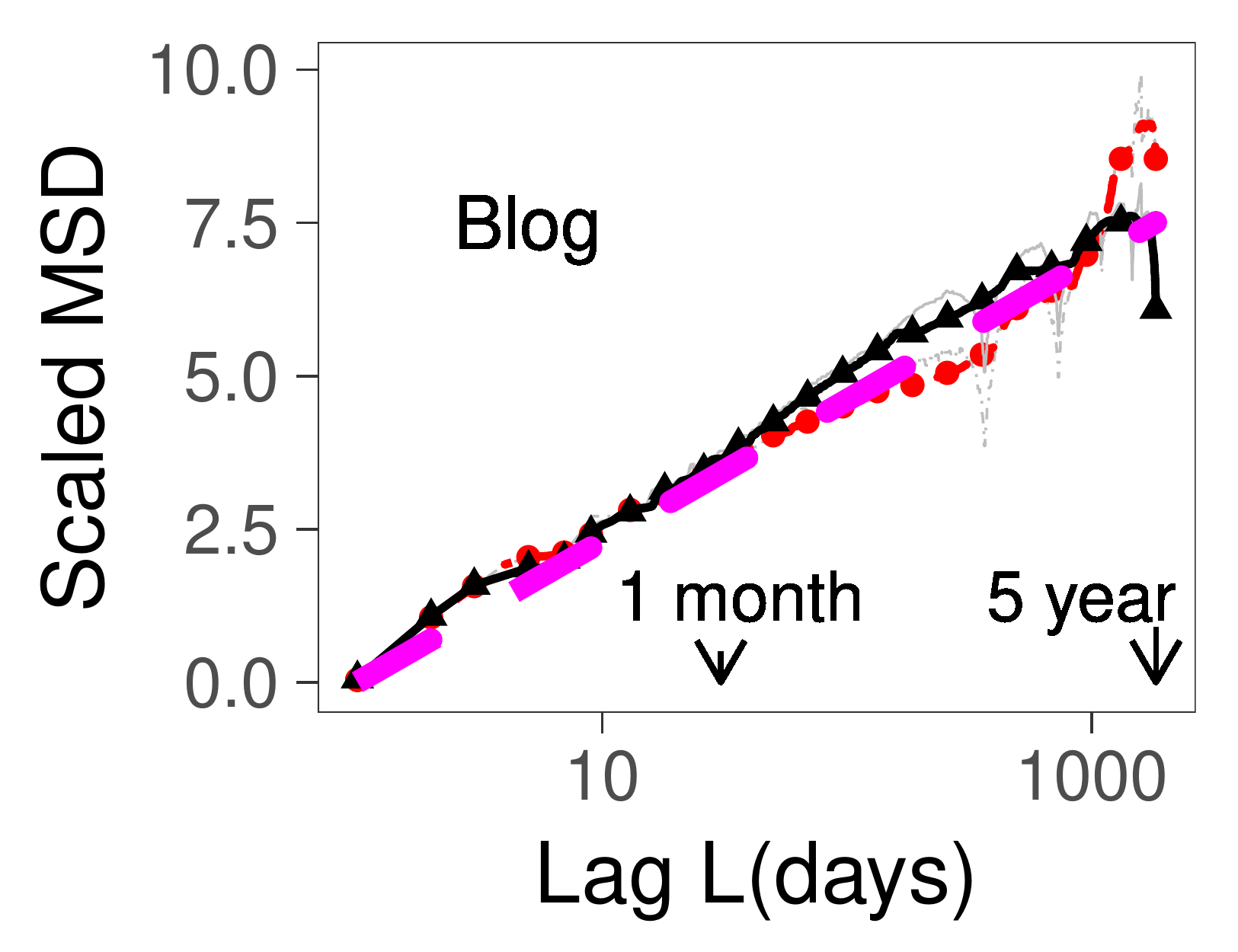}
				\put(45,113){(e)}
			\end{overpic}
		\end{minipage}
		\begin{minipage}{0.32\hsize}
			\begin{overpic}[scale=0.35,tics=5]
				{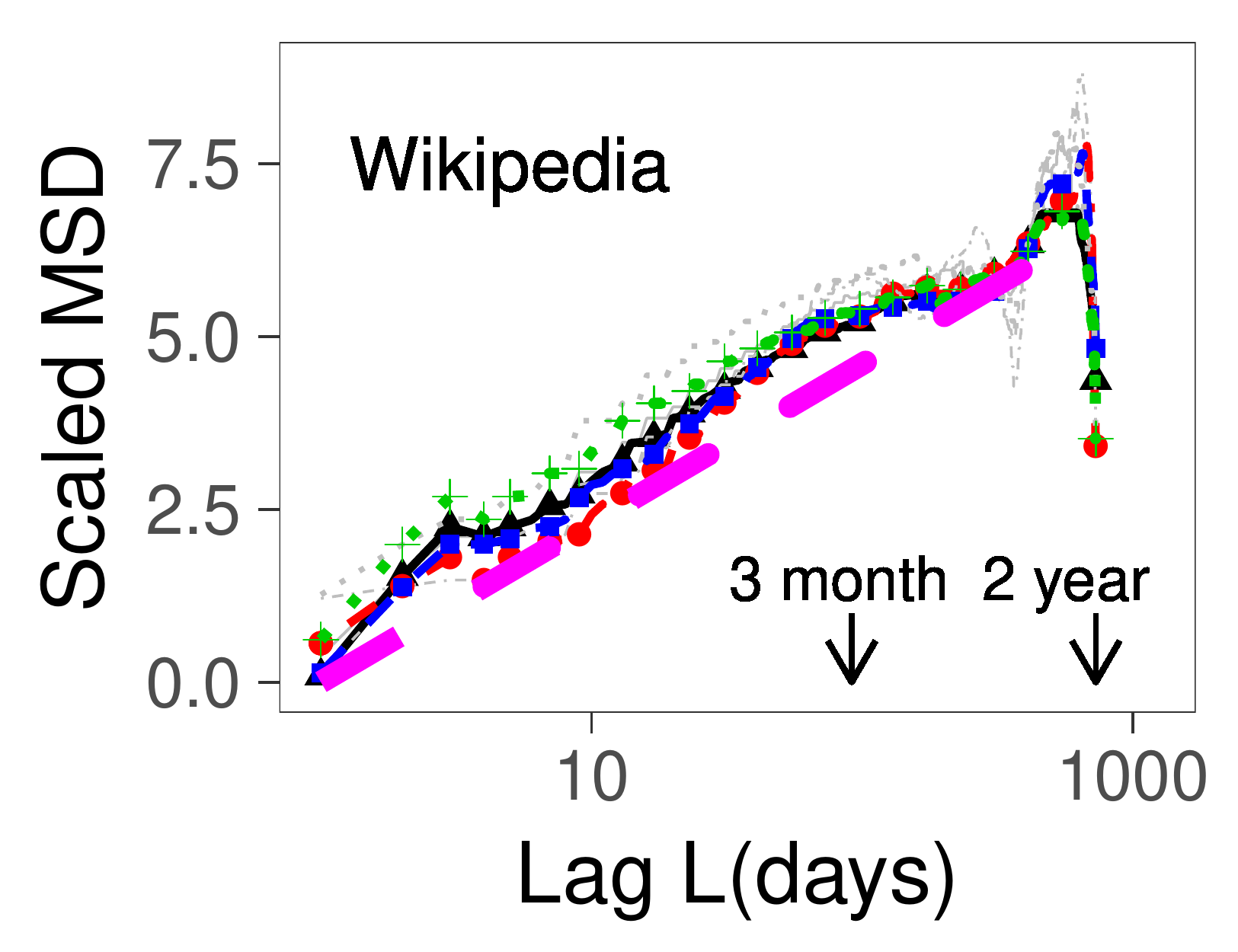}
				\put(42,113){(f)}
			\end{overpic}
		\end{minipage}
		\begin{minipage}{0.32\hsize}
			\begin{overpic}[scale=0.35,tics=5]
				{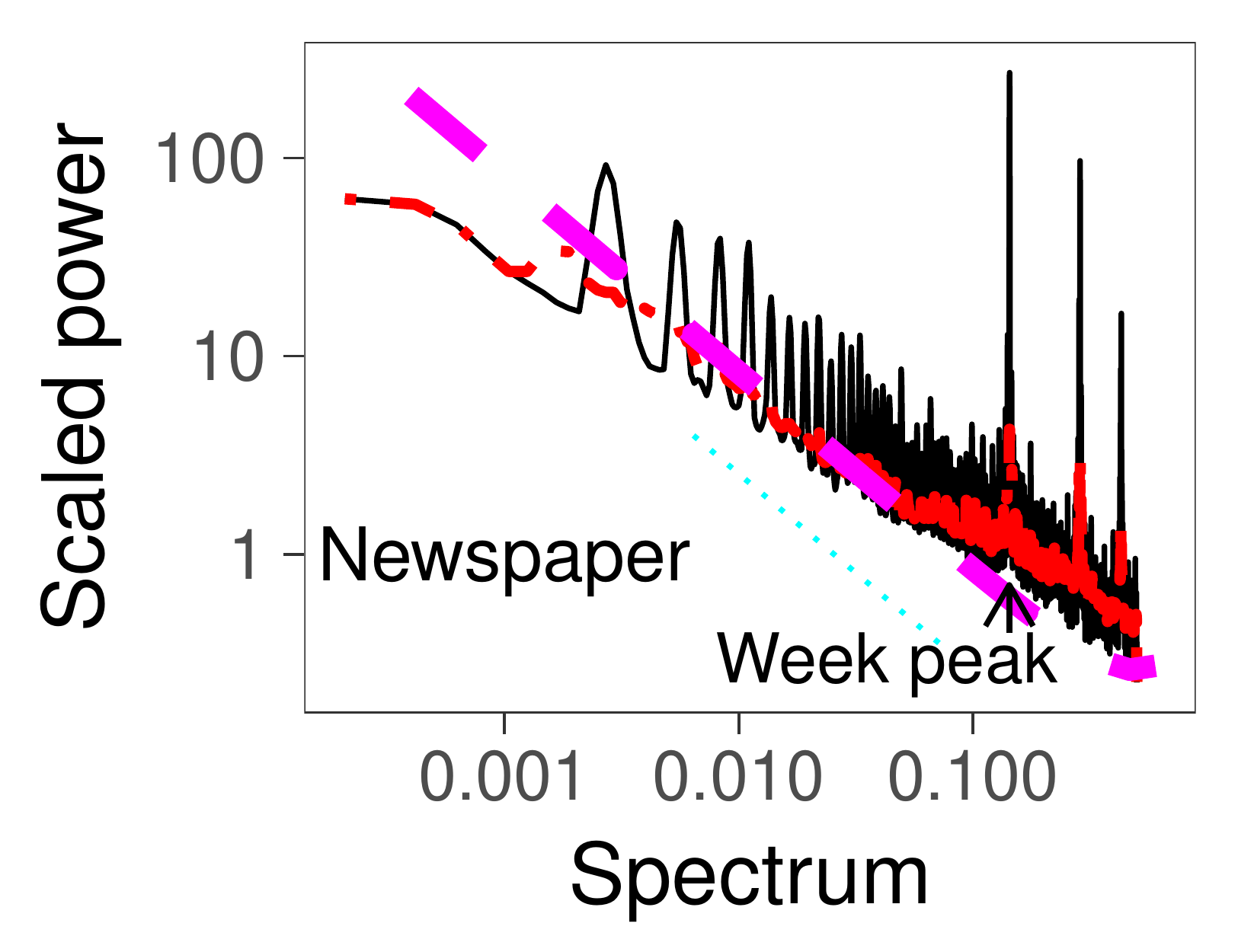}
				\put(64,113){(g)}
			\end{overpic}
		\end{minipage}
		\begin{minipage}{0.32\hsize}
			\begin{overpic}[scale=0.35,tics=5]
				{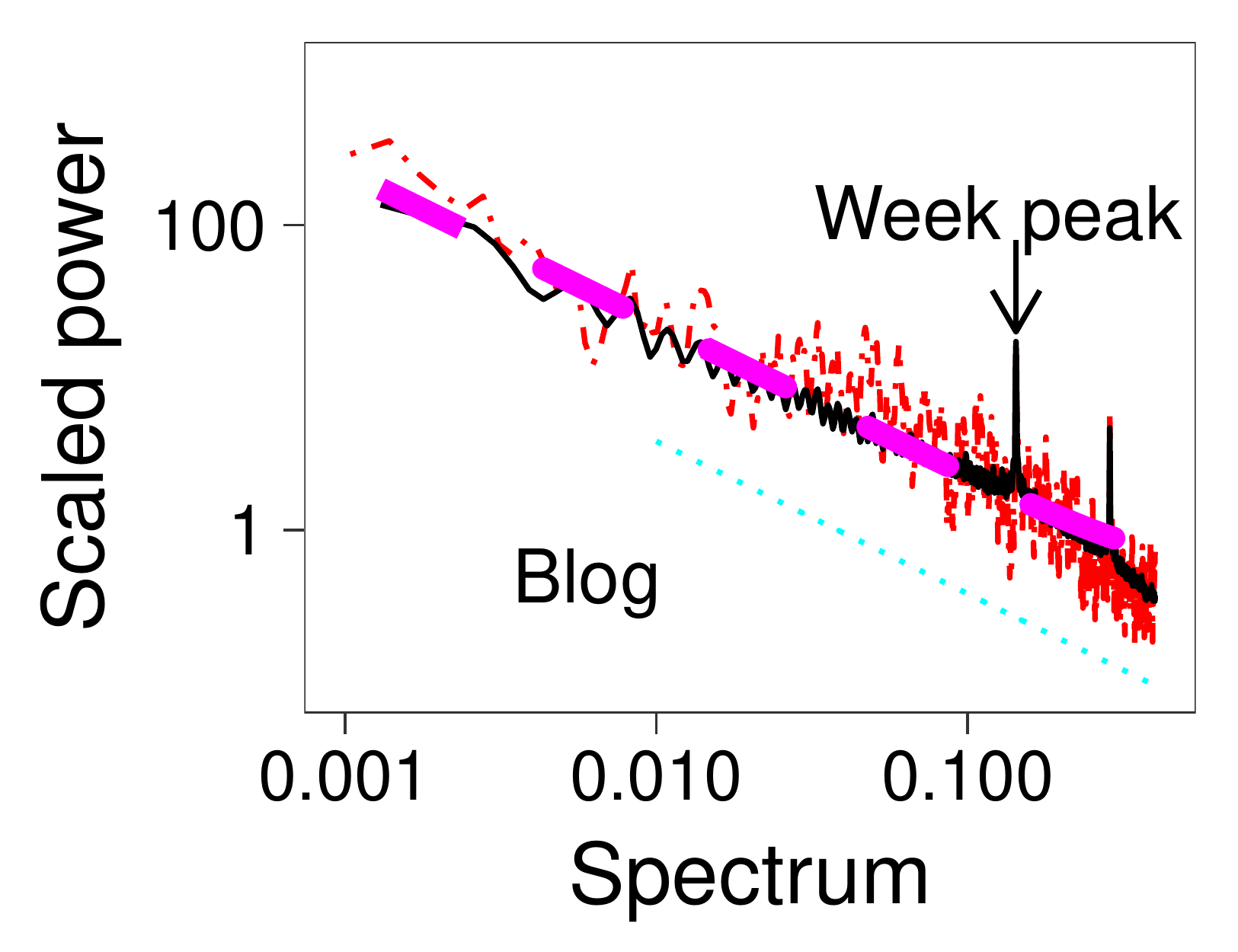}
				\put(50,112){(h)}
			\end{overpic}
		\end{minipage}
		\begin{minipage}{0.32\hsize}
			\begin{overpic}[scale=0.35,tics=5]
				{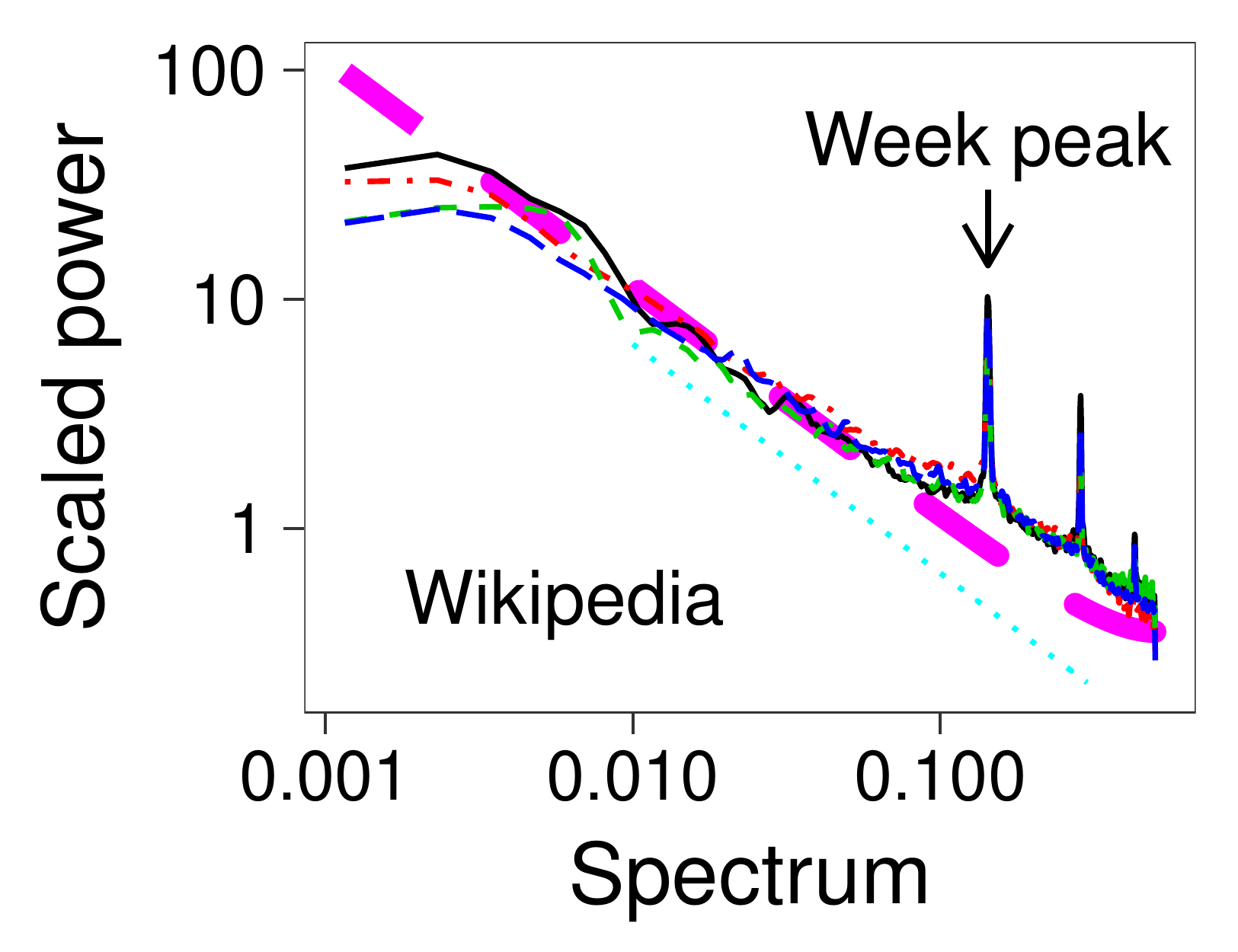}
				\put(60,113){(i)}
			\end{overpic}
		\end{minipage}
		\caption{(a-g) Ensemble median of the scaled MSD $\overline{\sigma'(L)}$ given by Eq. \ref{eq_msd_medi} for words with a mean $\hat{c}_j$ above 30 (excluding words with a small $\hat{c}_j$ because they have relatively large signal-to-noise ratios.). 
			The magenta thick dashed lines are the corresponding theoretical curves $\log(L)$. 
			(a)Newspaper data. The grey dots indicate the raw ensemble scaled MSD, the thin black solid line indicates the corresponding 7-day moving median, and the thick dash-dotted red line with circles is the 365-day moving median.
			(b)MSD for blog data. The grey thin line indicates the 7-day moving median of the ensemble scaled MSD for nouns (solid line) and for adjectives (dash-dotted line).
			The thick black line with triangles is the 365-day moving median for nouns and the red dash-dotted line with circles is that for adjectives. 
			(c)MSD for Wikipedia data. The grey thin lines indicate the 7-day moving median of the ensemble scaled MSD for English (solid line), French (dash-dotted line), Chinese (dotted line), and Japanese (long-dashed line).
			The thick lines indicate the 365-day moving median of the ensemble scaled MSD for English (black solid line with triangles), French (red dash-dotted line with circles), Chinese (green dotted line with plus signs), and Japanese (blue long-dashed line with squares).
			The results in these figures confirm that the theoretical curve substantially agrees with the corresponding empirical data. 
			Note that in the case of the newspaper, $\sigma_j^2(1)$ given in definition Eq. \ref{eq_msd_scale} was replaced with the corresponding 365-day moving median $MovingMedian365[{\sigma_j}^2](1)$ to avoid the effect of strong weekly and annual cycles.
			\\
			(h-k)Power spectral density analysis. The ensemble median of the word-independent normalised spectral density for words with a mean $\hat{c}_j$ above 30, $\overline{P_{f'}(\nu)}$ given by Eq. \ref{eq_def_ensemble_peri}. The magenta thick dashed lines are the theoretical curve given by Eq. \ref{eq_ensemble_peri} and the cyan guidelines are 
			$\propto 1/\nu$. (h) Newspaper data. The black solid line is the raw normalised spectral density $\overline{P_{f'}(\nu)}$ and the red dash-dotted line is the corresponding 31-point moving median. (i) Blog data. The line is the spectral density  $\overline{P_{f'}(\nu)}$ for nouns (black solid line) and for adjectives (red dash-dotted line). (j) Wikipedia data.  The line is the spectral density  $\overline{P_{f'}(\nu)}$ for English (black solid line), French (red dash-dotted line), Chinese (green short-dashed line), and Japanese (blue long-dashed line).  
		}
		\label{ensemble_msd}
	\end{figure*}

	\begin{figure*}
	\begin{minipage}{0.32\hsize}
	\begin{overpic}[scale=0.33,tics=5]
		{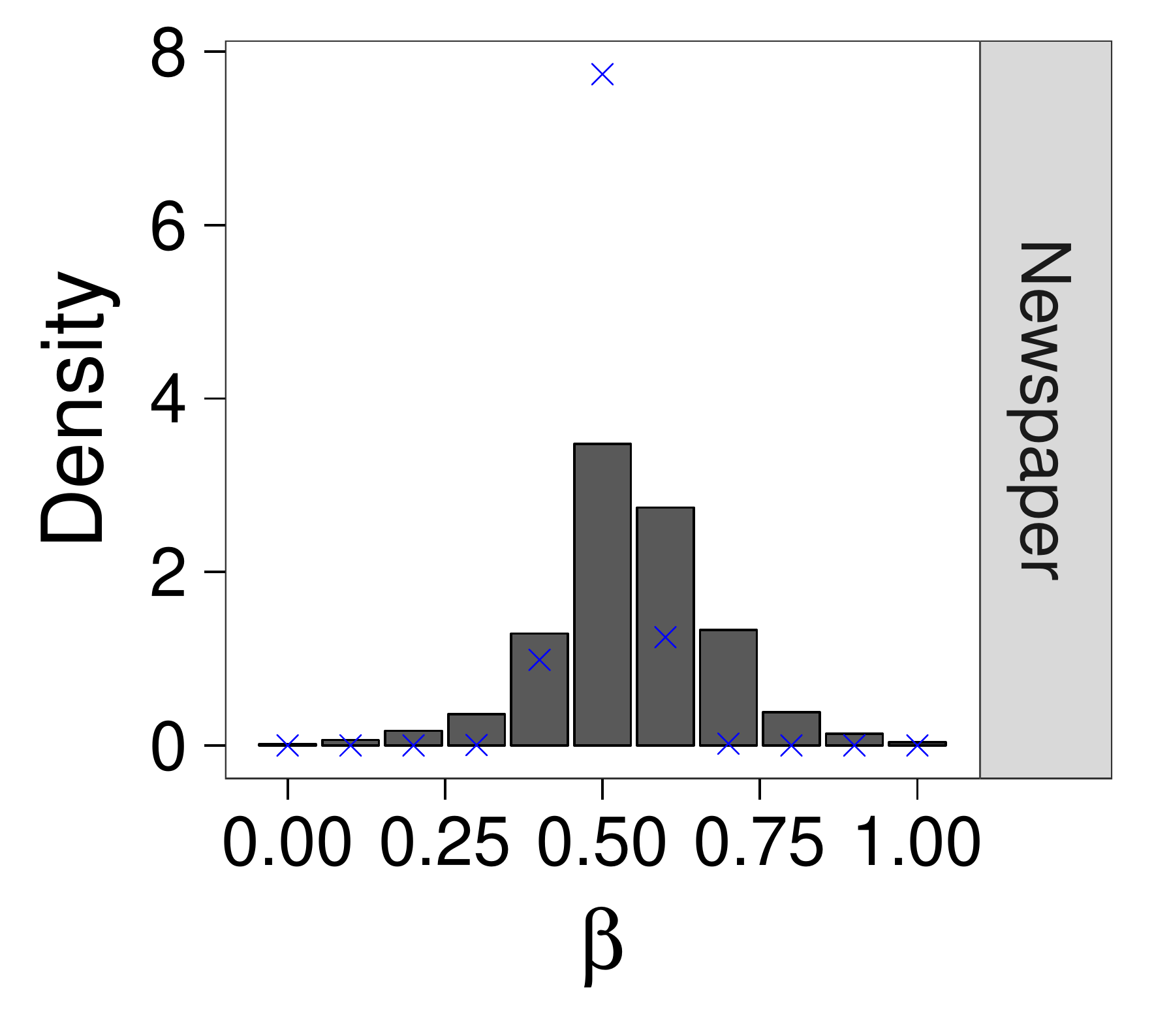}
		\put(37,135){(a)}
	\end{overpic}
	\end{minipage}
	\begin{minipage}{0.32\hsize}
		\begin{overpic}[scale=0.34,tics=5]
			{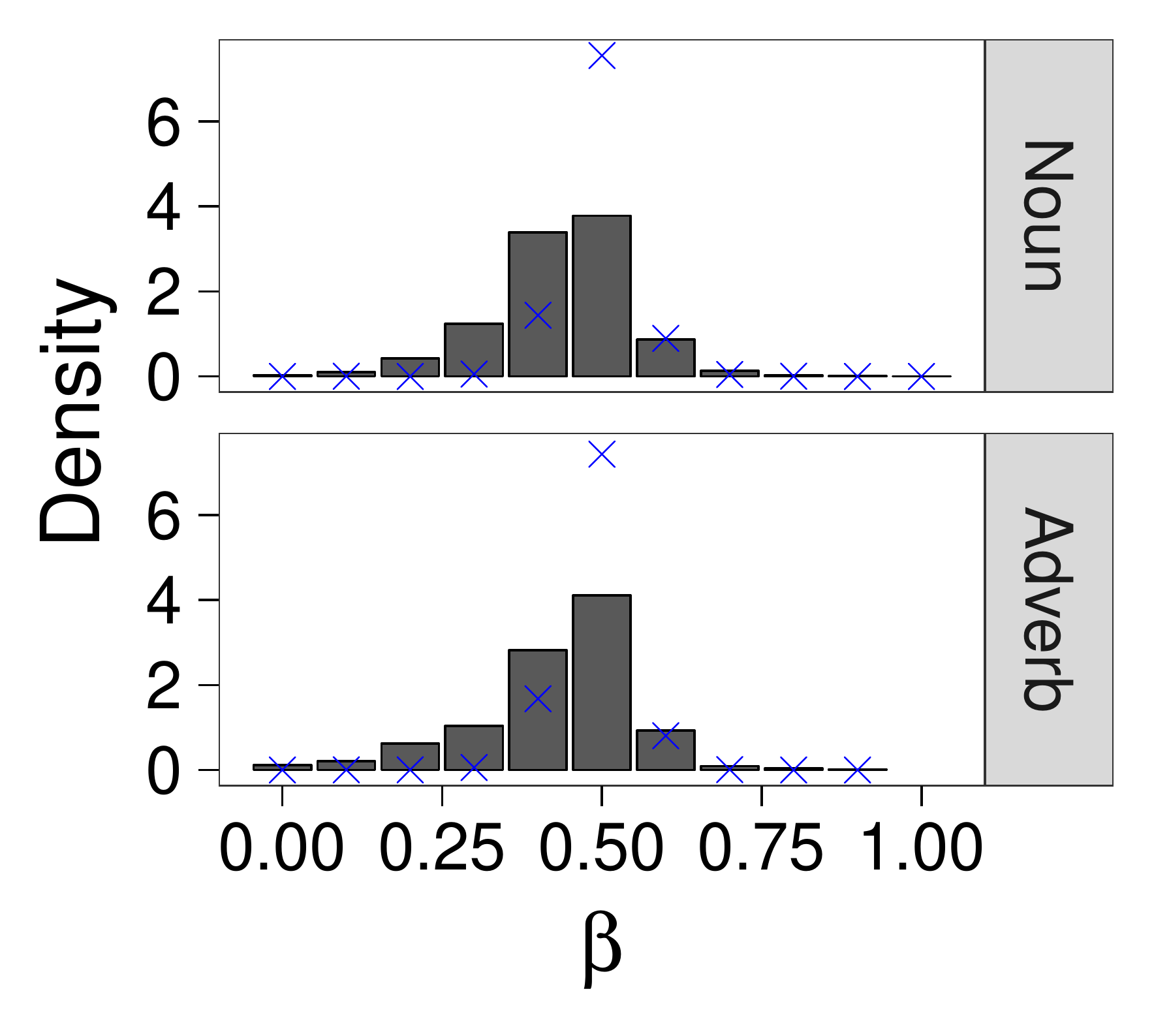}	
			\put(45,140){(b)}
		\end{overpic}
	\end{minipage}
	\begin{minipage}{0.32\hsize}
		\begin{overpic}[scale=0.37,tics=5]
			{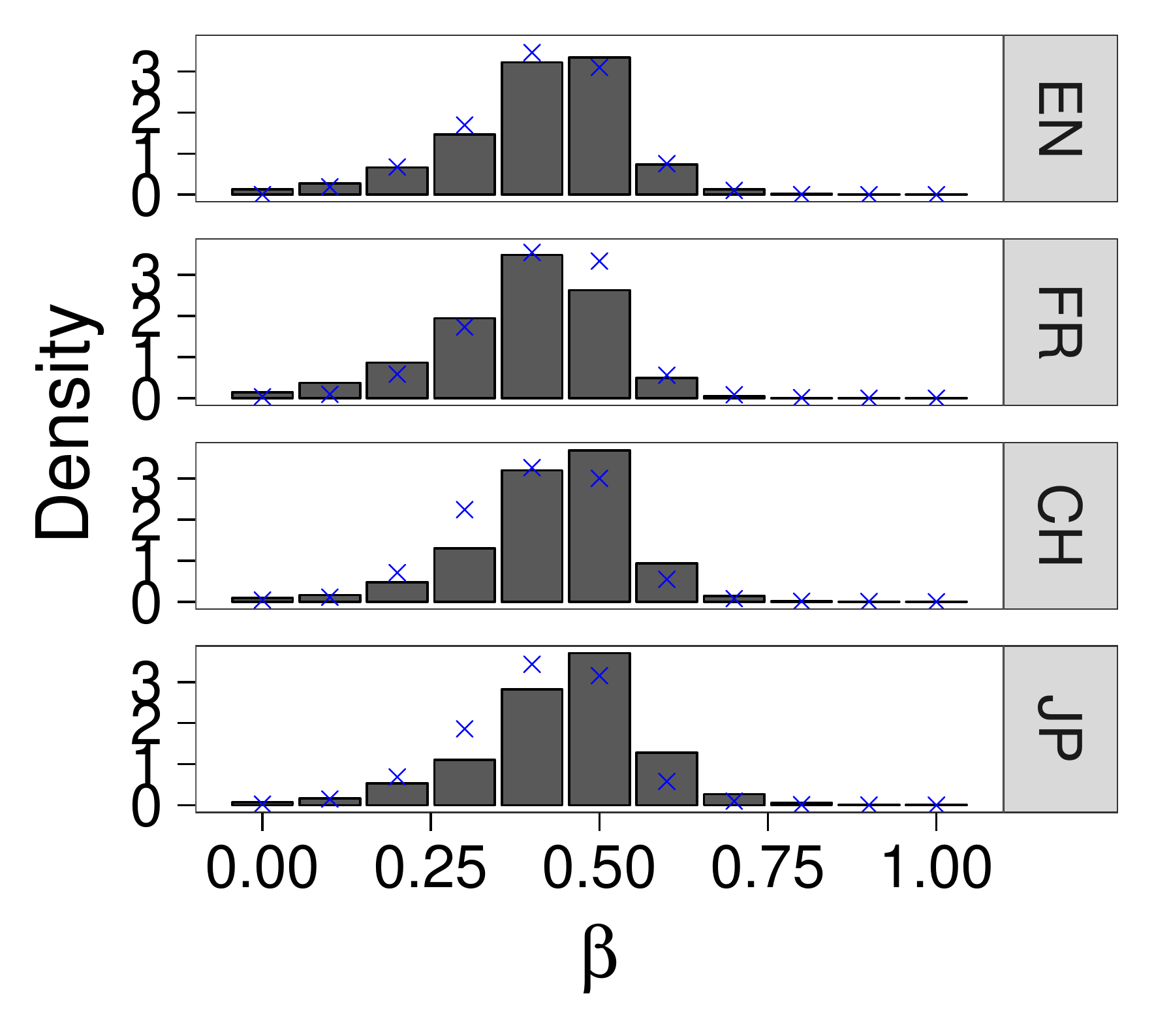}	
			\put(35,153){(c)}
		\end{overpic}
	\end{minipage}	
	\caption{\textcolor{black}{Histograms of estimated forgetting exponent $\hat{\beta}_j$ in the model described by Eq. \ref{eq_rw} and Eq. \ref{eq_rd}
	for words with a mean $\hat{c}_j$ above 30.  
	Herein we estimate $\hat{\beta}_j$ of individual words and subsequently construct the histograms. 
	 Details of the estimation method are provided in Appendix \ref{app_beta}. The data are shown in (a) for the newspaper, (b)the nouns (first row) and the adjectives (second row) for the blog, (c)English (first row), French (second row), Chinese (third row), and Japanese (fourth row) from Wikipedia articles, with all histograms standardized.  From these figures, we can confirm that the modes are approximately $\hat{\beta}_j \approx 0.5$ for all datasets. 
	 The blue crosses are the reference histograms of estimated $\hat{\beta}_j$ for the corresponding numerical simulations of the model given by Eqs. \ref{eq_rw} and \ref{eq_rd} in which the parameter are $\beta=0.5$, $\hat{c}_j=100$, $\hat{\eta}_j=0.09$ and $\hat{\delta}_j=0.07$  (the sampling number is 2000).	
	 }
	}
	\label{fig_beta}	
	\end{figure*}

	\begin{figure*}
		\begin{minipage}{0.32\hsize}
\begin{overpic}[scale=0.35,tics=5]
{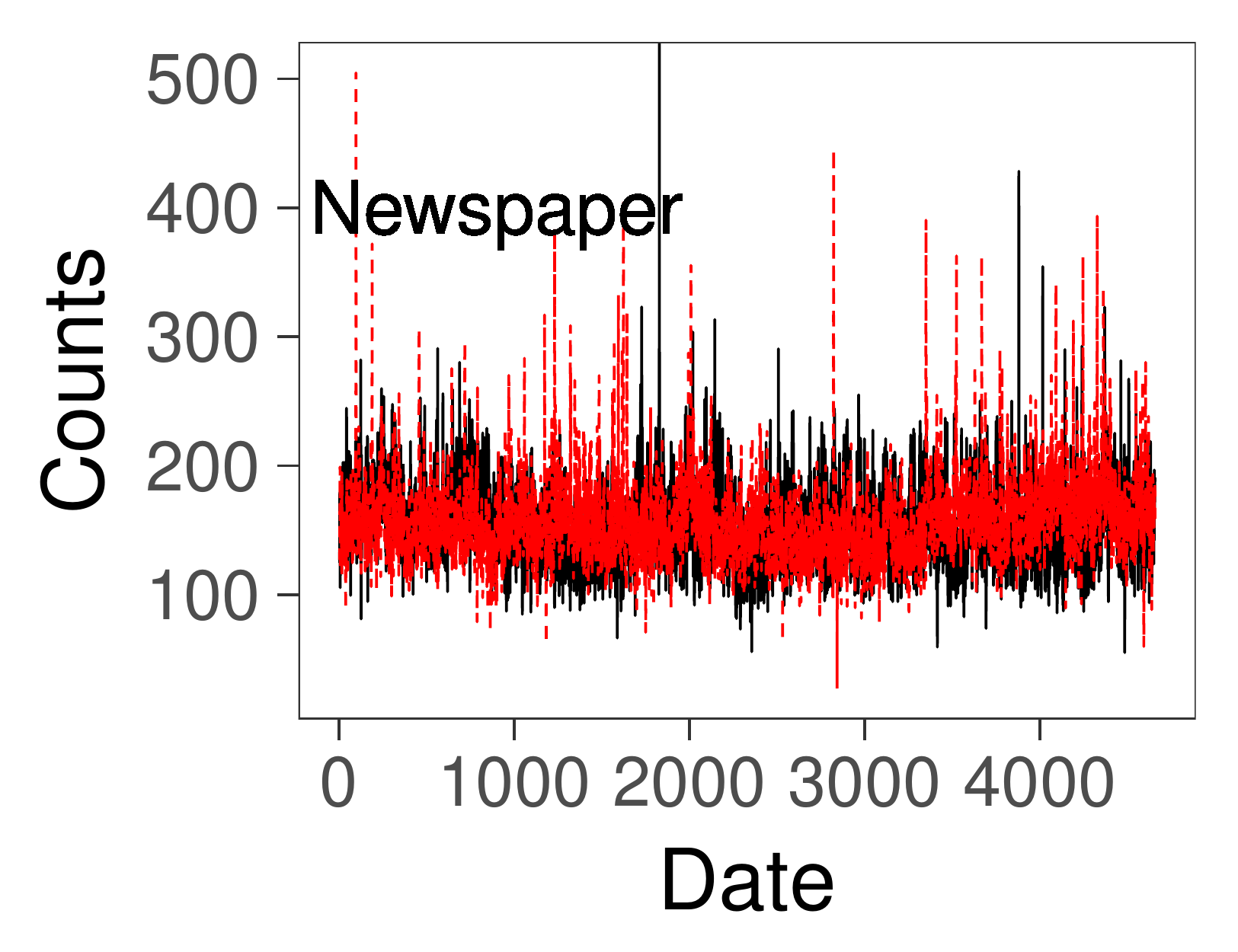}
\put(52,111){(a)}
\end{overpic}
\end{minipage}
\begin{minipage}{0.32\hsize}
\begin{overpic}[scale=0.35,tics=5]
{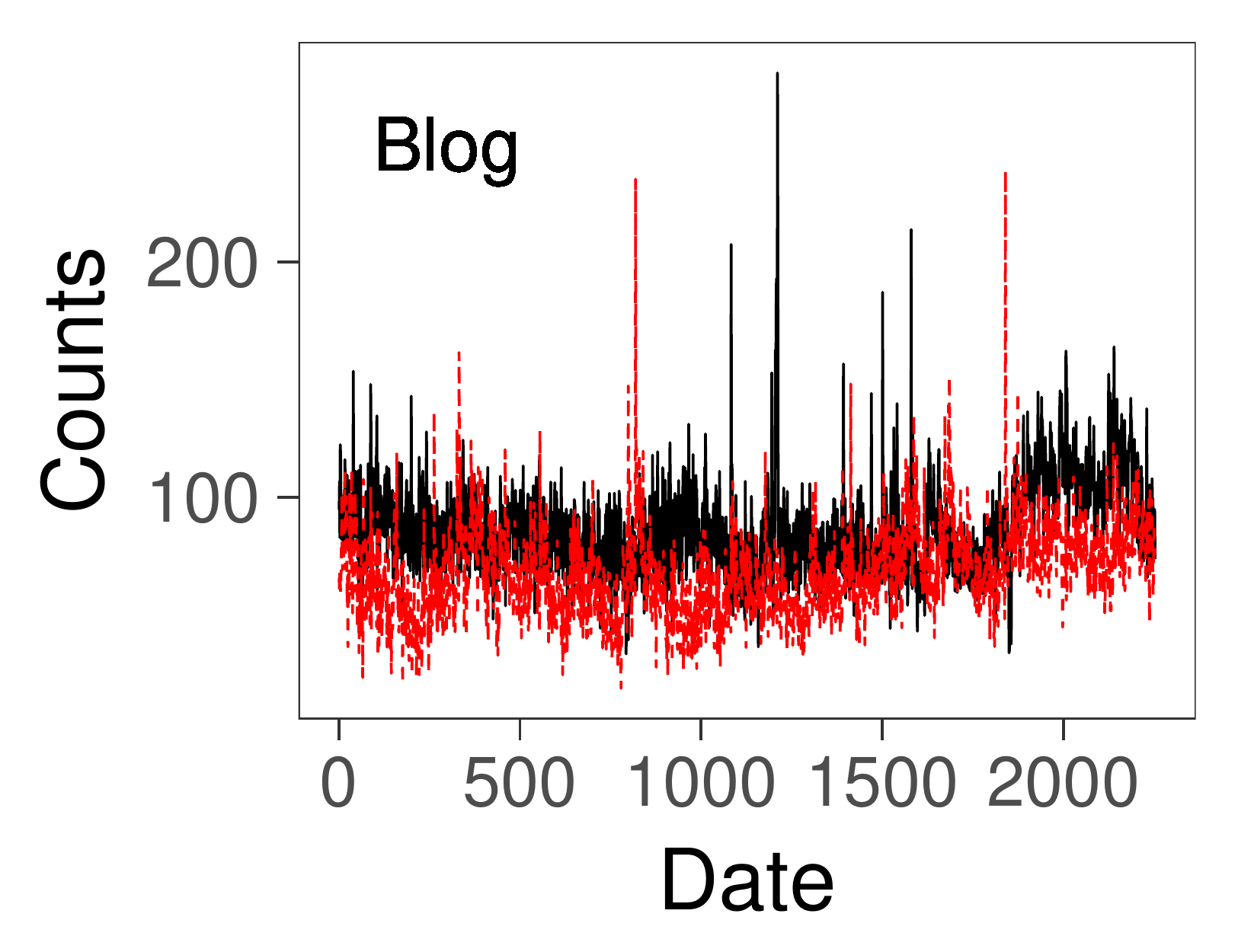}
\put(125,111){(b)}
\end{overpic}
\end{minipage}
\begin{minipage}{0.32\hsize}
\begin{overpic}[scale=0.35,tics=5]
{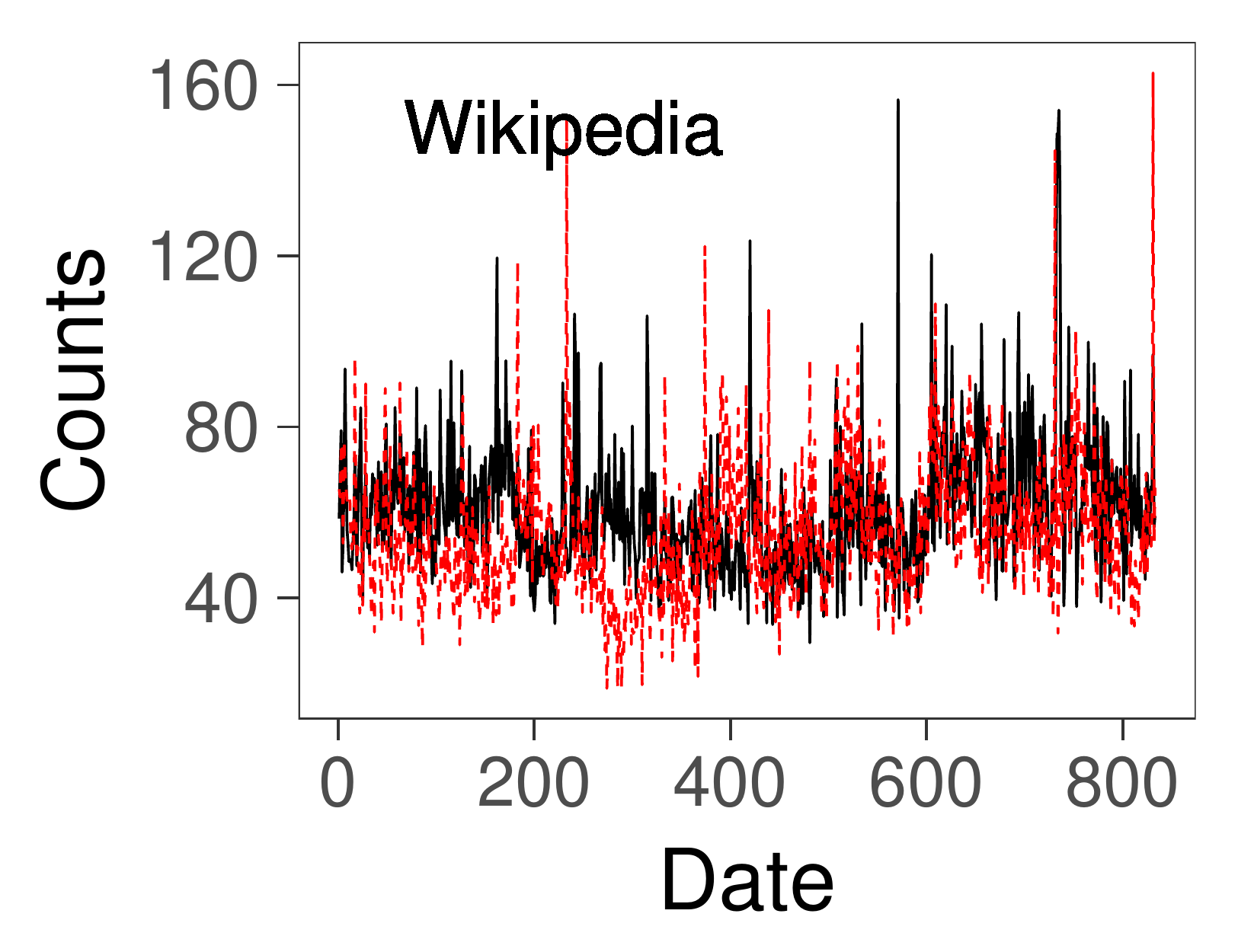}
\put(128,112){(c)}
\end{overpic}
\end{minipage}
\begin{minipage}{0.32\hsize}
\begin{overpic}[scale=0.35,tics=5]
{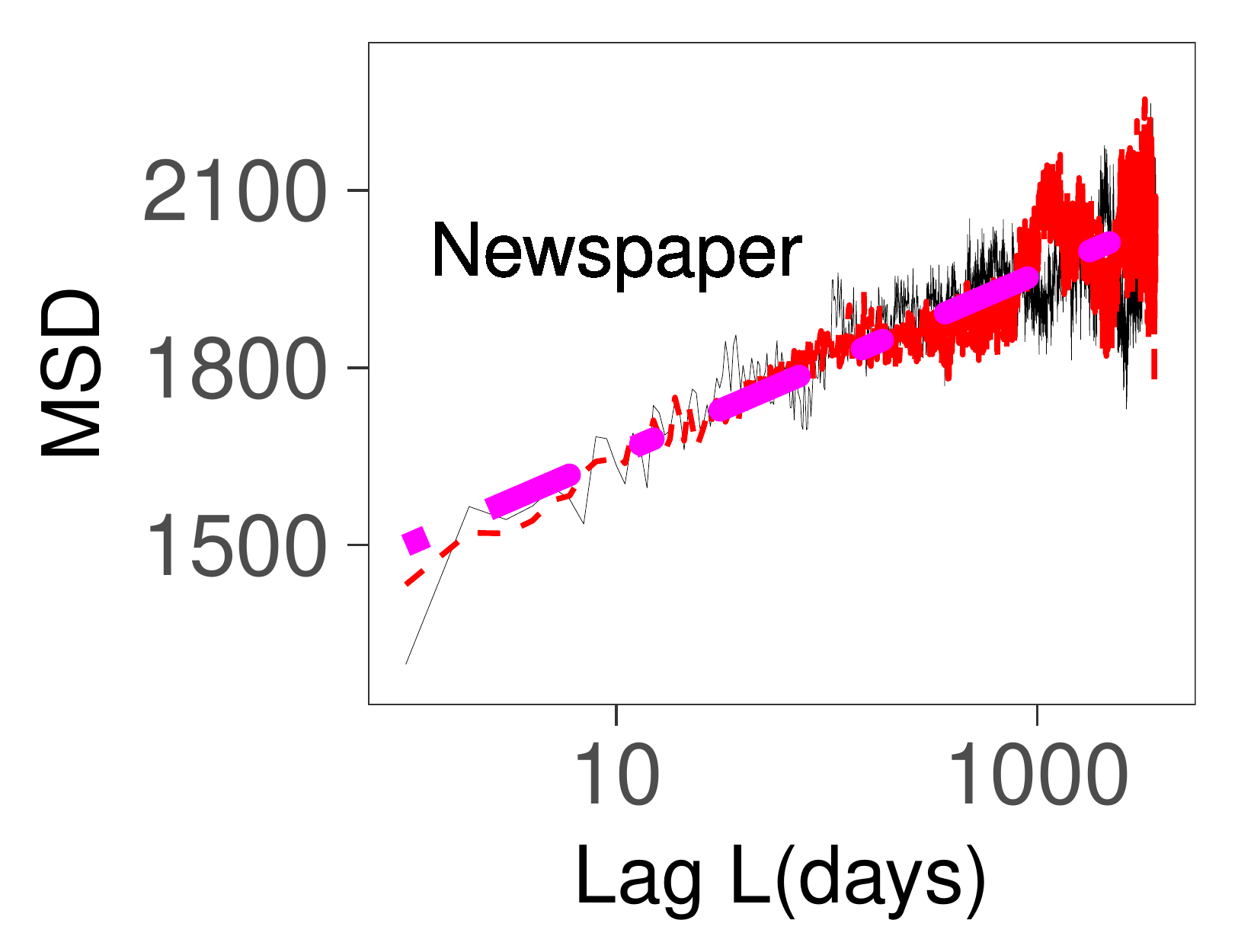}
\put(52,111){(d)}
\end{overpic}
\end{minipage}
\begin{minipage}{0.32\hsize}
\begin{overpic}[scale=0.35,tics=5]
{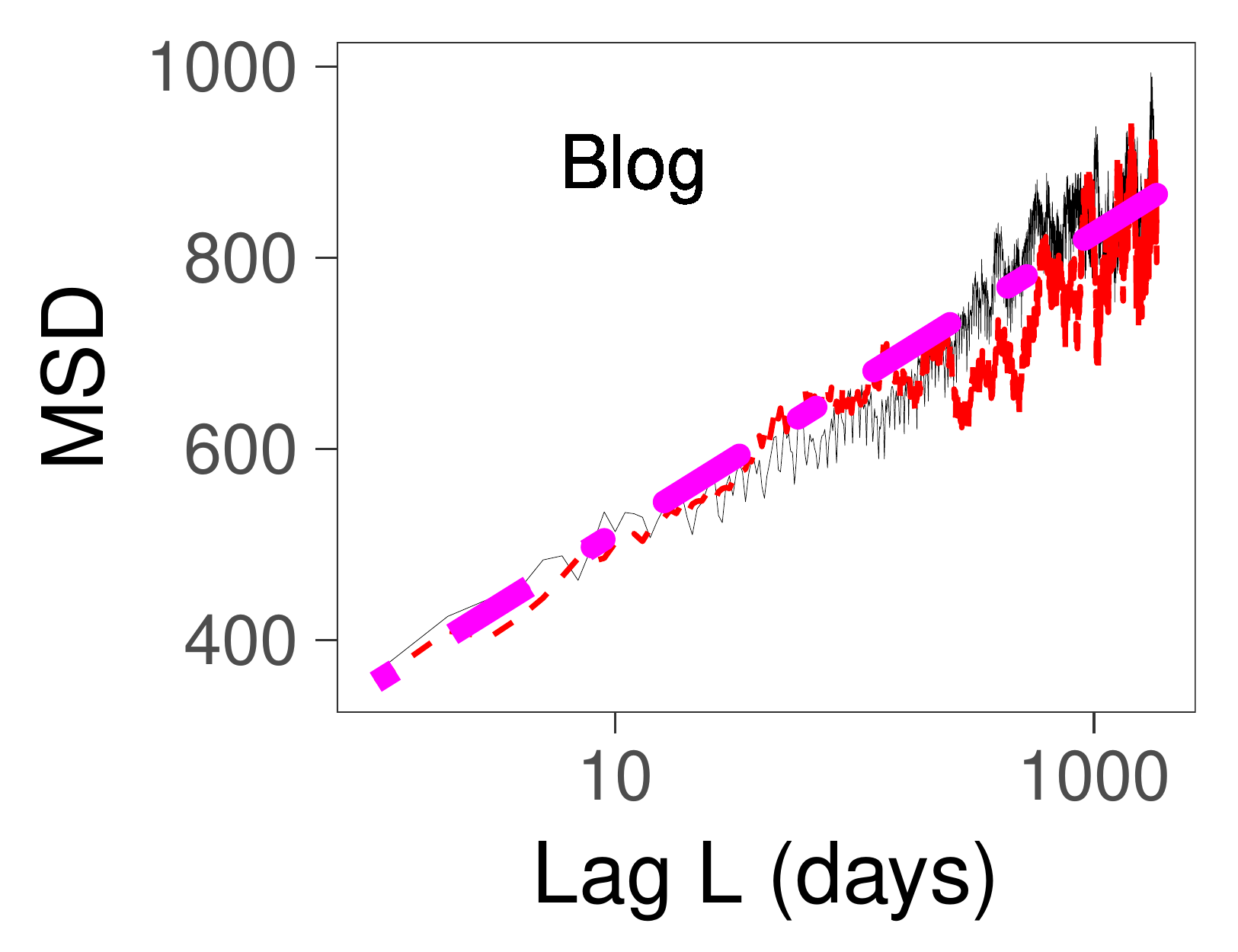}
\put(48,110){(e)}
\end{overpic}
\end{minipage}
\begin{minipage}{0.32\hsize}
\begin{overpic}[scale=0.35,tics=5]
{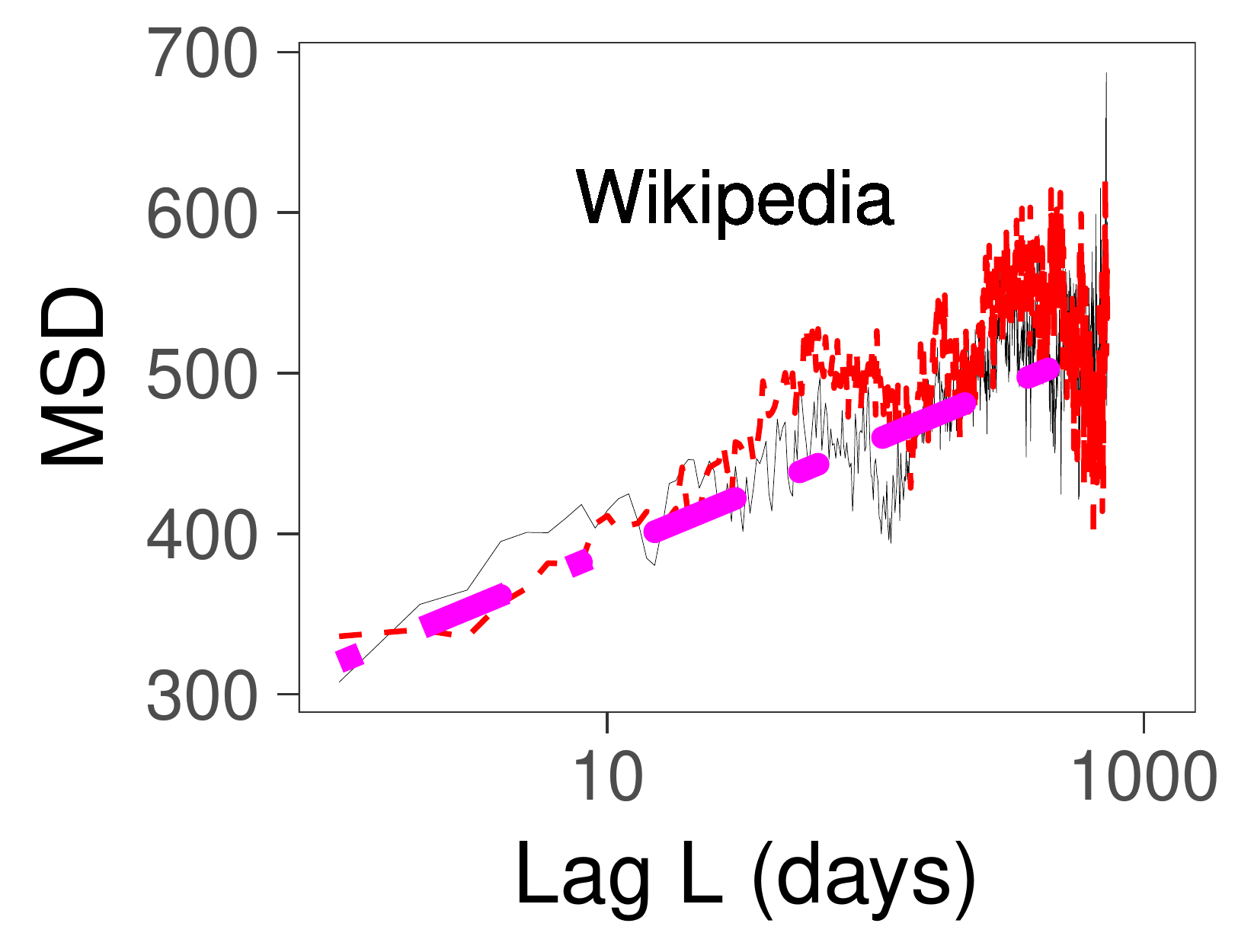}
\put(42,110){(f)}
\end{overpic}
\end{minipage}
\begin{minipage}{0.32\hsize}
\begin{overpic}[scale=0.35,tics=5]
{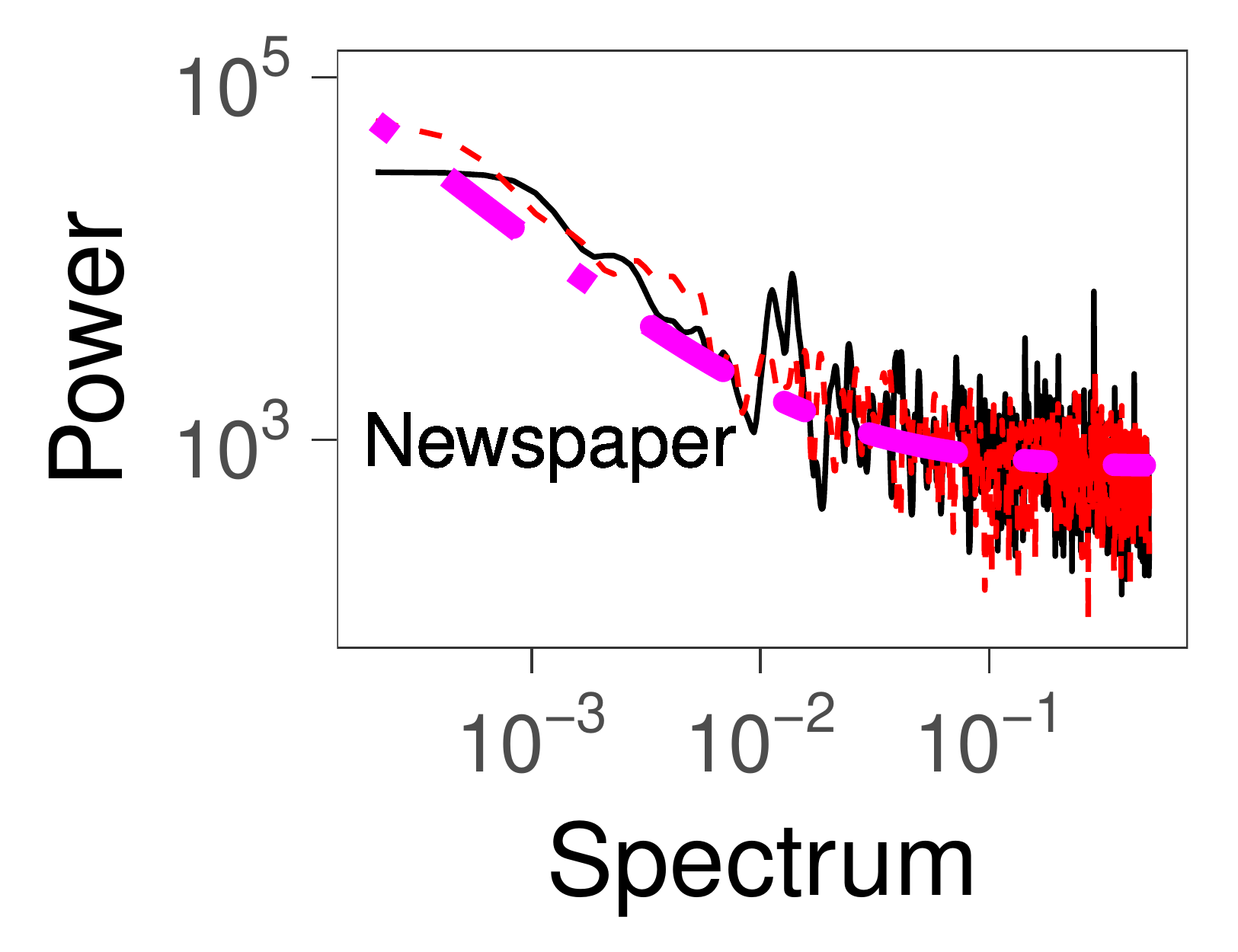}
\put(62,111){(g)}
\end{overpic}
\end{minipage}
\begin{minipage}{0.32\hsize}
\begin{overpic}[scale=0.35,tics=5]
{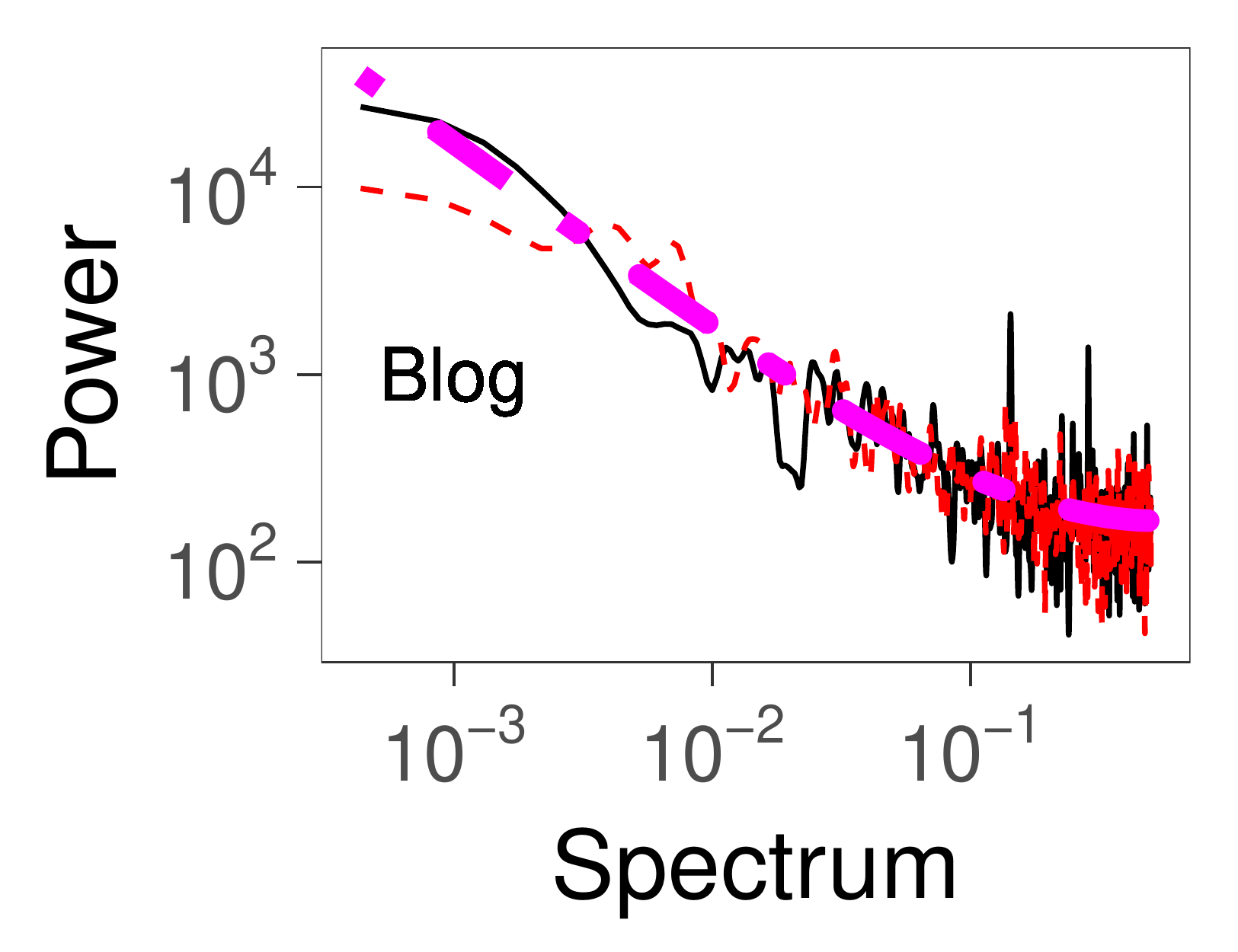}
\put(78,111){(h)}
\end{overpic}
\end{minipage}
\begin{minipage}{0.32\hsize}
\begin{overpic}[scale=0.35,tics=5]
{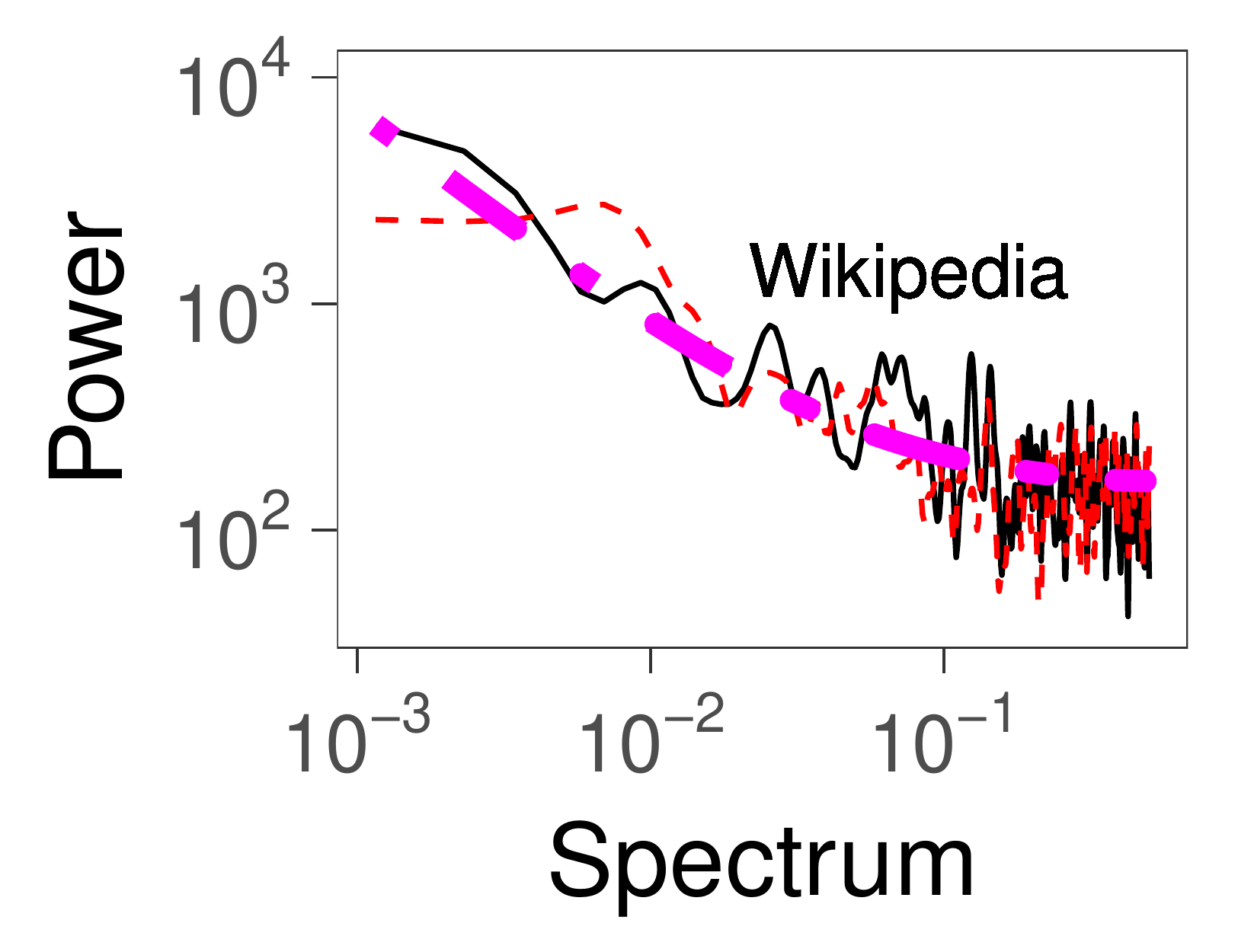}
\put(62,111){(i)}
\end{overpic}
\end{minipage}

\begin{minipage}{0.32\hsize}
\begin{overpic}[scale=0.35,tics=5]
{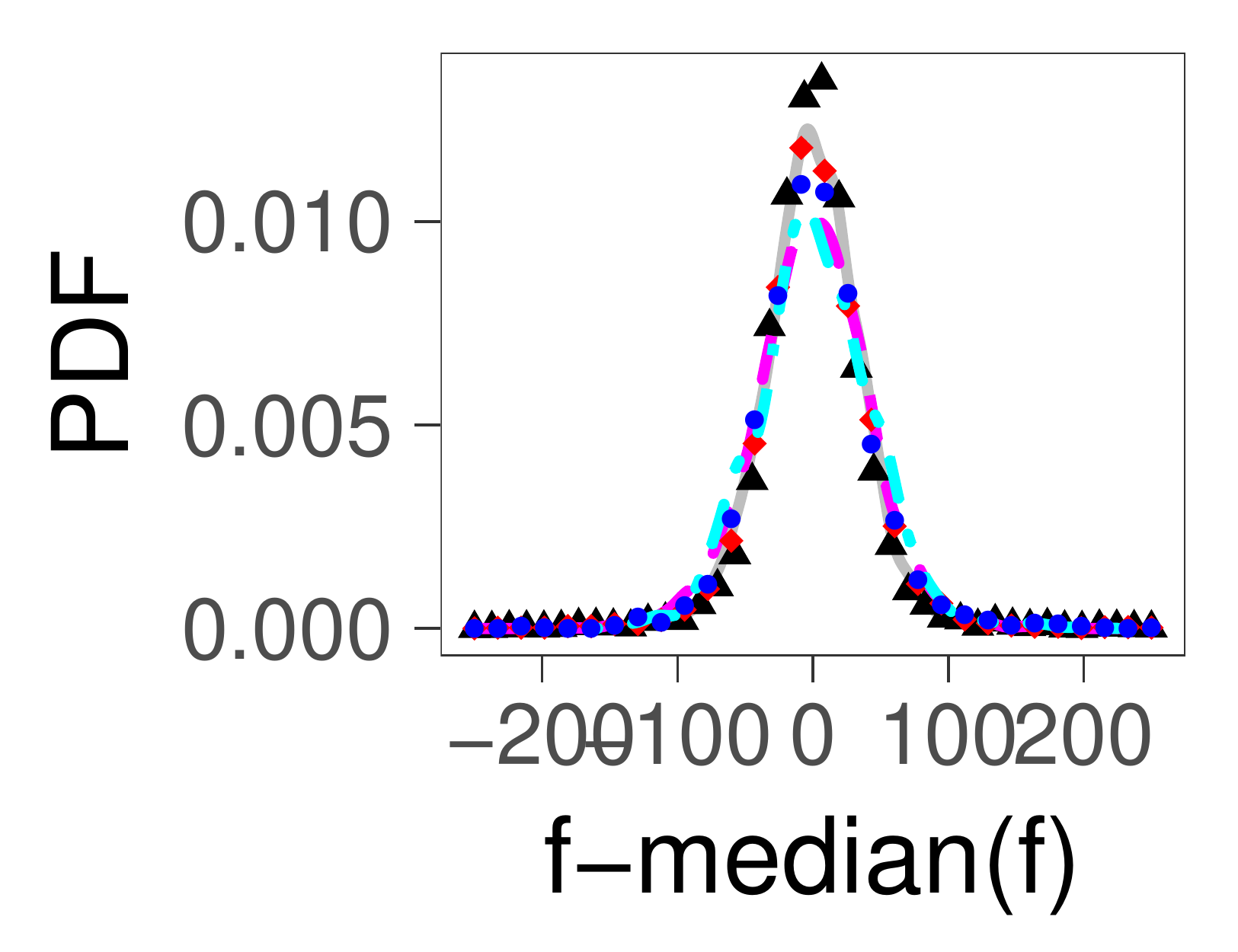}
\put(62,109){(j)}
\end{overpic}
\end{minipage}
\begin{minipage}{0.32\hsize}
\begin{overpic}[scale=0.35,tics=5]
{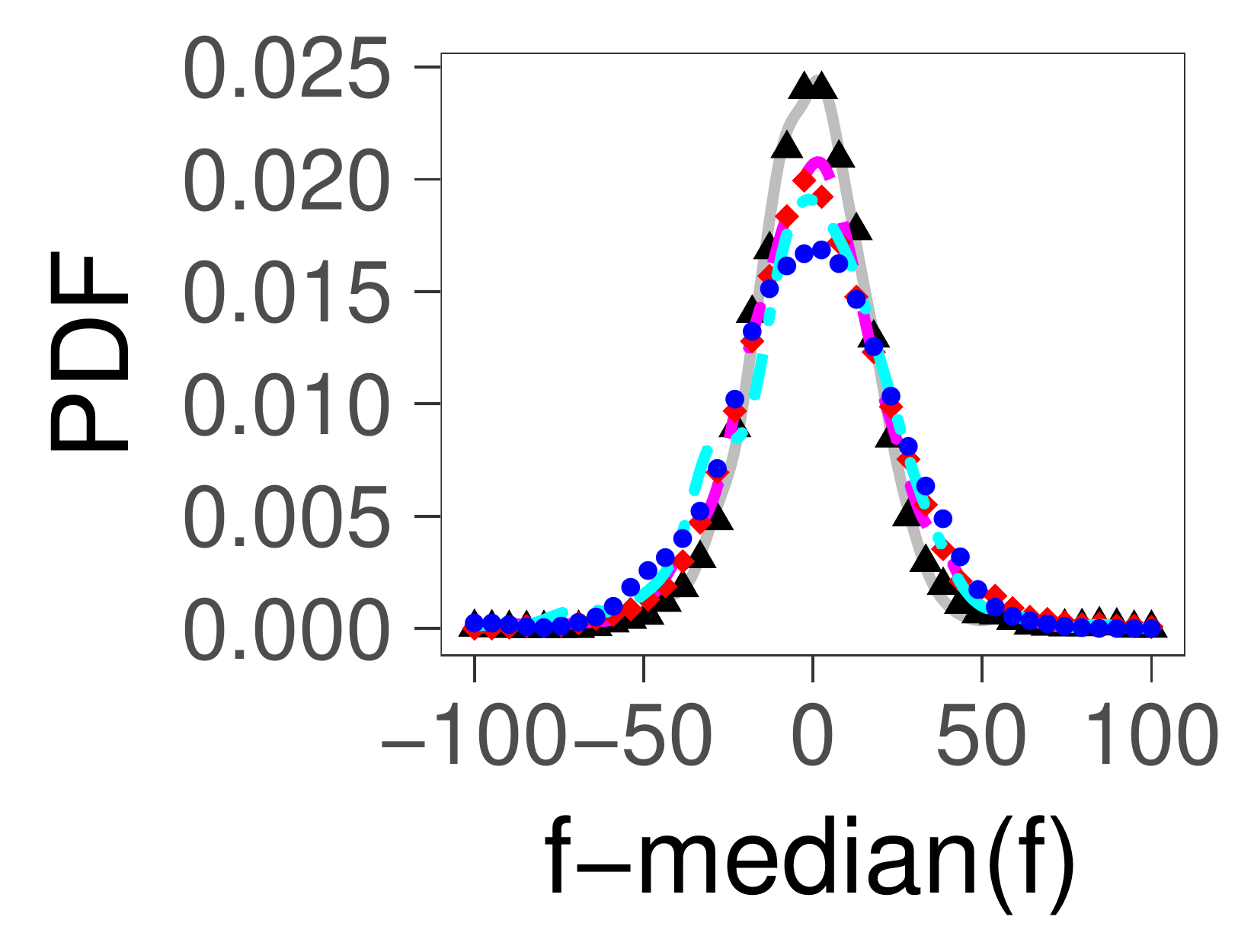}
\put(62,111){(k)}
\end{overpic}
\end{minipage}
\begin{minipage}{0.32\hsize}
\begin{overpic}[scale=0.35,tics=5]
{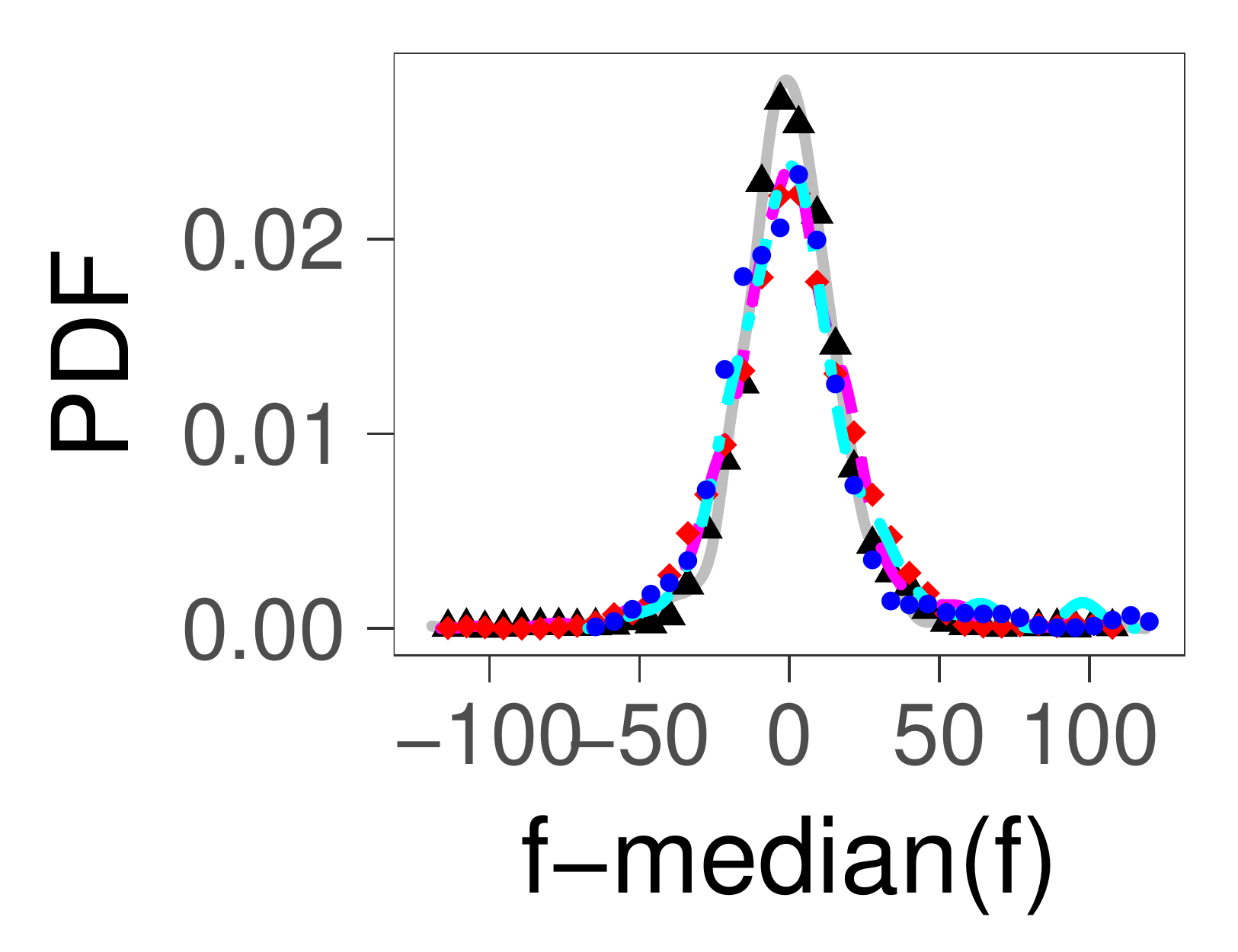}
\put(57,110){(l)}
\end{overpic}
\end{minipage}
		
		
		\caption{
			Comparison between empirical observations and model of the word counts given by Eqs. \ref{eq_rw} and \ref{eq_rd}. 
			\textcolor{black}{Subfigures in the first column indicate the newspaper data for ``Tachiba (i.e. position or standpoint in English;  $\hat{c}_j=150.3$, $\hat{\eta}=0.068$ and $\hat{\delta}=0.15$)'', the second column indicates the blog data for ``Sanada (i.e. well-known Japanese family name; $\hat{c}_j=83.16$, $\hat{\eta}=0.12$ and $\hat{\delta}=0.065$)'',  and the third column indicates Wikipedia data for ``Handle ($\hat{c}_j=58.2$, $\hat{\eta}=0.12$, $\hat{\delta}=0.14$)''.}
			%
			%
			%
			Panels (a), (b), and (c) are the normalised word counts $f_j(t)$. The empirical data are shown as a black solid line and the results of 
			the numerical simulation are shown as a red dashed line.
			Panels (d), (e), and (f) are the corresponding MSDs given by \textcolor{black}{Eq. \ref{def_msd0}} and the magenta thick dashed-dotted line is the theoretical curves of the model expressed by Eq. \ref{eq_msd_f}. 
			Panels (g), (h), and (i) are the corresponding power spectrum densities \textcolor{black}{given by Eq. \ref{def_psd0}} and the magenta thick dashed-dotted line is the corresponding theoretical curve obtained by Eq. \ref{eq_psd_f0}.
			Panels (j), (k), and (l) compare the probability density functions (PDFs) of the empirical data and numerical simulations \textcolor{black}{given by Eq. \ref{def_pdf0}.} 
			The black triangles indicate the empirical data for $L=1$,
			red diamonds for $L=30$, and  blue circles for $L=365 \times 10$ (newspapers), $L=365 \times 5$ (blogs) or $L=365 \times 2$ (Wikipedia).
			In addition, the grey solid lines show the corresponding numerical simulation for $L=1$, the magenta dashed line for $L=30$ and the cyan dash-dotted lines for $L=365 \times 10$ (newspapers), $L=365 \times 5$ (blogs) or $L=365 \times 2$ (Wikipedia).
			The results in these figures confirm that the theoretical model is almost consistently in accordance with the empirical data.  \textcolor{black}{ Note that the p-values of the two-sample Kolmogorov-Smirnov test (KS test) in panels (j-l): [Black triangles empirical distribution vs  grey solid line simulation distribution, red diamonds distribution vs magenta dashed line distribution, blue circles distribution vs cyan dash-dotted line distribution] are [0.89, 0.46, 0.46] for the newspapers, [0.94, 0.30, 0.74] for the blogs and [0.98, 0.92, 0.97] for Wikipedia.} \textcolor{black}{In this statistical test, we check whether the samples obtained by the empirical data and those obtained by the numerical simulation come from the same distribution, where samples are $T-L$ points differences of $f_j(t)$ of data and the corresponding simulation results, namely $\{f_j(1+L)-f_j(1),f_j(2+L)-f_j(2),\cdots f_j(T)-f_j(T-L)\}$. The KS test requires the IID samples, but our data have a weak autocorrelation; hence, these p-values are approximated values.}
		} 
		\label{fig_sim}
	\end{figure*}
	
\begin{figure*}
		\begin{minipage}{0.32\hsize}
\begin{overpic}[scale=0.32,tics=5]
{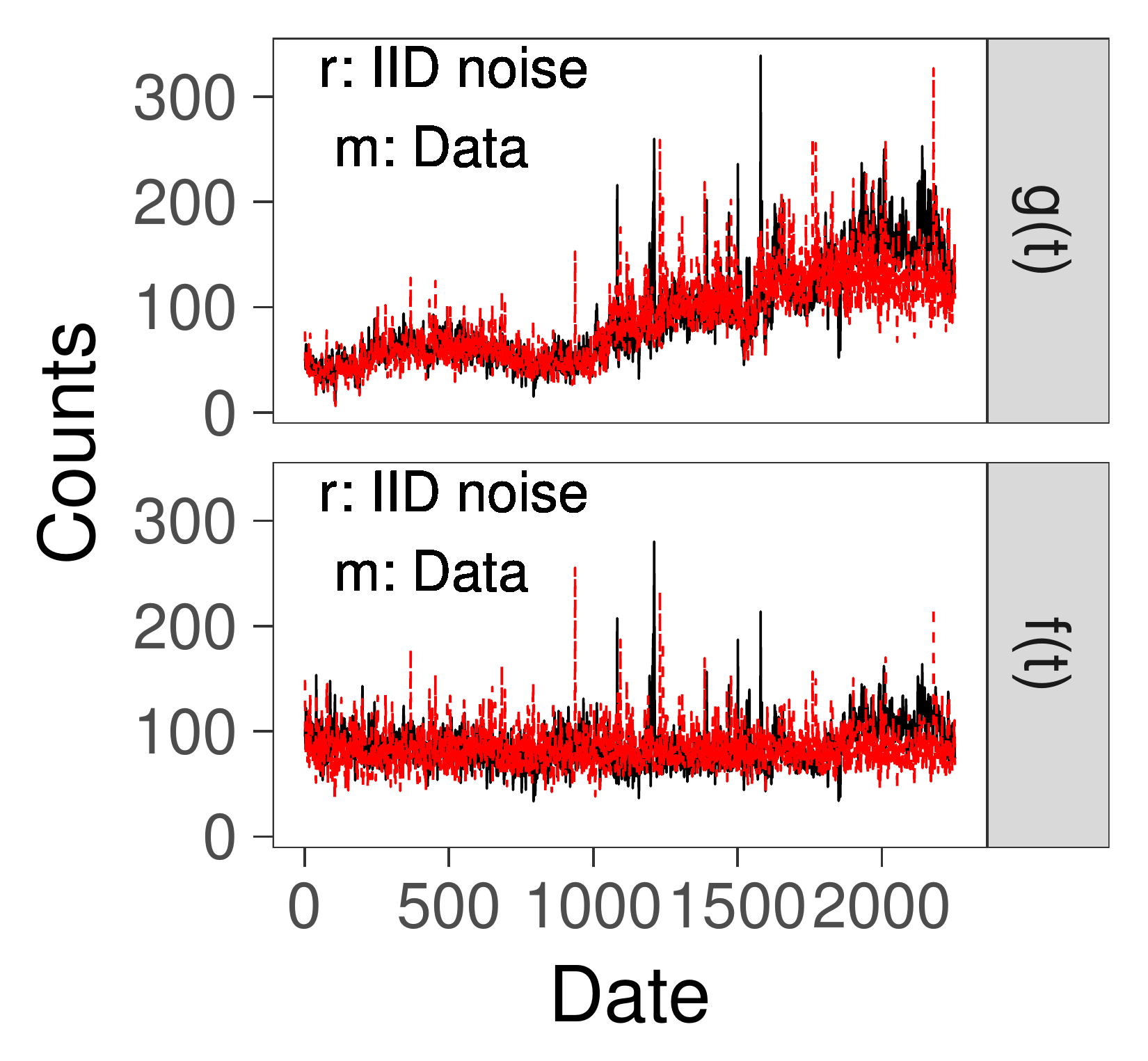}
\put(110,128){(a)}
\end{overpic}
\end{minipage}
\begin{minipage}{0.32\hsize}
\begin{overpic}[scale=0.32,tics=5]
{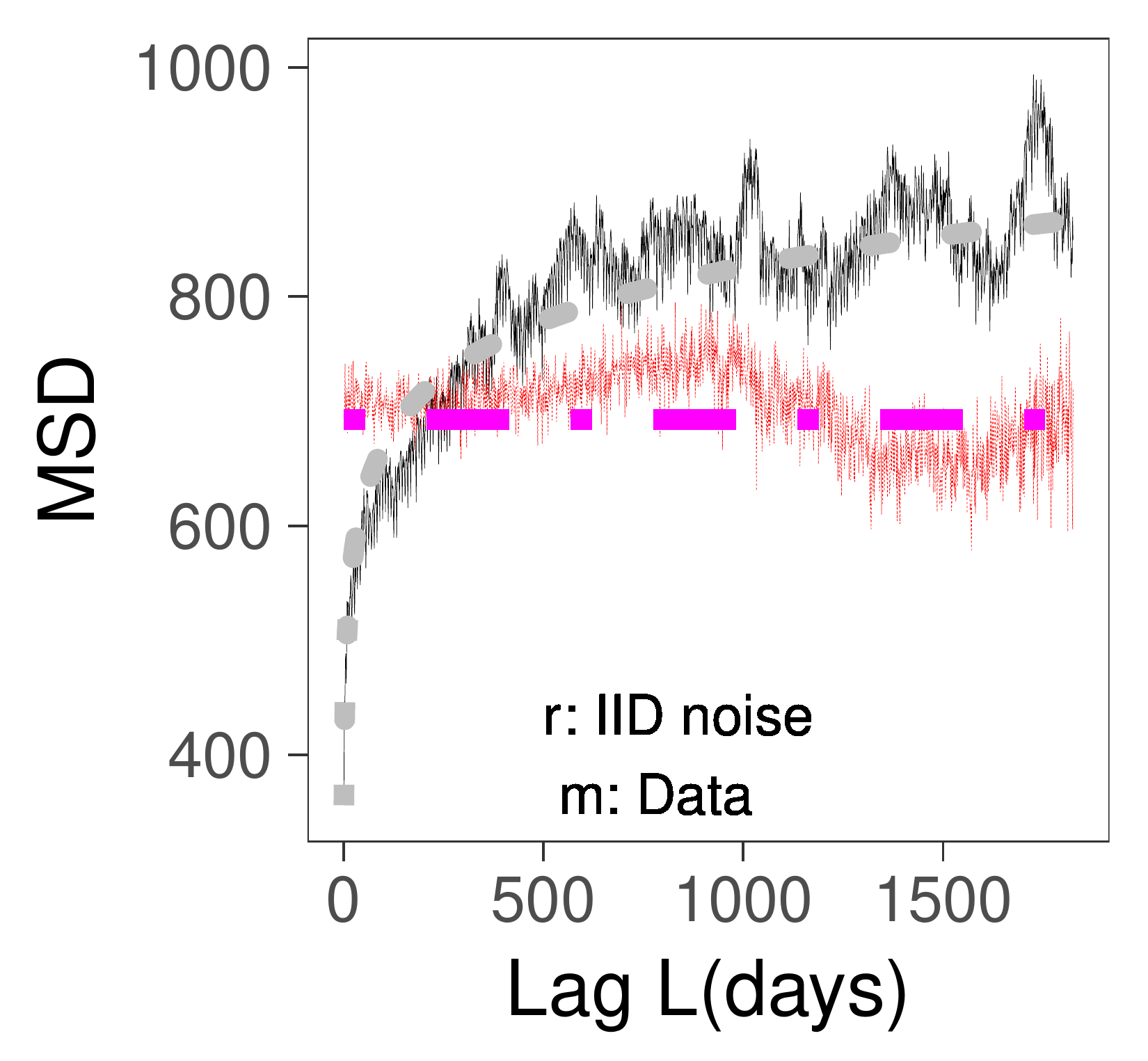}
\put(48,127){(b)}
\end{overpic}
\end{minipage}
\begin{minipage}{0.32\hsize}
\begin{overpic}[scale=0.32,tics=5]
{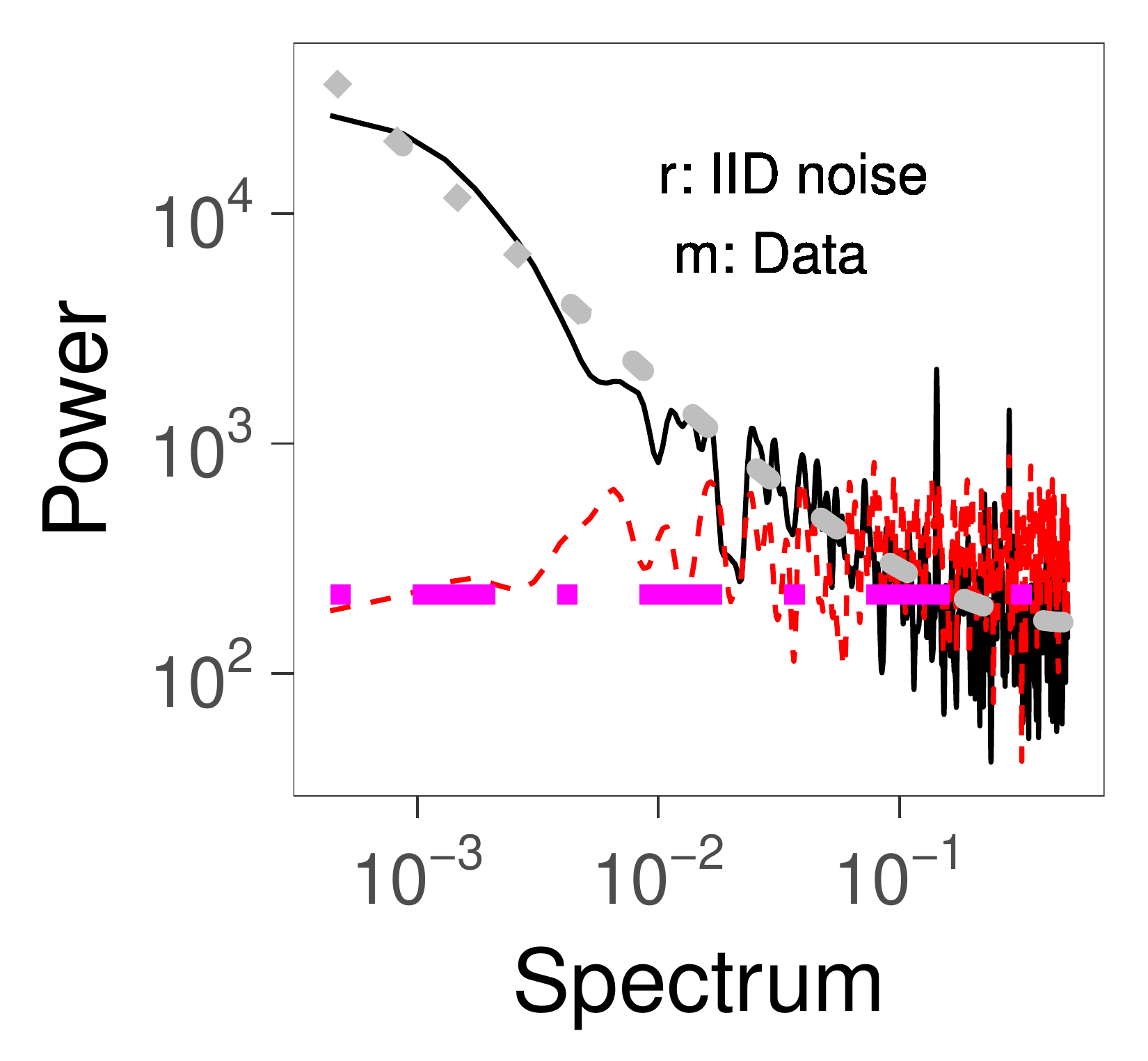}
\put(60,126){(c)}
\end{overpic}
\end{minipage}
		\begin{minipage}{0.32\hsize}
\begin{overpic}[scale=0.32,tics=5]
{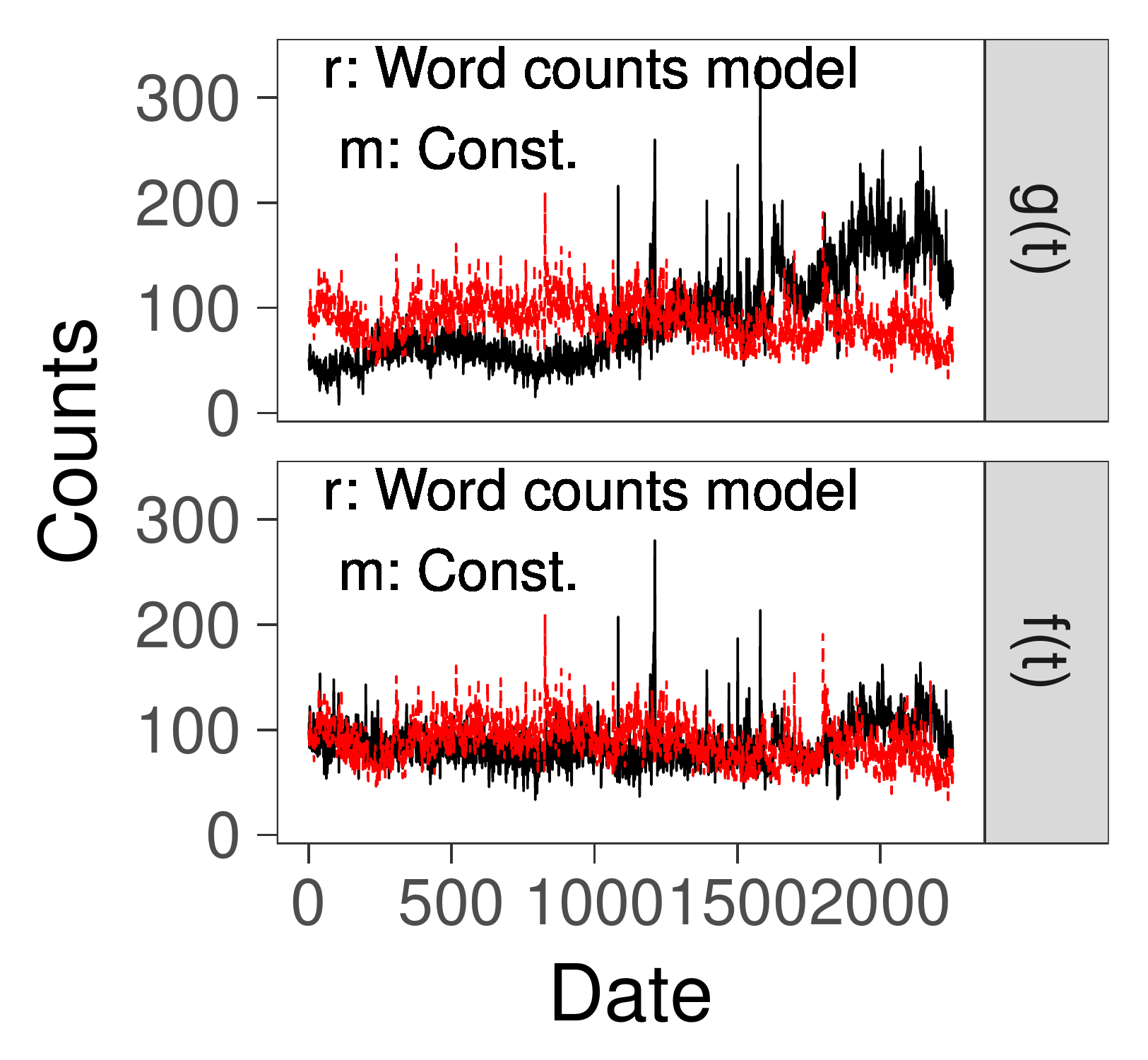}
\put(118,128){(d)}
\end{overpic}
\end{minipage}
\begin{minipage}{0.32\hsize}
\begin{overpic}[scale=0.32,tics=5]
{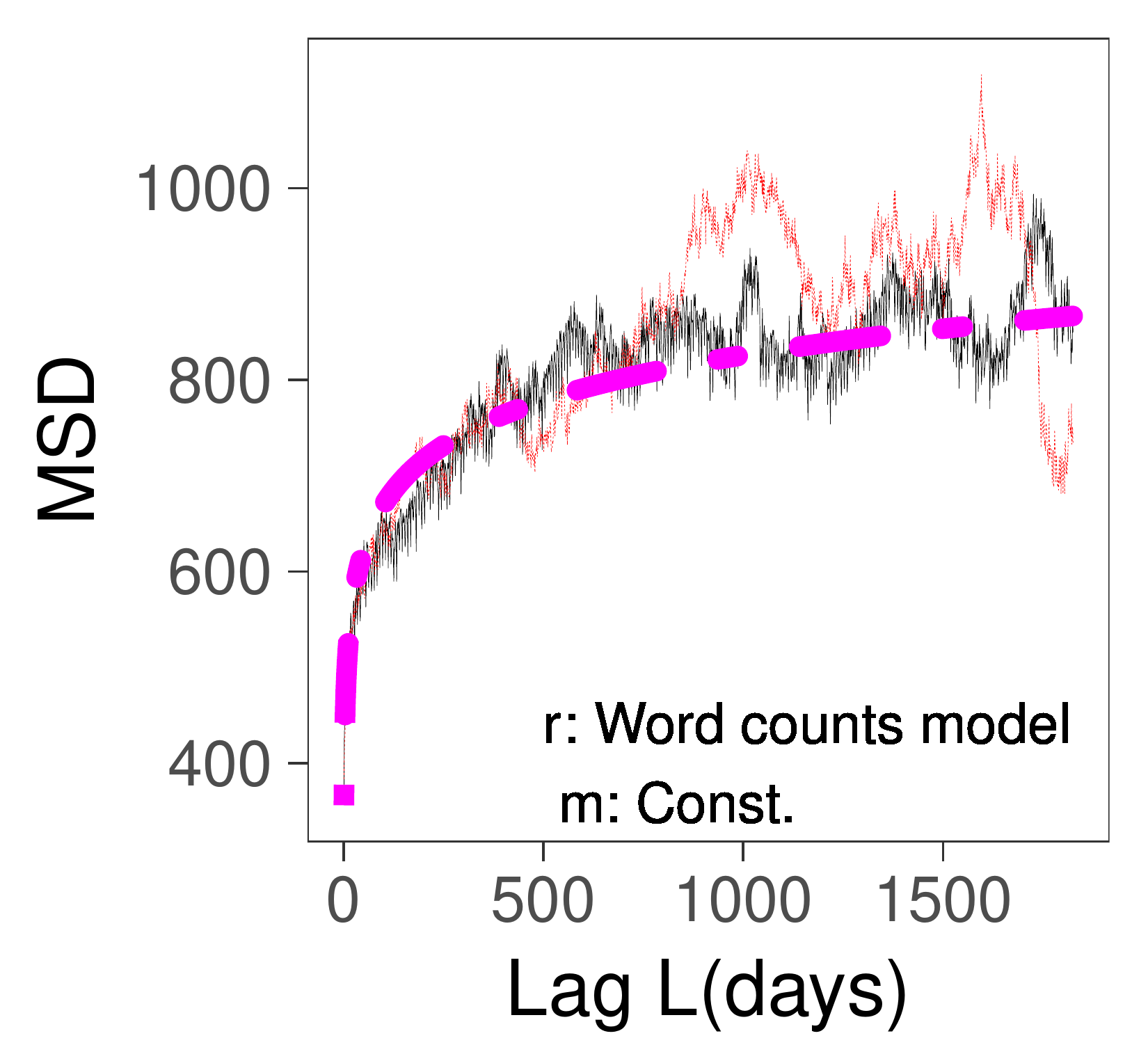}
\put(50,127){(e)}
\end{overpic}
\end{minipage}
\begin{minipage}{0.32\hsize}
\begin{overpic}[scale=0.32,tics=5]
{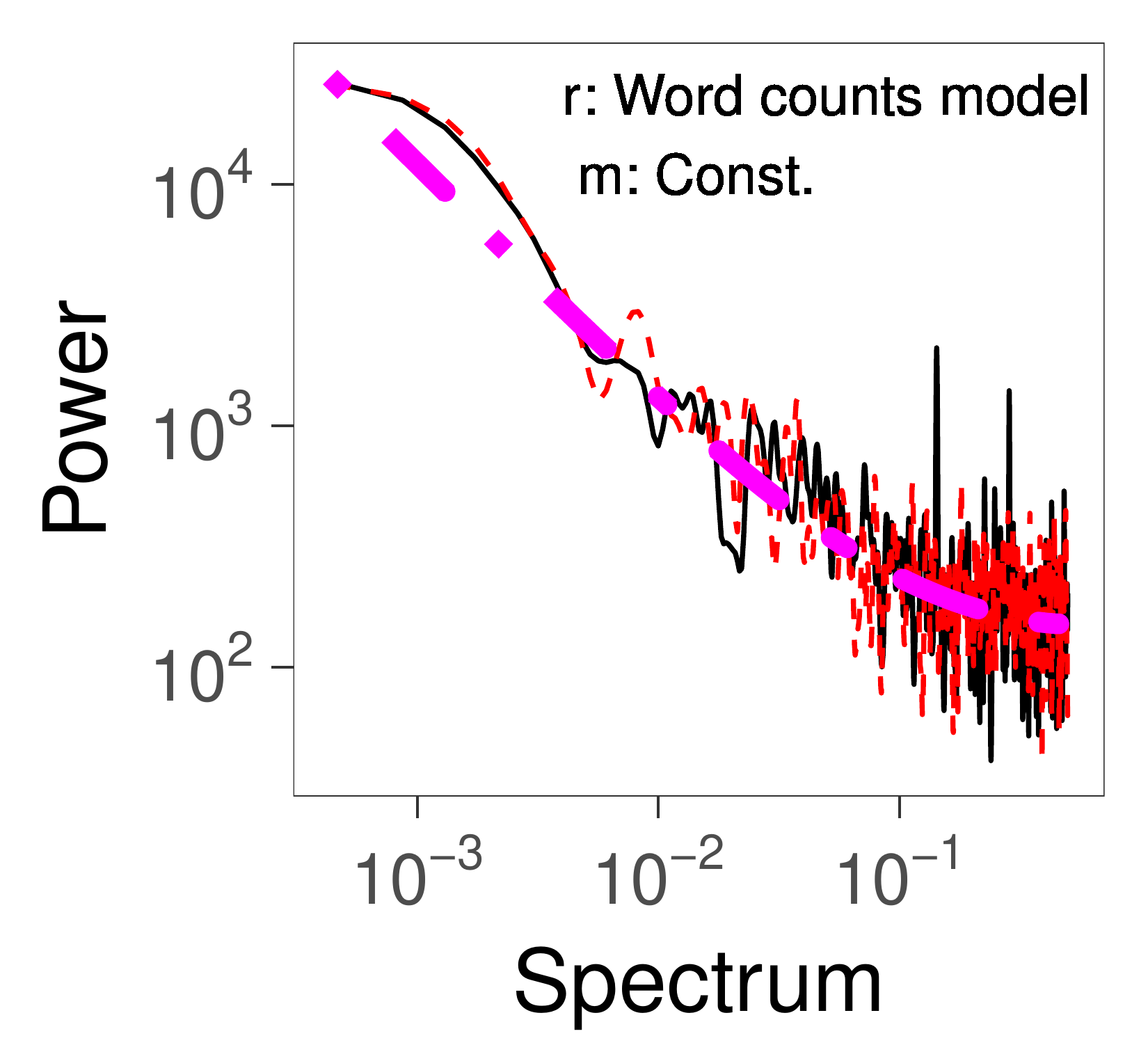}
\put(48,103){(f)}
\end{overpic}
\end{minipage}
\begin{minipage}{0.32\hsize}
\begin{overpic}[scale=0.32,tics=5]
{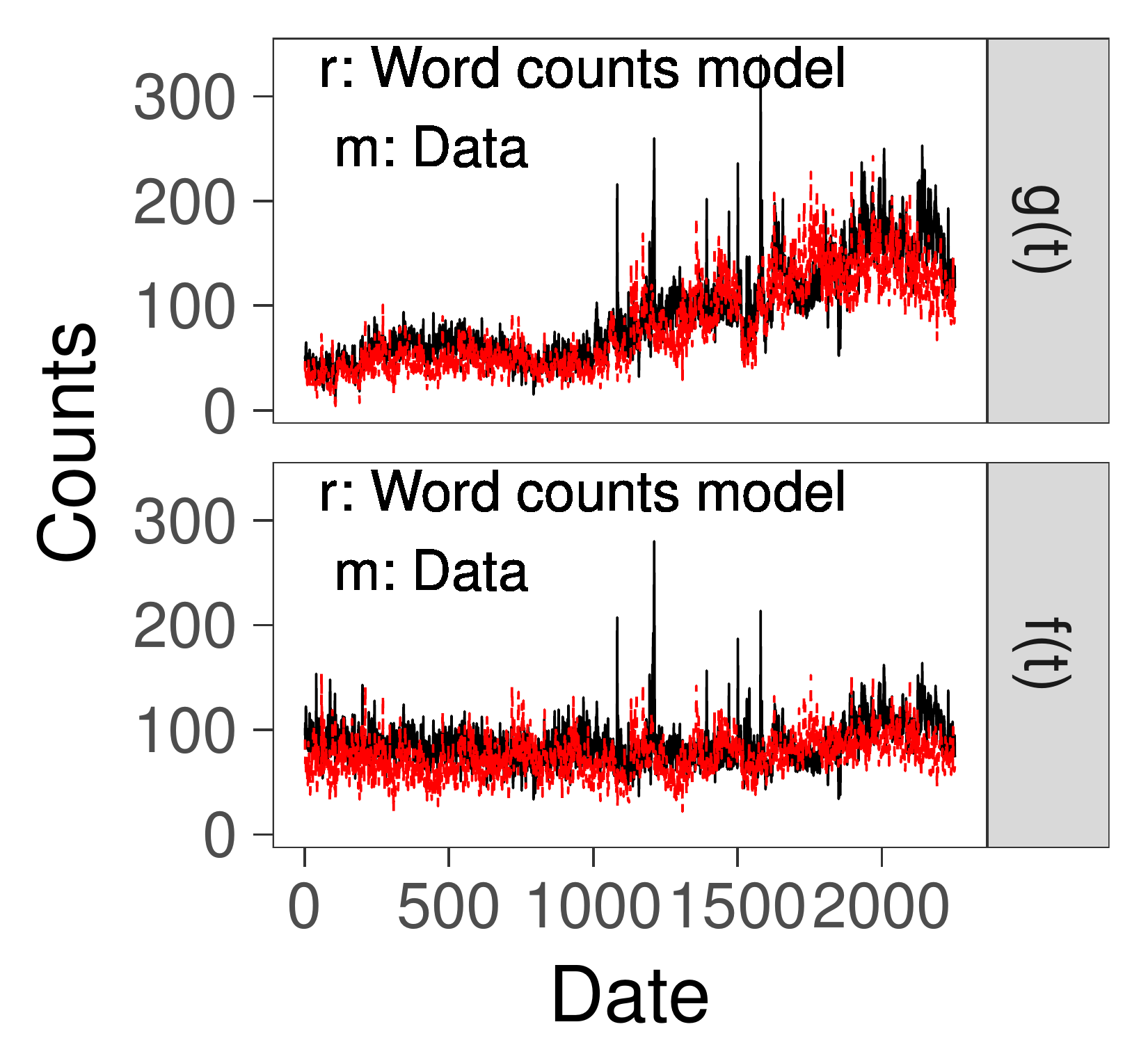}
\put(118,128){(g)}
\end{overpic}
\end{minipage}
\begin{minipage}{0.32\hsize}
\begin{overpic}[scale=0.32,tics=5]
{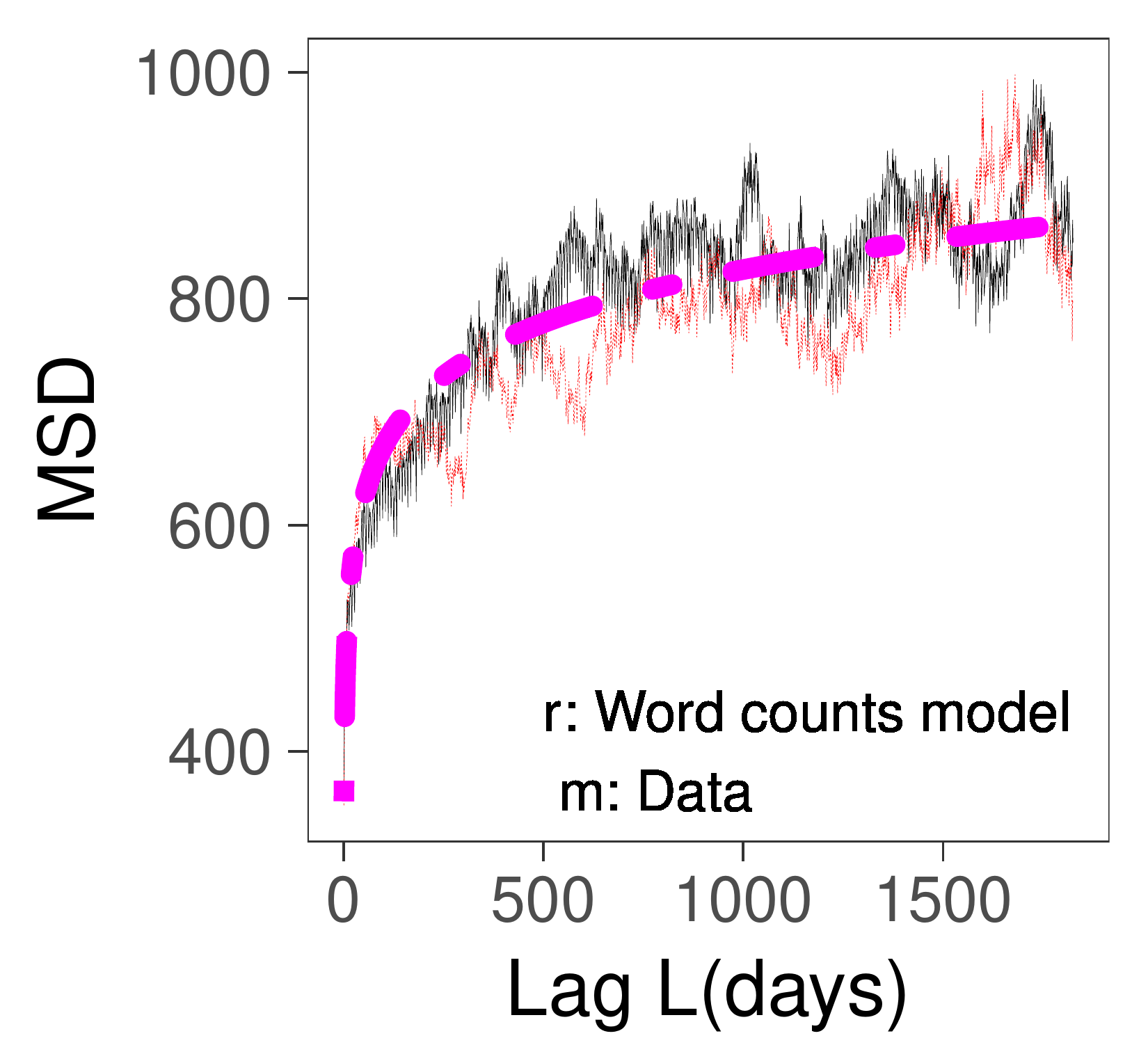}
\put(50,127){(h)}
\end{overpic}
\end{minipage}
\begin{minipage}{0.32\hsize}
\begin{overpic}[scale=0.32,tics=5]
{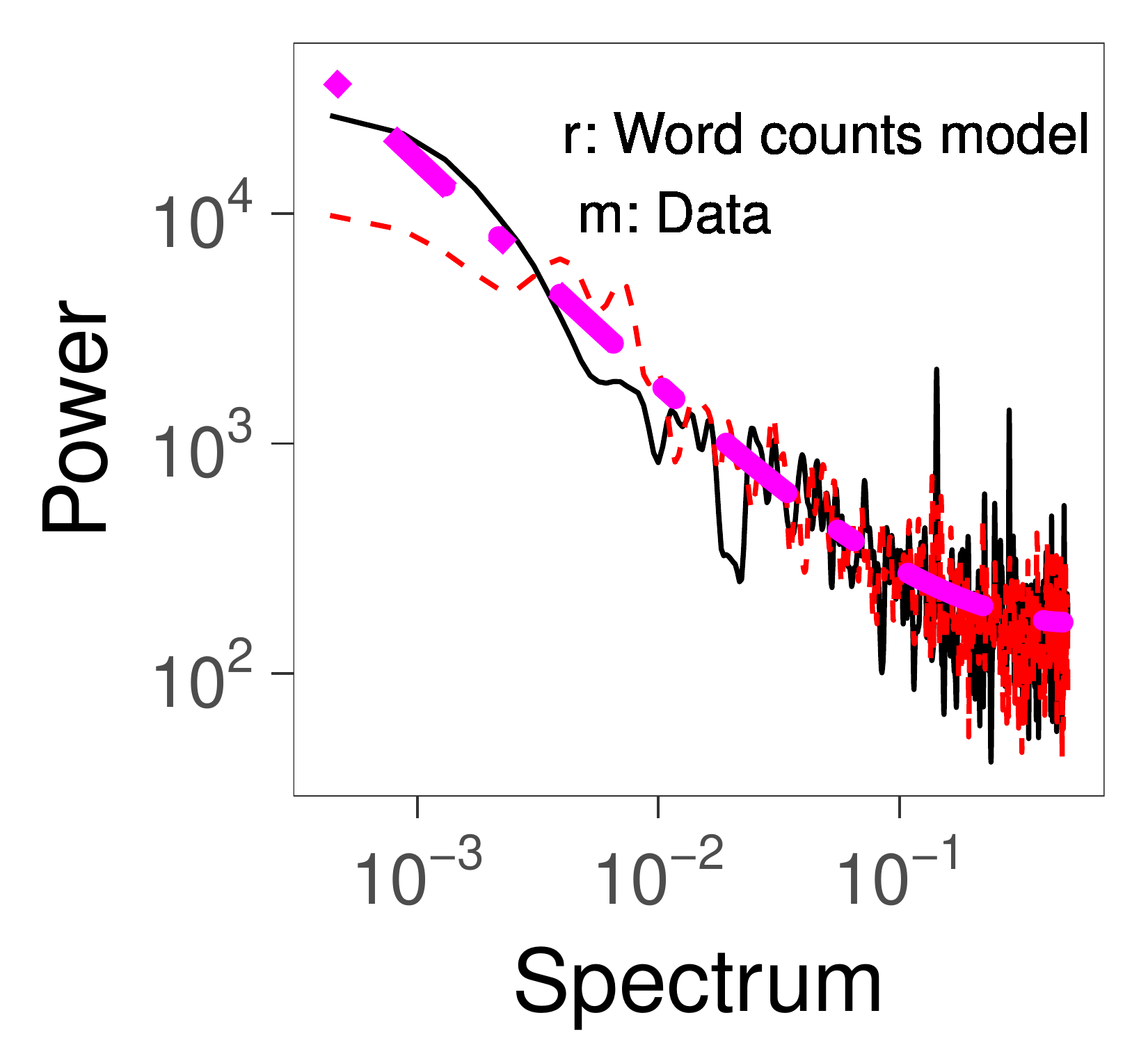}
\put(50,95){(i)}
\end{overpic}
\end{minipage}
		\begin{minipage}{0.32\hsize}
\begin{overpic}[scale=0.32,tics=5]
{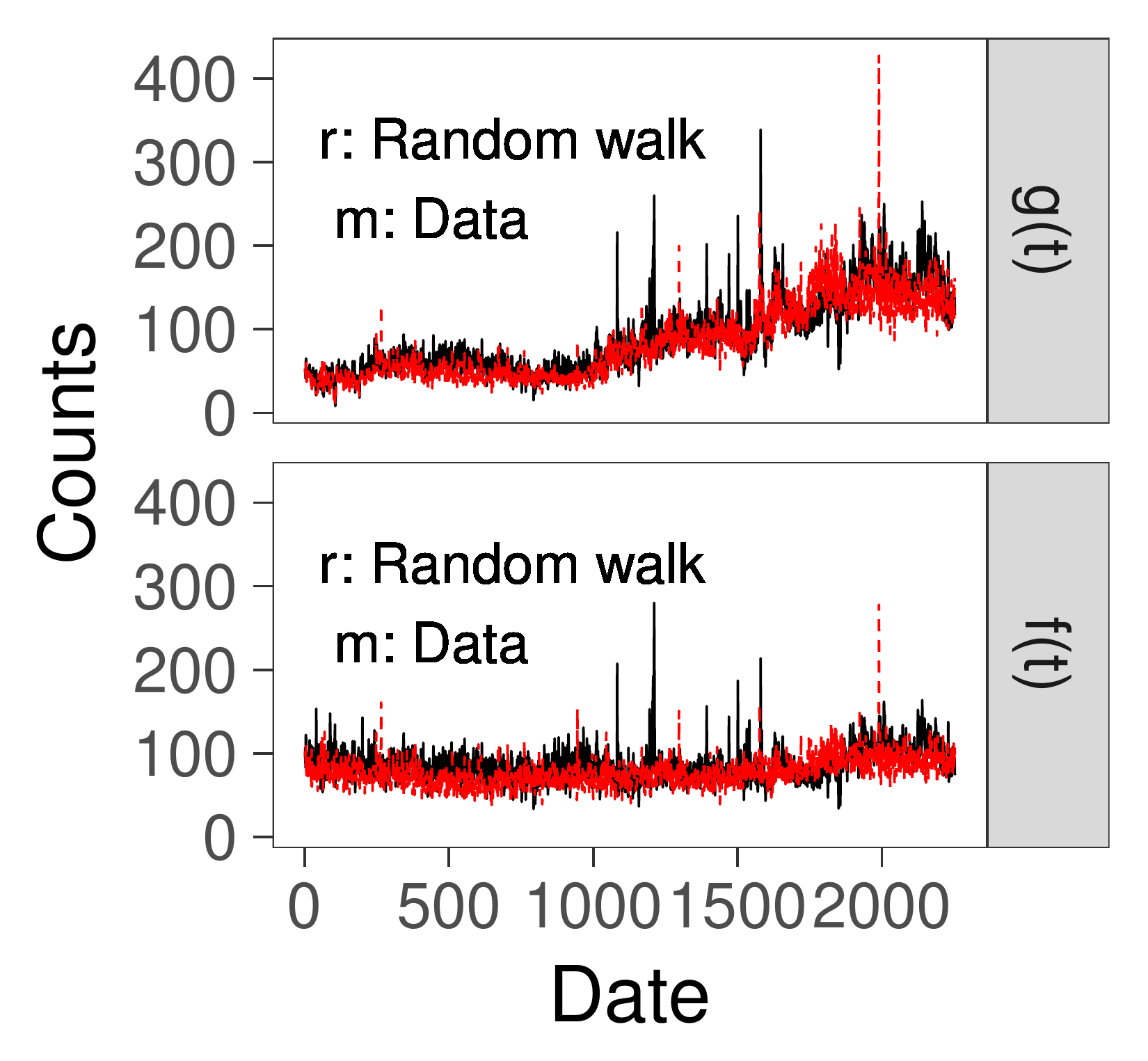}
\put(120,128){(j)}
\end{overpic}
\end{minipage}
\begin{minipage}{0.32\hsize}
\begin{overpic}[scale=0.32,tics=5]
{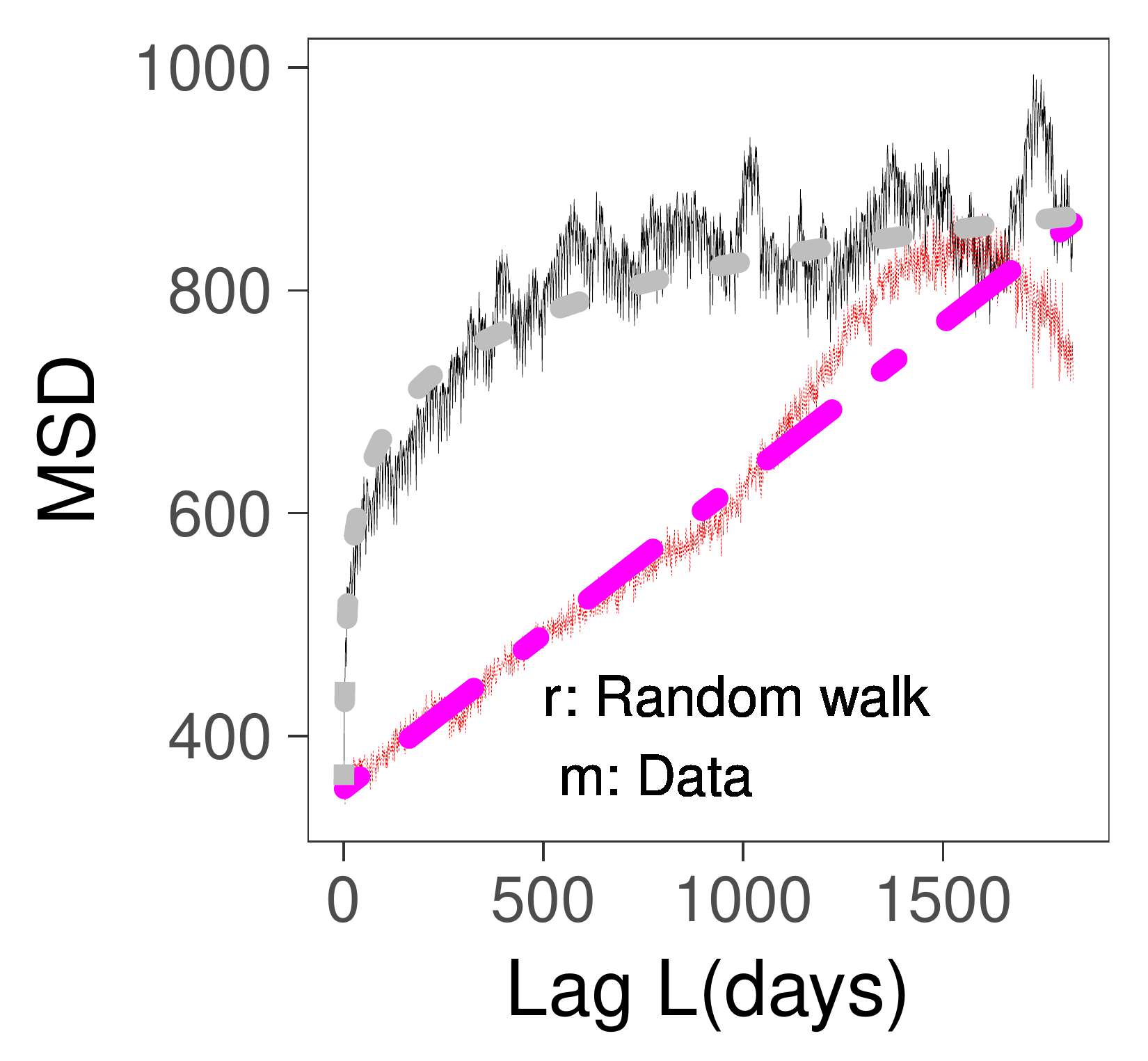}
\put(50,127){(k)}
\end{overpic}
\end{minipage}
\begin{minipage}{0.32\hsize}
\begin{overpic}[scale=0.32,tics=5]
{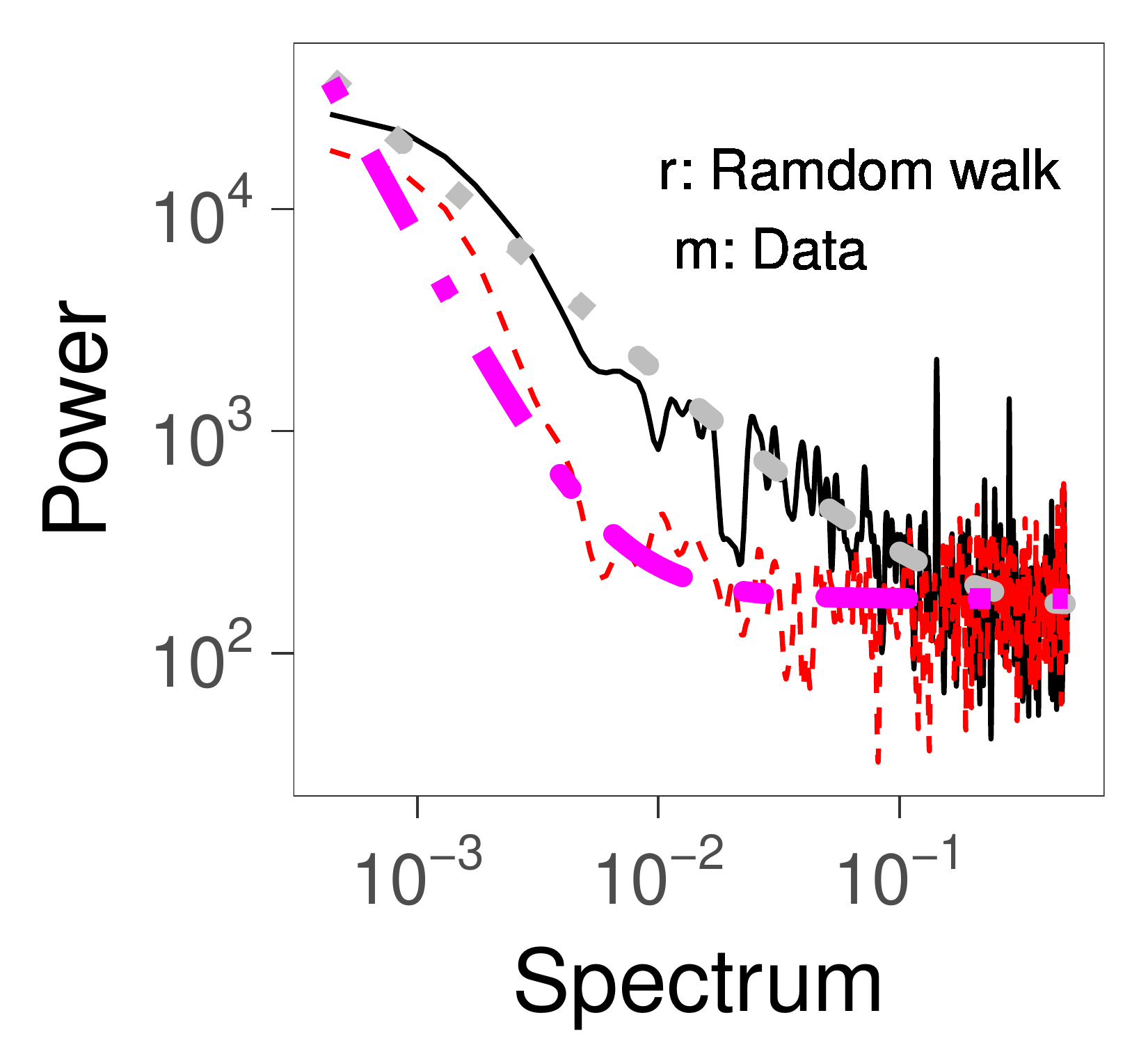}
\put(45,95){(l)}
\end{overpic}
\end{minipage}

		\caption{\textcolor{black}{The comparison of the simulation results of the models given by Eqs. \ref{eq_rw} and \ref{eq_rd} between different $\beta$ (the speeds of forgetting) and $m(t)$ (database size). The subfigures in the first row indicate the results of the case where $r_j(t)$ is sampled from the IID noise ($\beta \to 1; \hat{\eta}=0.12, \hat{\zeta}=0.065$). The third row indicates the results of the case where $r_j(t)$ is sampled from the word count model ($\beta = 0.5; \hat{\eta}=0.12,\hat{\zeta}=0.065$). The fourth row indicates the results of the case where $r_j(t)$ is sampled from the random walk model ($\beta = 1.0$; $\hat{\eta}=0.0063,\hat{\zeta}=0.12$;). Herein, the scaled database size $m(t)$ (i.e. scaled total number articles) is estimated from data (Appendix \ref{app_m}). The subfigures in the second row present the simulation, in which the database size is conserved $m(t)=1$, and $r_j(t)$ is the same as the third row ($\beta = 0.5$). For all data, $\check{c}_j=83.1$.
		The column corresponds to the properties of these models. Panels (a), (d) ,(g) and (j) (first column) show the word counts $g_j(t)$ and $f_j(t)$. 
		Panels (b), (e) , (h) and (k) (second column) are the MSD of $f_j(t)$, and panels (c), (f), (i) and (l) (third column) are the PSD of $f_j(t)$.
		 The red thin dashed lines indicate the results of the numerical simulations, while the thick purple dash dotted lines denote the corresponding theoretical curves given by Eq. \ref{eq_msd_f_app0} for MSD and Eq. \ref{eq_app_psd_f0} for PSD. The black solid lines represent the empirical data for ``Sanada (i.e. well-known Japanese family name)'' in the blog data. The gray thick dotted lines are the corresponding theoretical curves.
	These figures confirm that the random walk model and the IID noise cannot reproduce the empirical results. In addition, from the second row, we can also confirm that $m(t)$ is not essential in reproducing the empirical properties $f_j(t)$.}}
	%
		\label{fig_sim2}
	\end{figure*}
		
	\section{Data analysis: Ultraslow-like diffusion in the empirical data}
	\label{section_data_ana}
	We next calculate the MSD of the actual data.
	We use the following temporal MSD for data analysis, 
	\begin{equation}
	\left<f_j^2(L)\right> \approx \frac{ \sum_{t=1}^{T-L} (f_j(t+L)-f_j(t))^2 }{T-L} \equiv \sigma_j^2(L), \label{eq_msd_base}
	\end{equation}
\textcolor{black}{where $L$ is the time lag (e.g. for $L=7$, it corresponds to a weekly difference; for $L=30$, it corresponds to almost a monthly difference; and for $L=365$, it corresponds to a yearly difference.).} 
	\textcolor{black}{Thus, the MSD quantifies the changes of word counts of the focused keyword growth in $L$ days.
	Note that the statistics has a meaning when the differential $\{(f_j(t+L)-f_j(t))\}$ is steady. $\{f_j(t)\}$ (the normalised counts by the scaled database size) sampled from our mathematical model (described subsequently) do not contradict with this condition (Appendix \ref{app_msd_f}), and the majority of corresponding empirical data approximately satisfies this condition, although the raw counts $\{g_j(t)\}$ do not always satisfy this condition because of effects, such as increasing database size (Fig. \ref{tseries}).}
	\par
	Fig. \ref{single_msd} shows examples of the MSDs of typical words for the Japanese newspaper (a), Japanese blogs (b), and English Wikipedia page views (c).
	The results in these figures confirm that all growth of MSDs is essentially approximated by the logarithmic function, 
	\begin{equation}
	\left<f_j^2(L)\right> \approx a_j \log(L) + b_j.  \label{eq_msd_single}
	\end{equation}
	\par
	Next, we verify the validity of the above result by calculating the ensemble median of (temporal) scaled MSD by using all words with a large frequency on the respective databases. 
	If we assume Eq. \ref{eq_msd_single}, 
	the scaled MSD $\sigma_j'^2(L)$ has a word-independent curve, 
	\begin{equation}
	\sigma_j'^2(L) \equiv \frac{\sigma_j^2(L)-\sigma_j^2(1)}{Medi_l[\sigma_j^2(l)-\sigma_j^2(1)]/Medi_l[\log(l)]} \approx \log(L) \label{eq_msd_scale}
	\end{equation}
	where $Medi_l[x(l)]$ is the temporal median of the set $\{x(1)\,x(2),x(3),\cdots,x(L_{max})\}$ and $L_{max}$ is the maximum lag which we use to make a graph.
	Thus, we can use the ensemble over words $\overline{\sigma'}^2(L)$ and the ensemble median obeys the logarithmic function,  
	\begin{equation}
	\overline{\sigma'}^2(L) \equiv Medi_e[{\sigma'}_j^2(L)] \approx \log(L), \label{eq_msd_medi}
	\end{equation}
	where $Medi_e[x_j(t)]$ is the median over the words set $\{x_1(t),x_2(t),\cdots x_W(t)\}$ and $W$ is the size of the set. 
	We take the median over the set of the mean frequency over 30, $\hat{c}_j=\sum^{T}_{t=1}f_j(t)/T \geq 30$. 
	We exclude words with a small mean because they have relatively large signal-to-noise ratios (see Eq. \ref{eq_msd_f}).
	Figs. \ref{ensemble_msd} (a)-(f) show that the logarithmic curve is approximately observed for all data sets, namely newspapers, blogs, and Wikipedia content (English, French, Chinese, and Japanese). 
	Here, because there are words with a non-negligible weekly or annual cycle, the raw ensemble of MSD also has these cycles (grey dots or grey thin lines). Thus, we can observe the logarithmic curve by using the 365-day moving median, which cancels these cycles. 
	Note that by replacing the ensemble median with the ensemble mode in Eq. \ref{eq_msd_medi}, we can also obtain the essentially same logarithmic diffusion. This logarithmic diffusion is not in conflict with our intuition that languages are basically stable but change constantly.  \par
	\section{Model}
	\textcolor{black}{
	This section explains the properties of word counts by the \textcolor{black}{combination of two probabilistic models:(i) the} random walk model with the power-law forgetting and (ii) the random diffusion model (i.e, a kind of the Poisson point process).
  The random walk model describes the latent concerns of the focused word $r_j(t)$ and it can explain the ultraslow diffusion essentially. Besides, the random diffusion model expresses the connection between the latent concern $r_j(t)$ described above-mentioned random walk model and the observable word counts $g_j(t)$ or $f_j(t)$. 
Here, first we introduce and discuss the random walk model, and next, we introduce the word counts model which is the combination of the random walk model and random diffusion model. }
	\subsection{Model: Relation with the random walk} 	
	\textcolor{black}{Here}, we present the extent to which the empirical result and a random walk correspond with the power-law forgetting, which is one of the most representative standard explanations of anomalous diffusion in previous studies. This approach is also equivalent to the fractional dynamics approach (in our case, the fractional Langevin equation approach). \par
	The random walk model with the power-law forgetting is given by 
	\begin{equation}
	r_j(t)=\frac{1}{\Gamma(1-\beta)} \sum_{s=0}^{\infty} (s+d_\beta)^{-\beta} \cdot \eta_j(t-s), \label{eq_rw}
\end{equation}
where $d_\beta=\Gamma(1-\beta)^{-1/\beta}$, $\beta$ is a constant used to characterise the forgetting speed and $\eta_j(s)$ is independent and identically distributed noise where the mean takes zero and the standard deviation is $\hat{\eta}_j$, that is, we can write ${\eta}_j(s)=\hat{\eta}_j \times \eta^{(0)}_j(s)$. Here, $\eta^{(0)}_j(s)$ is independent and identically distributed noise where the mean takes zero and the standard deviation is 1.

This model is an extension of the normal random walk model, namely, the model corresponds to the random walk for $\beta=0$ and to the steady IID noise for $\beta \to 1$. 
For the time-series of the word counts, the model is interpreted by considering that the social concern of the $j$-th word at the time $t$, $r_j(t)$ is determined by the summation of received outer shocks $\eta(t)$ until the time $t$ in the case of $\beta=0$.
In the case of $\beta>0$, $r_j(t)$ (i.e. the social concern) is determined by both the above-mentioned summation effects and the effects of forgetting shocks in a power-law manner. 
\par
From the Appendix \ref{MSD_forget}, the MSD of this model is calculated by  
\textcolor{black}{
\begin{eqnarray}
&&\sigma(L;r_j)^2 \equiv \sum_{t=1}^{T-L}\frac{(r_j(t+L)-r_j(t))^2}{T-L}= \label{def_msd_r} \\
&&\left<(r_j(t+L)-r_j(t))^2\right> \propto  
\begin{cases}
L^{1-2\beta}  & (0 \leq \beta <0.5) \\
\log(L) &(\beta=0.5)\\
O(1) &(\beta > 0.5) \\
\end{cases}.
\label{th_msd}
\end{eqnarray}
}
This formula implies $\beta \approx 0.5$ corresponds to our empirical results, that is, the logarithmic-like diffusion. \par
We also verify the validity of the model by comparing the power spectrum density (PSD) between the data and the model.
\textcolor{black}{The PSD of the model Eq. \ref{eq_rw} is approximated by 
\begin{eqnarray}
P(\nu;r_j) &\approx&  \hat{\eta}_j^2 (2 \sin(2 \pi \nu/2))^{-2(1-\beta)} \\
&=& \hat{\eta}_j^2 P_r(\nu), \label{spec1} 
\end{eqnarray}
where $P_r(\nu) \equiv (2 \sin(2 \pi \nu/2))^{-2(1-\beta)}$.}
\textcolor{black}{We use herein the formula of the PSD of $ARFIMA(0,1-\beta,0)$ \cite{granger1980introduction}, by which our model was approximated (Appendix \ref{app_arfima})} 
\textcolor{black}{and the empirical PSD of the time series $\{r_j(t)\}$ is calculated as follows:
\begin{equation}
	P(\nu;r_j) = \frac{1}{T} \left|\sum_{t=1}^{T} \exp(- i 2 \pi \nu t) r_j(t)  \right|^2, \label{def_psd_r}
\end{equation}
where $\nu$ is the frequency [1/days].
}
\textcolor{black}{$ARFIMA$ is the abbreviation for autoregressive fractionally integrated moving average model, which is a well-known time-series model that describes a time-series with a long memory, in the field of statistics \cite{burnecki2014algorithms}. $ARFIMA(0,1-\beta,0)$ is defined by Eq. \ref{arfima}.}
For $\nu \to 0$, this formula is also approximated by
\begin{equation}
P(\nu;r_j) \propto \nu^{-2(1-\beta)}.
\end{equation}
Thus, for $\beta=0.5$ the power spectrum is approximated by the simple power law, $1/\nu$. \par
\textcolor{black}{
Because the concern of word $r_j(t)$ is directly observed from the actual word counts data $f_j(t)$ as mentioned (see the section \ref{sec_rdmodel}), alternatively we use the normalised power spectrum of word counts for $\beta=0.5$, $P(\nu;f_j)'$ ,    
\begin{eqnarray}
&&P(\nu;f_j)' \\ \nonumber
&& \equiv \frac{P(\nu;f_j)-Min[P(\nu;f_j)]}{\int^{\nu_{min}}_{1/2}[P(\nu;f_j)-Min[P^(\nu;f_j)]]d \nu } \\
&&=\frac{(P_r(\nu)-Min[P_r(\nu)])}{\int^{\nu_{min}}_{1/2}[P_r(\nu)-Min[P_r(\nu)]]d \nu } \\
&&\approx  \frac{   (2 \sin(2\pi \nu/2))^{-1}   }{  (\int^{\nu_{min}}_{1/2}( 2 \sin( 2\pi \nu/2) )^{-1} d \nu )   }.
\label{eq_normal_peri}
\end{eqnarray}
where $\nu_{min}$ is the minimum $\nu$ in the observation, and we used the assumption
\begin{equation}
P(\nu;f_j)=v_j \times P_r(\nu)+w_j, \label{eq_p_f}
\end{equation}
where $v_j$ and $w_j$ are constants depending on the word $j$.}
Hence, we can obtain the information of $P_r(\nu)$ from the observable $P(\nu,f_j)$, which we estimate by using a periodogram
in this study.
The validity of this assumption is discussed in section \ref{sec_rdmodel}.  
\textcolor{black}{
Figs. \ref{ensemble_msd}(g-i) show the ensemble median of the normalised power spectrum of word counts $P(\nu;f_j)'$ given by 
Eq.\ref{eq_normal_peri} over the word sets, 
\begin{eqnarray} 
&&\overline{P_{f'}(\nu)} \equiv Medi_e[P(\nu;f_j)'], \label{eq_def_ensemble_peri} \\
&&\approx \frac{   (2 \sin( 2\pi \nu/2))^{-1}   }{  (\int^{\nu_{min}}_{1/2}( 2 \sin(2\pi \nu/2) )^{-1} d \nu )   }. 
\label{eq_ensemble_peri}
\end{eqnarray}
where, for data analysis, we take the median over the set in which the mean $\hat{c}_j$ is above 30.}
The results in these figures confirm that Eq. \ref{eq_ensemble_peri} is in agreement with $\overline{P_{f'}(\nu)}$ of actual data given by Eq. \ref{eq_def_ensemble_peri} for all data sets.  \par
In order to check the plausibility of $\beta \approx 0.5$, in addition, we estimate $\beta$ directly from data, with respect to individual words. Herein we use the model described by  Eq. \ref{eq_rw} and Eq. \ref{eq_rd} (outlined subsequently) and details of the estimation method are provided in the Appendix \ref{app_beta}.   Fig. \ref{fig_beta} shows 
the histogram of estimated $\beta$ for the newspaper data, blog data, and Wikipedia data.  \textcolor{black}{This figure confirms that the mode of estimated $\beta$ takes the value of approximately 0.5 for all datasets. 
} 
%
\subsubsection{Relation to the fractional dynamics}
Here, we address the relation between the fractional dynamics and the random walk model. 
\textcolor{black}{From Appendix \ref{app_arfima}}, the continuous version of Eq. \ref{eq_rw} corresponds to the fractional Langevin equation, 
which is the expansion of the Langevin equation \cite{eab2011fractional,magdziarz2007fractional}, 
\begin{equation}
\frac{d^{1-\beta}}{dt^{1-\beta}} r(t) =  \eta(t), \label{eq_frac_lan}
\end{equation}
where, on the condition that $\beta=0$, this equation is the normal Langevin equation. 
Here the Riemann-Liouville fractional derivative operator of $\frac{d^{\alpha}}{dt^{\alpha}}$ is defined by 
\begin{equation}
\frac{d^{\alpha}}{dt^{\alpha}}x(t) = \frac{d}{dt}\frac{1}{\Gamma(1-\alpha)} \int^{t}_{-\infty} x(t) (t-s)^{-\alpha} ds.
\end{equation}
This operator is satisfied with $(d^{^{1/\alpha}}/dt^{1/\alpha})^\alpha f(x)=(d/dt) f(x)$.
For example, in the case of $\alpha=1/3$, the operator is $1/3$ times the derivative operator,that is, three operations of $\frac{d^{1/3}}{{dt}^{1/3}}$ mean one normal derivative, 
\begin{equation}
\left(\frac{d^{1/3}}{dt^{1/3}}\right)^3 x(t)=\frac{d^{1/3}} {dt^{1/3}}\frac{d^{1/3}}{dt^{1/3}}\frac{d^{1/3}}{dt^{1/3}} x(t) =\frac{d}{dt} x(t).
\end{equation}
Therefore, in the case of the word counts, namely, $\beta=0.5$, we can obtain the half-order fractional Langevin equation,  
\begin{equation}
\sqrt{\frac{d}{dt}}r(t) \approx  \eta (t), \label{eq_sqrd_rw}
\end{equation}
where $\sqrt{\frac{d}{dt}}$ is the half derivative operator, $\sqrt{\frac{d}{dt}}=\frac{d^{1/2}}{dt^{1/2}}$.
\textcolor{black}{Thus, the properties of the word count time series are right-in-the-middle dynamics between the IID noise (zero-order differentiation) and the normal random walk (first-order differentiation).}
 \par
\textcolor{black}{Table \ref{tableA} provides a summary of the properties of the model given by Eq. \ref{eq_rw}.}
\par
\begin{table*}
\centering
\begin{tabular}{lccccc}
\hline
& Random walk  & &\multicolumn{1}{c}{Word counts}  & & IID noise  \\
\hline
$\beta$ & $\beta=0$    & $(0<\beta<0.5)$ & \multicolumn{1}{c}{$\beta=0.5$}       &  $(0.5<\beta<1.0)$ & $\beta=1$ \\
\multicolumn{6}{l}{\textcolor{black}{(i) Dynamics} statistics} \\ 
Stability & \multicolumn{3}{c|}{Unsteady} &  \multicolumn{2}{c}{Steady} \\
MSD $\sigma(L;r_j)^2$ [Eq. \ref{def_msd_r}] & \multicolumn{1}{c}{Normal-dif.}  &  \multicolumn{1}{|c}{Sub-dif.} &  \multicolumn{1}{|c|}{Ultraslow-dif.} & \multicolumn{2}{c}{Steady} \\
\textcolor{black}{$ \quad {}^{*}L$: Time-lag [days]} & $\propto L$ &\multicolumn{1}{|c|}{$\propto  L^{1-2/\beta}$} & \multicolumn{1}{|c|}{$\propto \log(L)$ } & \multicolumn{2}{c}{$\propto O(1)$ }\\
\textcolor{black}{PSD $P(\nu;r)$ [Eq. \ref{def_psd_r}}]  &  \multicolumn{5}{c}{$\propto \nu^{-2(1-\beta)}$} \\ 
\textcolor{black}{$ \quad {}^{*}\nu$: Frequency [1/days]}& $ \propto \nu^{-2}$ & & $\propto  \nu^{-1}$ &  & $\propto \nu^{0}$  \\  
\multicolumn{6}{l}{\textcolor{black}{(ii) Time} evolution} \\ 
&  \multicolumn{5}{c}{ $(d/dt)^{1-\beta} r_j(t) = \eta(t)$} \\ 
&  $\frac{d}{dt}r_j(t)=\eta(t)$ && $\sqrt{\frac{d}{dt}} r_j(t) = \eta(t)$ &&  $r_j(t)=\eta(t) $ \\
\hline
\end{tabular}
\caption{Summary of the model properties obtained by Eq. \ref{eq_rw}} 
\label{tableA}
\end{table*}
\subsection{Model of word counts}
\label{sec_rdmodel}
In the previous section,  we confirmed that the logarithmic diffusion in word counts can be explained by the random walk with power-law forgetting given by Eq. \ref{eq_rw} essentially. 
However, this random walk model cannot explain all the statistical properties of word counts we observed in this paper. For example, we cannot explain: 
\textcolor{black}{(i) the} discreteness of the row word counts $g_j(t)$ and (ii) the word-dependent constants $(a_j, b_j)$ in Eq. \ref{eq_msd_single} and $(v_j,w_j)$ in Eq. \ref{eq_p_f}. 
Thus, lastly, we discuss the connection between the essential dynamics of the concern of word $r_j(t)$ given by Eq. \ref{eq_rw} (i.e. the latent value) and the time series of word counts $g_j(t)$ or $f_j(t)$ (i.e. the observed value). \par
Here, we use the random diffusion model (RD model) introduced in \cite{PhysRevLett.100.208701,PhysRevE.87.012805,sano2009,RD_base} to sample $g_j(t)$ or $f_j(t)$.
The RD model is a kind of point process, which can be deduced from the simple model of the writing activity of independent bloggers \cite{RD_base}.
\textcolor{black}{In this model, values are sampled from the Poisson distribution of which the rate (or intensity) function is determined by a random variable or a stochastic process (i.e. the doubly stochastic Poisson process \cite{lowen2005fractal}).} In the case of blogs, the rate function is connected to the latent concern of word $r_j(t)$.
Particularly, the RD model is given by \cite{RD_base}
\begin{equation}
g_j(t) \sim Poi( \Lambda_j(t)), \label{eq_rd}
\end{equation}
and its rate function of the Poisson distribution (denoted by $Poi(.)$), $\Lambda_j(t)$ is determined by the following definition of the product:  
\begin{equation}
\Lambda_j(t) \sim  m(t) \times \hat{c_j} \times r_j(t)  \times (1+\hat{\delta}_j \times \Delta^{(0)}_j(t;\vector{\hat{\theta}})),\label{Lambda} 
\end{equation}
where 
\begin{itemize}
\item $m(t)$ is the (normalised) scale of the database such as the total number of blogs, assuming that $\sum_{t=1}^{T}m(t)/T \approx 1$ for normalisation (see Fig.\ref{tseries}(b)), where $m(t)$ is estimated by the ensemble median of the number of words at time t, as described in the Appendix \ref{app_m}. 
\item $\hat{c}_j$ is the scale of the $j$-th word, namely, the temporal means of the $j$-th word, where we estimate the mean of the raw word count of data $\hat{c}_j \approx \sum_{t=1}^{T}g_j(t)/T$. 
\item $r_j(t)$ is the scaled time-variation of concern of the $j$-th word sampled from Eq. \ref{eq_rw}, where we set the $r_j(1)=1$ (This sampling using Eq. \ref{eq_rw} is the no-verity of this study in comparison with Ref. \cite{RD_base}.) . 
\item $\hat{\delta}$ is the magnitude (i.e. the standard deviation) of the ensemble fluctuation, which may be related to the magnitude of the heterogeneity of bloggers \cite{RD_base}. 
\item $\Delta^{(0)}_j(t;\vector{\hat{\theta}})$ is the normalised ensemble fluctuation, which is sampled from the system-dependent random variable with a mean $0$, standard deviation $1$, and parameters $\vector{\hat{\theta}}$ that characterize the distribution.
\end{itemize}
\par
Note that in the previous study Ref. \cite{RD_base}, we estimated $r_j(t)$ from data directly by using the moving average for data analysis or used the assumption $r_j(t+1)-r_j(t) \approx 0$ for analytical calcultion. 
Thus, we could not discuss the properties of the dynamics as such in Ref \cite{RD_base} (The model only describes the fluctuation when dynamics of $r_j(t)$ are given).  
However, in this study we introduce the time-evolution model $r_j(t)$ given by Eq. \ref{eq_rw}, enabling us to calculate the basic dynamics of already popular words (The model describes not only the fluctuation but also the dynamics.). 
\par 
\textcolor{black}{
Also note that the word counts are confined from zero to the size of databases (i.e. the total number of articles for the newspaper and blog data and the total number of Wikipedia users for the Wikipedia data) in the actual data. However, our model does not consider this limitation, and this problem is substantially neglectable in our situation.
The main reasons for this are as follows:
\begin{itemize}
\item The time evolution is very slow (i.e. logarithmic diffusion) on the condition the initial value being $r_j(1)=1$ and the finite time step (a maximum of approximately 10 years); hence, $r_j(t)$ walks on around 1.
The cases taking a negative value (for which the rate function $\Lambda_j(t)<0$ and Eq. \ref{eq_rd} become meaningless) and a very large value (related to the limitation of the total number of articles ) were not sampled practically.
\item Almost all words take a very small temporal mean of the word counts $\hat{c}_j$ to be affected by the limitation of the total number of articles $M(t)$ ($\hat{c}_j << M(t)$).
\end{itemize}
Though in our case, we were almost able to avoid these problems of constraints without a special treatment, in general situations, such as infinite time step ($T \to \infty$), we may have to extend the model to explicitly describe these constraints.
} \par
Fig. \ref{fig_sim} compares the statistical properties of the word counts time-series between the empirical data and the numerical simulation of the RD model driven by the random walk model with the power-law forgetting given by Eq. \ref{eq_rd} and Eq. \ref{eq_rw} with respect to the (i) MSD, (ii) PSD, and (iii) (temporal) probability density function (PDF) of the differential $f_j(t+L)-f_j(t)$ 
(see Eq. \ref{def_pdf0}).
The results in these figures confirm that the numerical simulations are almost in accordance with the empirical observations in the newspaper, blog, and Wikipedia data, respectively. \par
In these simulations, $\eta^{(0)}_j(t)$ and $\Delta^{(0)}_j(t;\hat{\vector{\theta}})$ are sampled from the normalised noncentral t-distribution \textcolor{black}{$Z^{(0)}(4,2)$}. 
The normalised noncentral t-distribution $Z^{(0)}(\xi,\mu)$, of which the mean is zero and the standard deviaton 1, is the shifted and scaled noncentral t-distribution $Z^{(0)}(\xi,\mu)$. 
\textcolor{black}{The noncentral t-distribution $Z^{(0)}(\xi,\mu)$ is a skewed heavy tail distribution,  
the tail parameter $\xi$ determine the heaviness of tail and the noncentrality parameter $\mu$ determine the skewness of the distribution. On the condition that $\mu=0$, the noncentral t-distribution corresponds to (normal or no-skew) t-distribution.
The detail of this distribution given by Appendix \ref{app_tdist}. }
\par
%
\textcolor{black}{In the figure, we use the word-dependent or system-dependent parameters, namely, the mean frequency (scale) of word counts $\hat{c}_j$, the speed of the diffusion (or mean strength of outer shocks) $\hat{\eta}_j$ and the magnitude of ensemble fluctuation $\hat{\delta}_j$ (maybe related to the heterogeneity of bloggers) as follows: $\hat{c}_j=150.3$, $\hat{\eta}_j=0.068$ and $\hat{\delta}_j=0.15$ for ``Tachiba (i.e. position or standpoint in English)'' on the newspaper data, $\hat{c}_j=83.16$, $\hat{\eta}_j=0.12$ and $\hat{\delta}_j=0.065$ for ``Sanada (i.e. well-known Japanese family name)'' on the blog data,  and $\hat{c}_j=58.2$, $\hat{\eta}_j=0.12$ and $\hat{\delta}_j=0.14$ for ``Handle'' on the English Wikipedia data.} \par
Lastly, we show the relation between the RD model and the word-dependent constant $a_j$, $b_j$, in Eq. \ref{eq_msd_single} and $v_j$, $w_j$ in Eq. \ref{eq_p_f}. 
From Appendix \ref{app_msd_f}, the (mean of the temporal) MSD of $f(t)$ is written by 
\begin{eqnarray}
&& \left<f_j(t+L)-f_j(t) \right> \nonumber \\ 
&\approx& \left<\sigma_j(L)^2 \right> \\ 
&\approx&  a_j \log(L)+b_j \quad  (L\gg1)
\label{eq_msd_f}
\end{eqnarray}
where $a_{j}=2  \hat{c}_j^2 \hat{\eta}_j^2 /\Gamma(1/2)^2$, $b_j= \hat{c}_j^2 \hat{\eta}_j^2 (-2\log(4)-2\psi^{(0)}(\Gamma(1/2)^{-2}))/\Gamma(1/2)^2 +2 \hat{c}_j^2 \hat{\delta_j}^2+2 \hat{c}_j$, $\psi^{(0)}(x)$ is the digamma function and this curve is shown in the magenta thick dash-dotted lines in Figs. \ref{fig_sim} in (d-f). \par
\textcolor{black}{In addition, from Appendix \ref{app_p_f} the power spectrum density of $f_j(t)$ is written by, 
\begin{eqnarray}
&&P(\nu;f_j) 
\approx v_j \times (2 \sin(2 \pi \nu/2))^{-1} +w_j  . \label{eq_psd_f0}
\end{eqnarray}
where $v_{j}=\hat{c}_j^2  \hat{\eta}_j^2$, $w_j=\hat{c}_j^2 \hat{\delta}_j^2 +\hat{c}_j$ and this curve is shown in the magenta thick dash-dotted lines in Figs. \ref{fig_sim} in (g-i).
}
 \par
\textcolor{black}{We also verified that the model cannot reproduce the statistical properties of the empirical data 
 on the condition that $\beta$ does not take around $0.5$. 
 Fig. \ref{fig_sim2} shows the results in which we compare the empirical data with the numerical simulations for different $\beta$ (the speeds of forgetting) and $m(t)$ (database size): (i) the IID noise ($\beta \to 1$), (ii) the simple random walk model ($\beta=0$) and (iii) the case where the database size is constant ($m(t)=1$ and $\beta=0.5$).
 This figure confirms that the IID noise ($\beta \to 1$) and the random walk model ($\beta=0$) cannot reproduce the empirical properties. In addition, $m(t)$ is not essential in reproducing the empirical properties $f_j(t)$ (see panels in the second line).}
\par 
Note that the model given by Eq. \ref{eq_rd} and Eq. \ref{eq_rw} can also explain the ``fluctuation scaling'', which is known as the other property of word counts on social media such as blogs \cite{eisler2008fluctuation,sano2009, RD_base}.  
\textcolor{black}{The relation between the empirical fluctuation scaling and the model given by Eq. \ref{eq_rd} and  Eq. \ref{eq_rw} will be discussed in our next paper.}
\section{Conclusion and discussion}
In this paper, from the viewpoint of the diffusion of complex systems, we investigated the stability of the time-series of word counts of already popular words (i.e. ``mature phase words``) on some nationwide language data sets (newspaper articles, blog articles, and Wikipedia page views). \par  
Firstly, by analysing the data, we commonly observed a logarithmic-like diffusion (i.e. an ultraslow-like diffusion) in word counts between different data sets. 
Although ultraslow diffusion has been extensively studied by using theories or mathematical models, few empirical observations have been reported.
Moreover, this logarithmic-like diffusion from observed facts is not in conflict with the intuition in which languages are basically stationary but change constantly. \textcolor{black}{This intuition may be related to the empirical studies of the stability of word count statistics: (i) more frequent words change slower \cite{lieberman2007quantifying,gerlach2016similarity}, and (ii) some observations implied small stable core (kernel) vocabularies as distinguished from other many vocabularies for specific communications which are not shared by all people \cite{ferrer2001two,ferrer2017origins}.} \par
Secondly, we show that the logarithmic diffusion of word counts is essentially explained by the random walk model with forgetting in the power law. This random walk model corresponds with the fractional Langevin equation, which is a typical mathematical model in previous theoretical studies of anomalous diffusions. 
\textcolor{black}{
The speed of forgetting characterized by the power-law exponent $\beta \approx 0.5$ in Eq. \ref{eq_rw} has the following meanings:
\begin{itemize}
\item The border (or thresholds) between the stationary and the nonstationary (Eq. \ref{th_msd}), and
\item Right-in-the-middle dynamics between IID noise and the normal random walk (Eq. \ref{eq_frac_lan}).
\begin{equation}
\begin{cases}
\frac{d^{0}}{{dt}^{0}}r_j(t)=r_j(t) =  \eta(t) & \text{(IID noize; $\beta=1$)} \\
\frac{d^{1/2}}{dt^{1/2}}r_j(t)=\sqrt{\frac{d}{dt}} r_j(t) =  \eta(t)  & \text{(Word counts; $\beta=1/2$)} \\
\frac{d^{1}}{dt^{1}}r_j(t)=\frac{d}{dt} r_j(t) =  \eta(t) & \text{(Random walk, $\beta=0$)}, \\
\end{cases}
\end{equation}
\end{itemize}
which are summarized in Table. \ref{tableA}.} \par
Thirdly, we confirmed the RD model driven by the random walk model with forgetting given by Eqs. \ref{eq_rw} and \ref{eq_rd} in the power law can almost reproduce the empirical properties time-series of typical words \textcolor{black}{(Fig. \ref{fig_sim})}: (i) MSD, (ii) PSD, and (iii) PDF.  \par
\par
Although our model can explain the dynamical properties of the word counts time series, our framework cannot explain the model parameter $\beta \approx 0.5$ in Eq. \ref{eq_rw}. 
This special value, $\beta \approx 0.5$, which is the threshold between steady dynamics and unsteady dynamics, is observed detail-independently (i.e. words, languages, and media independent) as far as we investigated. Thus, clarifying the origin of the parameter $\beta \approx 0.5$ may provide a clue to understand the fundamental dynamical and memory properties of human systems or societies as complex systems. 
\textcolor{black}{
 In the micro-level study, namely, the study of single documents, the power law of the forgetting process with the word-dependent exponents which are distributed approximately around 0.5, is used to explain the empirical stretched exponential distribution of the recurrence distance of words (e.g., for the phrase ``This cat is big. That cat is small.'', the recurrence distance of ``cat'' is 4.)  \cite{altmann2009beyond}. This quantitative similarity of the power law of forgetting dynamics between data of micro- (single document) and macro-level (nation-wide collective behavior datasets) studies might provide important suggestions to understand the origin of the 0.5th exponent obtained in our macro-level study from micro-level human behavior. 
}

\begin{acknowledgments}
The authors would like to thank Hottolink, Inc. for providing the data. This work was supported by JSPS KAKENHI, Grant Number \textcolor{black}{JP17K13815}.
\end{acknowledgments}
%
\bibliography{adj6c_pnas}

\begin{thebibliography}{10}

\bibitem{altmann2016statistical}
E.~G. Altmann and M. Gerlach, {\em Creativity and Universality in Language}
  (Springer, ADDRESS, 2016), pp.\ 7--26.

\bibitem{metzler2000random}
R. Metzler and J. Klafter, Physics reports {\bf 339},  1  (2000).

\bibitem{da2014ultraslow}
M.~A.~A. da~Silva, G.~M. Viswanathan, and J.~C. Cressoni, Physical Review E
  {\bf 89},  052110  (2014).

\bibitem{bouchaud1990anomalous}
J.-P. Bouchaud and A. Georges, Physics reports {\bf 195},  127  (1990).

\bibitem{burov2011single}
S. Burov, J.-H. Jeon, R. Metzler, and E. Barkai, Physical Chemistry Chemical
  Physics {\bf 13},  1800  (2011).

\bibitem{lowen2005fractal}
S.~B. Lowen and M.~C. Teich, {\em Fractal-based point processes} (John Wiley \&
  Sons, ADDRESS, 2005), Vol.~366.

\bibitem{sinai1983limiting}
Y.~G. Sinai, Theory of Probability \& Its Applications {\bf 27},  256  (1983).

\bibitem{godec2014localisation}
A. Godec {\it et~al.}, Journal of Physics A: Mathematical and Theoretical {\bf
  47},  492002  (2014).

\bibitem{sanders2014severe}
L.~P. Sanders {\it et~al.}, New Journal of Physics {\bf 16},  113050  (2014).

\bibitem{bodrova2015ultraslow}
A.~S. Bodrova, A.~V. Chechkin, A.~G. Cherstvy, and R. Metzler, New Journal of
  Physics {\bf 17},  063038  (2015).

\bibitem{cherstvy2013population}
A.~G. Cherstvy and R. Metzler, Physical Chemistry Chemical Physics {\bf 15},
  20220  (2013).

\bibitem{eab2011fractional}
C.~H. Eab and S.~C. Lim, Physical Review E {\bf 83},  031136  (2011).

\bibitem{song2010modelling}
C. Song, T. Koren, P. Wang, and A.-L. Barab{\'a}si, Nature Physics {\bf 6},
  818  (2010).

\bibitem{boyer2011non}
D. Boyer, M.~C. Crofoot, and P.~D. Walsh, Journal of The Royal Society
  Interface  rsif20110582  (2011).

\bibitem{matan2002crumpling}
K. Matan, R.~B. Williams, T.~A. Witten, and S.~R. Nagel, Physical Review
  Letters {\bf 88},  076101  (2002).

\bibitem{richard2005slow}
P. Richard {\it et~al.}, Nature materials {\bf 4},  121  (2005).

\bibitem{petersen2012statistical}
A.~M. Petersen, J. Tenenbaum, S. Havlin, and H.~E. Stanley, Scientific reports
  {\bf 2},    (2012).

\bibitem{gerlach2013stochastic}
M. Gerlach and E.~G. Altmann, Physical Review X {\bf 3},  021006  (2013).

\bibitem{lieberman2007quantifying}
E. Lieberman {\it et~al.}, Nature {\bf 449},  713  (2007).

\bibitem{michel2011quantitative}
J.-B. Michel {\it et~al.}, science {\bf 331},  176  (2011).

\bibitem{link1}
E.~G. Altmann and M. Gerlach, Physicists' papers on natural language from a
  complex systems viewpoint,
  http://www.pks.mpg.de/mpi-doc/sodyn/physicist-language/.

\bibitem{abrams2003linguistics}
D.~M. Abrams and S.~H. Strogatz, Nature {\bf 424},  900  (2003).

\bibitem{altmann2015statistical}
E.~G. Altmann and M. Gerlach, arXiv:1502.03296  (2015).

\bibitem{cong2014approaching}
J. Cong and H. Liu, Phys Life Rev. {\bf 11},  598  (2014).

\bibitem{i2005zipf}
R.~F. i~Cancho, Eur. Phys. J. B {\bf 47},  449  (2005).

\bibitem{nikkei}
{\relax Nikkei Inc.} and {\relax Nikkei Business Publications, Inc.}, Shinbun
  trend (web system), http://ntrend.nikkei.co.jp/.

\bibitem{wikitionaly}
Wiktionary:Frequency lists,
  https://en.wiktionary.org/wiki/Wiktionary:Frequency\_lists.

\bibitem{RD_base}
H. Watanabe, Y. Sano, H. Takayasu, and M. Takayasu, Physical Review E {\bf 94},
   052317  (2016).

\bibitem{idadic}
A. Masayuki and M. Yuji, User's manual (ipadic),
  http://chasen.naist.jp/snapshot/ipadic/ipadic/doc/ipadic-ja.pdf, 2003.

\bibitem{wikipedia_pageview}
Analytics/AQS/Pageviews,
  https://wikitech.wikimedia.org/wiki/Analytics/AQS/Pageviews.

\bibitem{granger1980introduction}
C.~W. Granger and R. Joyeux, Journal of time series analysis {\bf 1},  15
  (1980).

\bibitem{burnecki2014algorithms}
K. Burnecki and A. Weron, Journal of Statistical Mechanics: Theory and
  Experiment {\bf 2014},  P10036  (2014).

\bibitem{magdziarz2007fractional}
M. Magdziarz and A. Weron, Studia Math {\bf 181},  47  (2007).

\bibitem{PhysRevLett.100.208701}
S. Meloni, J. G\'omez-Garde\~nes, V. Latora, and Y. Moreno, Phys. Rev. Lett.
  {\bf 100},  208701  (2008).

\bibitem{PhysRevE.87.012805}
Y. Sano {\it et~al.}, Phys. Rev. E {\bf 87},  012805  (2013).

\bibitem{sano2009}
Y. Sano, K.~K. Kaski, and M. Takayasu,  in {\em Proc. Complex '09} (Springer,
  Berlin, Germany, 2009), No.~2, pp.\ 195--198.

\bibitem{eisler2008fluctuation}
Z. Eisler, I. Bartos, and J. Kertesz, Adv. Phys. {\bf 57},  89  (2008).

\bibitem{gerlach2016similarity}
M. Gerlach, F. Font-Clos, and E.~G. Altmann, Phys. Rev. X {\bf 6},  021009
  (2016).

\bibitem{ferrer2001two}
R. Ferrer~i Cancho and R.~V. Sol{\'e}, Journal of Quantitative Linguistics {\bf
  8},  165  (2001).

\bibitem{ferrer2017origins}
R. Ferrer-i Cancho and M.~S. Vitevitch, arXiv preprint arXiv:1801.00168
  (2017).

\bibitem{altmann2009beyond}
E.~G. Altmann, J.~B. Pierrehumbert, and A.~E. Motter, PLOS one {\bf 4},  e7678
  (2009).

\bibitem{abramowitz1964handbook}
M. Abramowitz and I.~A. Stegun, {\em Handbook of mathematical functions: with
  formulas, graphs, and mathematical tables} (Courier Corporation, ADDRESS,
  1964), Vol.~55.

\bibitem{hypergeom}
http://functions.wolfram.com/HypergeometricFunctions/.

\bibitem{johnson1994continuous}
N. Johnson, {\em Continuous univariate distributions} (Wiley, New York, 1994).

\end{thebibliography}


\begin{thebibliography}{10}

\bibitem{song2010modelling}
Chaoming Song, Tal Koren, Pu~Wang, and Albert-L{\'a}szl{\'o} Barab{\'a}si.
\newblock Modelling the scaling properties of human mobility.
\newblock {\em Nature Physics}, 6(10):818--823, 2010.

\bibitem{wikipedia_pageview}
Analytics/aqs/pageviews.
\newblock https://wikitech.wikimedia.org/wiki/Analytics/AQS/Pageviews.

\bibitem{RD_base}
Hayafumi Watanabe, Yukie Sano, Hideki Takayasu, and Misako Takayasu.
\newblock Statistical properties of fluctuations of time series representing
  appearances of words in nationwide blog data and their applications: An
  example of modeling fluctuation scalings of nonstationary time series.
\newblock {\em Physical Review E}, 94(5):052317, 2016.

\bibitem{wikitionaly}
Wiktionary:frequency lists.
\newblock https://en.wiktionary.org/wiki/Wiktionary:Frequency\_lists.

\bibitem{hock2009language}
Hans~Henrich Hock and Brian~D Joseph.
\newblock {\em Language history, language change, and language relationship: An
  introduction to historical and comparative linguistics}, volume 218.
\newblock Walter de Gruyter, 2009.

\bibitem{petersen2012statistical}
Alexander~M Petersen, Joel Tenenbaum, Shlomo Havlin, and H~Eugene Stanley.
\newblock Statistical laws governing fluctuations in word use from word birth
  to word death.
\newblock {\em Scientific reports}, 2, 2012.

\bibitem{lieberman2007quantifying}
Erez Lieberman, Jean-Baptiste Michel, Joe Jackson, Tina Tang, and Martin~A
  Nowak.
\newblock Quantifying the evolutionary dynamics of language.
\newblock {\em Nature}, 449(7163):713, 2007.

\bibitem{nikkei}
{\relax Nikkei Inc.} and {\relax Nikkei Business Publications, Inc.}
\newblock Shinbun trend (web system).
\newblock http://ntrend.nikkei.co.jp/.

\bibitem{idadic}
Asahara Masayuki and Matsumoto Yuji.
\newblock User's manual (ipadic).
\newblock http://chasen.naist.jp/snapshot/ipadic/ipadic/doc/ipadic-ja.pdf,
  2003.

\bibitem{link1}
Eduardo~G. Altmann and Martin Gerlach.
\newblock Physicists' papers on natural language from a complex systems
  viewpoint.
\newblock http://www.pks.mpg.de/mpi-doc/sodyn/physicist-language/.

\bibitem{abrams2003linguistics}
Daniel~M Abrams and Steven~H Strogatz.
\newblock Linguistics: Modelling the dynamics of language death.
\newblock {\em Nature}, 424(6951):900--900, 2003.

\bibitem{altmann2015statistical}
Eduardo~G Altmann and Martin Gerlach.
\newblock Statistical laws in linguistics.
\newblock {\em arXiv:1502.03296}, 2015.

\bibitem{cong2014approaching}
Jin Cong and Haitao Liu.
\newblock Approaching human language with complex networks.
\newblock {\em Phys Life Rev.}, 11(4):598--618, 2014.

\bibitem{i2005zipf}
R~Ferrer i~Cancho.
\newblock Zipf's law from a communicative phase transition.
\newblock {\em Eur. Phys. J. B}, 47(3):449--457, 2005.

\bibitem{gerlach2016similarity}
Martin Gerlach, Francesc Font-Clos, and Eduardo~G Altmann.
\newblock Similarity of symbol frequency distributions with heavy tails.
\newblock {\em Phys. Rev. X}, 6(2):021009, 2016.

\bibitem{ferrer2001two}
Ramon Ferrer~i Cancho and Ricard~V Sol{\'e}.
\newblock Two regimes in the frequency of words and the origins of complex
  lexicons: Zipffs law revisited?
\newblock {\em Journal of Quantitative Linguistics}, 8(3):165--173, 2001.

\bibitem{ferrer2017origins}
Ramon Ferrer-i Cancho and Michael~S Vitevitch.
\newblock The origins of zipf's meaning-frequency law.
\newblock {\em arXiv preprint arXiv:1801.00168}, 2017.

\bibitem{burnecki2014algorithms}
Krzysztof Burnecki and Aleksander Weron.
\newblock Algorithms for testing of fractional dynamics: a practical guide to
  arfima modelling.
\newblock {\em Journal of Statistical Mechanics: Theory and Experiment},
  2014(10):P10036, 2014.

\bibitem{granger1980introduction}
Clive~WJ Granger and Roselyne Joyeux.
\newblock An introduction to long-memory time series models and fractional
  differencing.
\newblock {\em Journal of time series analysis}, 1(1):15--29, 1980.

\end{thebibliography}
\appendix
\clearpage
\begin{center} {\bf Supporting information} \end{center}\par
\renewcommand{\theequation}{A\arabic{equation}}
\renewcommand{\thefigure}{A-\arabic{figure}}
\renewcommand{\thetable}{S-A\arabic{table}}
\setcounter{figure}{0}
\setcounter{equation}{0}
\section{Estimation of normalised scale of database $m(t)$ from the data}
\label{app_m}
We estimate the  normalised scale of database $m(t)$ such as the total number of blogs by using the moving median as follows: 
\textcolor{black}{
	\begin{enumerate}
	\item[Step 1.] We create a set $S$ consisting of the indexes of words such that $c_j$ takes a value larger than the threshold $\hat{c}_j(t) \geq 100$.
	\item[Step 2.] We estimate $m(t)$ as the median of $\{f_j(t)/\hat{c}_j:j \in S \}$ with respect to $j$.
	\item[Step 3.] For $t=1,2,3,\cdots,T$, we calculate $m(t)$ using step 2. 
	\end{enumerate}
} \par
Here, we use only words with $\hat{c}_j \geq 100$ in step 1 because we neglect the discreteness. In step 2, we apply the median because of its robustness to outliers.

\renewcommand{\theequation}{B\arabic{equation}}
\renewcommand{\thefigure}{B-\arabic{figure}}
\renewcommand{\thetable}{S-B\arabic{table}}
\setcounter{figure}{0}
\setcounter{equation}{0}
\section{Mean square displacement of power-law forgetting process}
\label{MSD_forget}
We calculate the MSD of the following power-law forgetting process given by Eq. \ref{r_forget}, 
\begin{equation}
r(t)=\sum_{s=0}^{\infty} \theta(s) \cdot \eta(t-s), 
\end{equation}
where 
\begin{equation} 
\theta(s)=\frac{(s+a(\beta))^{-\beta}}{Z(\beta)}, \label{theta}
\end{equation}
and
\begin{equation}
a \equiv a(\beta) \equiv  Z(\beta)^{-1/\beta}, 
\end{equation}
$Z(\beta)>0$ is an arbitrary constant.
The MSD can be calculated as 
\begin{eqnarray}
&&\left<(r(t+L)-r(t))^2 \right> \nonumber \\
&=&\left<(\sum^{\infty}_{s=0} \theta(s)  \eta(t+L-s) - \sum^{\infty}_{s'=0}\theta(s')  \eta(t-s'))^2 \right> \nonumber \\
&=&\left< \left \{\sum^{-1}_{s=-L} \theta(s+L) \eta(t-s) +\sum^{\infty}_{s=0} (\theta(s+L) \right.  \right. \nonumber \\
&-& \left. \left. \theta(s) ) \eta(t-s) \right\}^2 \right> \\ 
&=&\hat{\eta}^2  (\sum^{-1}_{s=-L} \theta(s+L)^2 +\sum^{\infty}_{s=0} (\theta(s+L) - \theta(s) )^2)  \\ 
&=&\hat{\eta}^2  (S_1+S_2),  
\label{ss}
\end{eqnarray}
where 
\begin{eqnarray}
S_1&=&\sum^{-1}_{s=-L} \theta(s+L)^2 \\
&=&\begin{cases}
\frac{1}{Z(\beta)^2}(\xi(2\beta,a)-\xi(2\beta,a+L)) & (\beta \neq 1/2, \beta \geq 0) \\ 
\frac{1}{Z(\beta)^2}(\psi^{(0)}(a+L)-\psi^{(0)}(a)) & (\beta=1/2) \\ 
\end{cases},   \nonumber \\
\label{s1_ans}
\end{eqnarray}
and
\begin{eqnarray}
S_2&=&\sum^{\infty}_{s=0} (\theta(s+L) - \theta(s) )^2 \nonumber \\
&=&\lim_{N \to \infty}\sum^{N}_{s=0} (\theta(s+L) - \theta(s) )^2 \\
&=&\lim_{N \to \infty}\sum^{N}_{s=0} \theta(s+L)^2+\sum^{N}_{s=0}  \theta(s)^2  -2 \sum^{N}_{s=0} \theta(s+L) \theta(s). \nonumber \\ \label{S_2}
\end{eqnarray}
\par
We calculate three terms in Eq. \ref{S_2}, respectively. 
The first term of Eq. \ref{S_2} is given by 
\begin{eqnarray}
&&\sum^{N}_{s=0} \theta(s+L)^2=  \nonumber \\
&&\begin{cases}
\frac{1}{Z(\beta)^2}(\zeta(2\beta,a+L)-\zeta(2\beta,a+L+N+1))  & \text{($\beta \neq 1/2$)}\\
\frac{1}{Z(\beta)^2}(\psi^{(0)}(a+L+N+1)-\psi^{(0)}(a+L)) & \text{($\beta = 1/2$)}
\end{cases} \nonumber \\
\end{eqnarray}
where $\zeta(\alpha,x)$ is the Hurwitz zeta function $\zeta(s,q)=\sum^{\infty}_{n=0}1/(q+n)^s$, and $\psi^{(0)}(x)$ is the digamma function, $\psi^{(0)}(x)=d/dx \log(\Gamma(x))$. 
The second term of Eq. \ref{S_2} is given by 
\begin{eqnarray}
&&\sum^{N}_{s=0} \theta(s)^2  \nonumber \\
&&=\begin{cases}
\frac{1}{Z(\beta)^2}(\zeta(2\beta,a)-\zeta(2\beta,a+N+1)) & \text{($\beta \neq 1/2$)}\\
\frac{1}{Z(\beta)^2}(\psi^{(0)}(a+N+1)-\psi^{(0)}(a)) & \text{($\beta = 1/2$)} 
\end{cases} \nonumber \\
\end{eqnarray}
For $N\gg1$, using the general formula $\zeta(x,\alpha)$ for $x\gg1$
\begin{equation}
\zeta(x,\alpha) \approx \frac{1}{\alpha-1}x^{-\alpha+1}+ \frac{1}{2} x^{-\alpha}+\cdots
\end{equation}
and
\begin{equation}
\psi^{(0)}(x) \approx \log(x)-\frac{1}{2x} + \cdots.
\end{equation}
we can obtain the approximation 
\begin{eqnarray}
&&\sum^{N}_{s=0} \theta(s+L)^2  \approx  \nonumber \\
&&
\begin{cases}
\frac{1}{Z(\beta)^2}(\zeta(2\beta,a+L)+\frac{(a+L+N+1)^{1-2\beta}}{(1-2\beta)})  & \text{($\beta \neq 1/2$)}\\
\frac{1}{Z(\beta)^2}(\log(a+L+N+1)-\psi^{(0)}(a+L)) & \text{($\beta = 1/2$)} 
\end{cases} \nonumber \\  \label{u3_1}
\end{eqnarray}
and
\begin{eqnarray}
\sum^{N}_{s=0} \theta(s)^2 
 \approx \begin{cases}
\frac{1}{Z(\beta)^2}(\zeta(2\beta,a)+\frac{(a+N+1)^{1-2\beta}}{(1-2\beta)}) & \text{($\beta \neq 1/2$)}\\
\frac{1}{Z(\beta)^2}(\log(a+N+1)-\psi^{(0)}(a)) & \text{($\beta = 1/2$)}
\end{cases} \nonumber \\ \label{u3_2} 
\end{eqnarray}
. \par
Lastly, we calculate the third term of Eq. \ref{S_2}.
Using the Euler-Maclaurin formula \cite{abramowitz1964handbook},  
\begin{eqnarray}
&&\sum^{b}_{k=a}g(k) \approx  \nonumber \\
&&\int^{b}_{a}g(x)dx+\frac{1}{2}(g(b)+g(a))+\frac{1}{12}(\frac{d}{dx}g(x)|_{b}-\frac{d}{dx}g(x)|_{a}) \nonumber \\
\label{euler}
\end{eqnarray}
 and $(x+N)^{-\beta} \to 0$, $(x+N)^{-\beta-1} \to 0$ for $N\gg1$, we can obtain
\begin{eqnarray}
&&\sum^{N}_{s=0} \theta(s+L) \theta(s) \approx \nonumber \\
&& \frac{1}{Z(\beta)^2}\{a^{-\beta} (a+L)^{-\beta} \nonumber \\
&& \frac{1}{Z(\beta)^2}\{\frac{1}{2} (a+1)^{-\beta} (a+1+L)^{-\beta} \nonumber \\
&&-\frac{(-\beta)}{12}((a+1)^{-\beta}(a+1+L)^{-\beta-1} \nonumber \\
&&+(a+1)^{-\beta-1}(a+1+L)^{-\beta}) \nonumber \\ 
&&+\int^{N}_{1}(x+a)^{-\beta} (x+a+L)^{-\beta}dx \}. \label{S2_0}
\end{eqnarray}
Here, we separate the term of $s=0$ from the summation to improve the accuracy.  \par
We calculate the integration term of Eq. \ref{S2_0}, 
\begin{eqnarray}
&&\int^{N}_{1}(x+a)^{-\beta} (x+a+L)^{-\beta}dx \\
&&=L^{-2\beta+1} \int^{\frac{N+a}{L}}_{\frac{a+1}{L}} x^{-\beta} (x+1)^{-\beta}dx. 
\end{eqnarray}
Executing the integration, 
\begin{eqnarray}
&&G(x,\beta)  \equiv \int x^{-\beta} (x+1)^{-\beta}dx  \nonumber \\
&&\begin{cases}
\text{($\beta \geq 0$, $\beta$ is a non-integer )} \\
\frac{x^{1-\beta}{}_{2}F_{1}(1-\beta,\beta,2-\beta,-x)}{1-\beta} \\
\text{$(\beta=1,2,3,4)$} \\ 
p^{(\beta)}_1\log(\frac{x+1}{x}) \\
+\sum^{\beta}_{k=2}\frac{p^{(\beta)}_k}{-k+1}((x+1)^{-k+1}+(-1)^kx^{-k+1}) \\
\end{cases}. \nonumber \\ \label{s2_int}
\end{eqnarray}
Here, we neglect an integral constant and  ${}_{2}F_{1}(a,b,c;x)$ is the Gaussian hypergeometric function \cite{hypergeom}, 
\begin{equation}
{}_{2}F_{1}(a,b,c;x)=\sum^{\infty}_{n=1}\frac{(a)_n(b)_n}{(c)_n}\frac{x^n}{n!}, \label{hyper_geom}
\end{equation} 
where $(A)_n=\Gamma(A+n)/\Gamma(A)$.
In additon, 
\begin{equation}
p^{(n)}_i=\frac{(-1)^n \Gamma(2n-i)}{\Gamma(n-i+1)\Gamma(n)}. \label{pp}
\end{equation}
is satisfied with a partial fraction decomposition
\begin{equation}
\frac{1}{(x+1)^n \cdot x^n}=\sum^{n}_{k=1}\frac{p^{(n)}_k}{(x+1)^k}+\frac{(-1)^k p^{(n)}_k}{x^k}.
\end{equation}
\par
For $x\gg1$, using the asymptotic formula of a Gaussian hypergeometric function \cite{hypergeom}, when $b-a$ and $c$ are non-integers, 
\begin{eqnarray}
{}_2F_1(a,b,c;x) \approx \frac{\Gamma(b-a) \Gamma(c) (-x)^{-a}}{\Gamma(b) \Gamma(c-a)}+\frac{\Gamma(a-b) \Gamma(c) (-x)^{-b}}{\Gamma(a) \Gamma(c-b)} \nonumber \\
\end{eqnarray}
or when $b=a$ and $c-a$ are positive integers,
\begin{eqnarray}
&&{}_2F_1(a,b,c;x) \approx  \nonumber \\
&& \Gamma(c) (\log(-x)-\psi^{(0)}(c-a)-\psi^{(0)}(a)-2 \gamma)(-x)^{a}, \nonumber \\
\end{eqnarray}
for $x\gg1$, the hypergeometric function in Eq. \ref{s2_int} is simplified 
\begin{eqnarray}
&&{}_2F_1(1-\beta,\beta,2-\beta,-x)= \nonumber \\ 
&&\begin{cases}
\frac{1-\beta}{1-2\beta} x^{-\beta}+\frac{\Gamma(2\beta-1)\Gamma(2-\beta)}{\Gamma(\beta)} x^{-(1-\beta)} & (0 < \beta < 1/2) \\ 
\frac{1}{2} (\log(x)+\log(4)) x^{-\beta} & (\beta=1/2) \\
\frac{\Gamma(2\beta-1)\Gamma(2-\beta)}{\Gamma(\beta)}x^{-(1-\beta)} & (\beta > 1/2)
\end{cases}. \nonumber \\
\end{eqnarray}
Substituting the results of these integrations into Eq. \ref{S2_0}, for $N\gg1$, 
we can obtain
\begin{eqnarray}
&&\sum^{N}_{s=0} \theta(s+L) \theta(s) \\
&\approx&  \frac{1}{Z(\beta)^2}\{ \nonumber \\
&& a^{-\beta} (a+L)^{-\beta} \nonumber \\
&&\frac{1}{2} (a+1)^{-\beta} (a+1+L)^{-\beta} \nonumber \\
&&+\frac{\beta}{12}((a+1)^{-\beta}(a+1+L)^{-\beta-1} \nonumber \\
&&+(a+1)^{-\beta-1}(a+1+L)^{-\beta}) \nonumber \\ 
&&-L^{1-2\beta}G((a+1)/L,\beta)+G_2(N,L,\beta) \} \nonumber \\ \label{u3_3}
\end{eqnarray}
where 
\begin{eqnarray}
&&G_2(N,L,\beta) \equiv \nonumber \\ 
&&\begin{cases}
\text{($\beta \neq 1/2, \beta>0$ $\beta$ is non-integer)} \\
\frac{N^{1-2\beta}}{1-2\beta}+L^{1-2\beta}\frac{\Gamma(2\beta-1)\Gamma(2-\beta)}{(1-\beta)\Gamma(\beta)}   \\
\text{ $(\beta=1/2)$ }\\
\log(N+a)-\log(L)+\log(4)  \\
  \text{($\beta=1,2,3,\cdots$)} \\
0 \\
\end{cases}. \nonumber \\ 
\end{eqnarray}
\par
From the results of Eq. \ref{S_2}, Eq. \ref{u3_1}, Eq. \ref{u3_2}, and Eq. \ref{u3_3}, $S_2$ is calculated as 
\begin{eqnarray}
&&Z(\beta)^2 S_2 \approx   \nonumber \\
&&- 2a^{-\beta}(a+L)^{-\beta} \nonumber \\
&&- (a+1)^{-\beta} (a+1+L)^{-\beta} \nonumber \\
&&-\frac{\beta}{6}((a+1)^{-\beta}(a+1+L)^{-\beta-1} \nonumber \\
&&+(a+1)^{-\beta-1}(a+1+L)^{-\beta}) \nonumber \\ 
&&+2L^{1-2\beta} G((a+1)/L,\beta)+ \nonumber \\
&&\begin{cases}
 \text{($\beta > 0$, $\beta \neq 1/2$, $\beta$ is non-integer )} \\
-2L^{1-2\beta} \frac{\Gamma(2\beta-1)\Gamma(2-\beta)}{(1-\beta)\Gamma(\beta)}+\zeta(2\beta,a)+\zeta(2\beta,a+L)  \\
\text{($\beta=1/2$)} \\
2\log(L)-2\log(4)-\psi^{(0)}(a)-\psi^{(0)}(a+L) \\
\text{($\beta=1,2,3,\cdots$)} \\
\zeta(2\beta,a)+\zeta(2\beta,a+L).
\end{cases} \nonumber \\
 \label{s2_ans}
\end{eqnarray} 
 \par
Conseqently, substituting Eq. \ref{ss} into Eq. \ref{s1_ans} and Eq. \ref{s2_ans}, we can obtain the MSD, 
\begin{eqnarray}
&&\left<(r(t+L)-r(t))^2 \right>  \approx \hat{\eta}^2/Z(\beta)^2  [ \nonumber \\
&&- 2a^{-\beta}(a+L)^{-\beta} \nonumber \\
&&- (a+1)^{-\beta} (a+1+L)^{-\beta} \nonumber \\
&&-\frac{\beta}{6}((a+1)^{-\beta}(a+1+L)^{-\beta-1} \nonumber \\
&&+(a+1)^{-\beta-1}(a+1+L)^{-\beta}) \nonumber \\ 
&&+2L^{1-2\beta} G((a+1)/L,\beta)+ \nonumber \\
&&\begin{cases}
 \text{($\beta > 0$, $\beta \neq 1/2$, $\beta$ is non-integer)} \\
-2L^{1-2\beta} \frac{\Gamma(2\beta-1)\Gamma(2-\beta)}{(1-\beta)\Gamma(\beta)}+2\zeta(2\beta,a) \\
\text{$(\beta=1/2)$ }
 \\2\log(L)-2\log(4)-2\psi^{(0)}(a) \nonumber \\
 \text{($\beta$ is integer)} \\
 2\zeta(2\beta,a)  \\
\end{cases} \nonumber \\
&&]. \label{diff_ans}
\end{eqnarray} 
For $L\gg1$, we can calculate the asymptotic form, 
\begin{eqnarray}
&&\left<(r(t+L)-r(t))^2 \right> \approx \hat{\eta}^2/Z(\beta)^2 \cdot \\
&&\begin{cases}
L & (\beta=0) \\
-2L^{1-2\beta} \frac{\Gamma(2\beta-1)\Gamma(2-\beta)}{(1-\beta)\Gamma(\beta)}+2 \zeta(2\beta,a) & (0<\beta<0.5) \\
2\log(L)-2\log(4)-2\phi^{(0)}(a) &(\beta=0.5)\\
2\zeta(2\beta,a) &(\beta>0.5) \\
\end{cases}, \nonumber \\  
\label{diff_ans1b}
\end{eqnarray}
which we use for $x<<1$ ${}_{2}F_{1}(1-\beta,\beta,2-\beta,-x) \approx 1$ and 
$x^{0.5}{}_{2}F_{1}(1-\beta,\beta,2-\beta,-x) =\log(\sqrt{x}+\sqrt{1+x}) \to 0$ $(x \to 0, \beta=0.5)$. 
 \par
Therefore, for $L\gg1$, the dominant term is 
\begin{eqnarray}
\left<(r(t+L)-r(t))^2 \right> \propto 
\begin{cases}
L^{1-2\beta}  & (0 \leq \beta <0.5) \\
\log(L) &(\beta=0.5)\\
O(1) &(\beta > 0.5) \\
\end{cases}.
\end{eqnarray}

\renewcommand{\theequation}{C\arabic{equation}}
\renewcommand{\thefigure}{C-\arabic{figure}}
\renewcommand{\thetable}{S-\arabic{table}}
\setcounter{figure}{0}
\setcounter{equation}{0}
\section{\textcolor{black}{Estimation of model parameters}}
\label{app_beta}
\textcolor{black}{
We estimate the parameters of the $j$-th word, $\hat{\beta}_j$, $\hat{\eta}_j$, $\hat{\delta_j}$ 
of the model given by Eqs. \ref{eq_rd} and  \ref{eq_rw}. 
For simplification, we omit the subscript $j$, which is used to distinguish between different words.} \par
\textcolor{black}{
In this paper, for estimation, we choose parameters that minimize the median squared divergence, considering both the MSD and the PSD,  
\begin{eqnarray}
&&(\hat{\beta},\hat{\eta},\hat{\delta})=\argmin_{\beta,\eta,\delta} L^{(d)}(\beta,\eta,\delta;D^{(d)}) \times L^{(s)}(\beta,\eta,\delta;D^{(s)}), \nonumber \\
\end{eqnarray}
where $L^{(d)}$ is the median-squared divergence for the MSD and $L^{(s)}$ is that for PSD. The definition of these values will be described subsequently.Here, $D^{(d)}$ is $m$ the empirical MSD data, $D^{(d)}=\{(x_1^{(d)},y_2^{(d)}),(x_1^{(d)},y_2^{(d)}),(x_1^{(d)},y_2^{(d)}), \cdots (x_m^{(d)},y_m^{(d)})\}$, where $x_i$ is the lag value and $y_i$ is corresponding the empirical MSD $y_i=\left<f(t-x_i)-f(t) \right>_t$. Similarly, $D^{(s)}$ is  the empirical PSD data, $D^{(s)}=\{(x_1^{(s)},y_2^{(s)}),(x_1^{(s)},y_2^{(s)}),(x_1^{(s)},y_2^{(s)}), \cdots (x_n^{(s)},y_n^{(s)})\}$, where $x_i^{(s)}$ is the spectrum and $y_i^{(s)}$ is the corresponding power density.} \par
\textcolor{black}{
Note that, because of robustness against outliers, we did not employ the mean squared divergence but used the median squared divergence.
In addition, we used both the MSD and the PSD, in order to ease the problem of multimodality, which strongly arises when we only use single statistics (the MSD or the PSD) for the estimation.}   \par
\paragraph{\textcolor{black}{Median squared divergence for the MSD $L^{(d)}$ } }
\textcolor{black}{
The median squared divergence for the MSD, $L^{(d)}$ 
is defined by 
\begin{eqnarray}
&& L^{(d)}(\beta,\eta,\delta;D^{(d)}) \nonumber \\
&&=Median_{\{(x_i,y_i) \in D^{(d)}\}}[\{\log(f^{(d)}(x_i;\beta,\eta,\delta))-\log(y_i)\}^2],\nonumber \\ 
\label{l_d}
\end{eqnarray}
where $f^{(d)}(x_i;\beta,\eta,\zeta)$ is the (corrected) theoretical function of the MSD given by Eq. \ref{cor_msd}. 
}
\par
\textcolor{black}{
 Many time series of word counts take into account seasonality and outliers. However, our theory does not consider seasonality and outliers, such as large breaking news. Thus, when we evaluate the Eq. \ref{l_d}, we need to apply the corrections for these effects as follows. 
\begin{itemize}
\item Before the calculation of the MSD, we remove the large outliers from word counts data $\{f_j(t)\}$ (to avoid the impact of large outliers on the MSD). In particular, we remove the large samples on the condition $x(t) > Median_t\{x(t)\}+5 \times IQR_t\{x(t)\}$
\item After the calculation of the MSD, we take the 365-days moving median of the raw MSD (see the section \ref{section_data_ana}) and use this value for the estimation, in order to cancel the seasonality and ease the effects of outliers).  
\end{itemize}
}
\par 
\textcolor{black}{In addition, we use the seasonality correction formula of the 365-days moving median of the MSD, 
\begin{equation}
f^{(g)}(x;\beta,\eta,\delta)=f_0^{(g)}(x;\beta,\eta,\delta)+2\hat{c}^2 \hat{\chi} \label{cor_msd}
\end{equation}
Here $f_0^{(g)}(x)$ is the original MSD of the model given by Eq. \ref{eq_rd} and  Eq. \ref{eq_rw}, and the other is the correction term of the seasonality.
} \par
 In particular, the original MSD of model is given by 
\begin{eqnarray}
f^{(g)}_0(x)=\hat{c}^2 \left<(r(t+x)-r(t))^2 \right>+2\hat{c}+2\hat{c}^2 \delta^2 
\end{eqnarray}
Here $\left<(r(t+x)-r(t))^2 \right>$ is given by Eq. \ref{diff_ans1b},  the mean word count $\hat{c}$ is estimated by $\hat{c}=\sum^{T}_{t=1}f(t)/T \approx Median_t\{f(t)\}$ ; 
this formula can be obtained through the same discussion as that presented in section \ref{app_msd_f}.
\par
\textcolor{black}{
$\hat{\chi}$ is the magnitude of relative seasonality estimated by, 
\begin{eqnarray}
\hat{\chi} \approx Var_u[\hat{\chi}(u)]=\sum^{365}_{u=1} \frac{(\hat{\chi}(u) - \sum^{365}_{u'=1}\hat{\chi}(u')/365 )^2}{365}, \nonumber \\
\end{eqnarray} 
where $\hat{\chi}(u)$ is weight of the seasonality of the day $u$ $(u=1,2,3,\cdots,365)$ estimated by  
\begin{eqnarray}
\hat{\chi}(u)=Median_{s}\{\frac{f(365 \times s+u)}{MovingMedian365\{f(365 \times s+u)\}}\}. \nonumber \\
\end{eqnarray} 
} \par
This correction formula Eq. \ref{cor_msd} was obtained through the following discussions.
First, we add the year-seasonality factor to Eq. \ref{Lambda}, 
\begin{eqnarray}
&&\Lambda(t) \sim m(t) \times c \times r(t)  \nonumber \\
&&\times \chi([(t-1) \bmod 365]+1) \times(1+\delta \times \Delta). 
\end{eqnarray}
Here $\chi(t)$ is the scaled year-seasonality factor, where we assumed $\sum_{s=1}^{365} \chi(s)/365 \approx 1$ and 
$\chi(s) >0$.
By using the same discussion as the section \ref{app_msd_f}, we can obtain the MSD of the new model, 
\begin{eqnarray}
&&\left<f(t+x)-f(t) \right> \approx \nonumber  \\
&& c^2 \left<(r(t+x)-r(t))^2 \right>+2c^2\delta^2+2c \nonumber \\
&&+c^2\left<(\chi([(t+x-1) \bmod 365] +1) \right. \nonumber \\
&&\left. -\chi([(t-1) \bmod 365] +1))^2 \right>, \label{f_t_chi}
\end{eqnarray}
where we use the approximation 
$\sum^{365}_{t=1}\chi(t)^2/365 \approx 1$. 
When we employee the 365-days moving median for analysis, we can use the following approximation 
\begin{eqnarray}
&&MovingMedian365_L\{\left<(\chi([(t+L-1) \bmod 365] +1) \right. \nonumber \\
&& \left. -\chi(t))^2 \right>_t\} \nonumber  \\
&&\approx MovingMean365_L\{\left<(\chi([(t+L-1) \bmod 365] +1) \right. \nonumber \\
&& \left. -\chi(t))^2 \right>_t\} \nonumber  \\ 
&&\approx 2 \cdot Var_t[\chi(t)], \label{chimedi}
\end{eqnarray}
Thus, we replace the term respect to $\chi(s)$ in \ref{f_t_chi} with the \ref{chimedi} and obtain the correction formula Eq. \ref{cor_msd}. \par
\textcolor{black}{Note that we sampled the $x_i$ in a logarithmic way (ex. $x_1=1, x_2=2, x_3=4, x_4=16 \cdots $)  to avoid that the contributions of data for the small $\{x_i \}$ are neglected substantially. } \par
\paragraph{\textcolor{black}{Median squared divergence for the PSD $L^{(s)}$}}
\textcolor{black}{The median squared divergence for the PSD, $L^{(s)}$ 
is defined by
\begin{eqnarray}
&& L^{(s)}(\beta,\eta,\delta;D^{(s)}) \nonumber \\
&&=Median_{\{(x_i,y_i) \in D^{(s)}\}}[\{\log(f^{(s)}(x_i;\beta,\eta,\delta))-\log(y_i)\}^2], \nonumber \\
\end{eqnarray}
Here, the theoretical formula of the PSD given by, 
\textcolor{black}{
\begin{eqnarray}
f^{(s)}(x;\beta,\eta,\delta)= c^2 \eta^2 (2 \sin(2\pi x/2))^{-2 (1-\beta)}+c+c^2 \delta^2,  \nonumber \\ 
\end{eqnarray}
}
where we use the discussion in the section \ref{app_p_f} and Eq. \ref{spec1}.} \par
\textcolor{black}{
In the same way as the MSD, we ease the effects of outliers as follows:  
\begin{itemize}
\item Before we calculate the PSD, we remove the large outliers from word count data $\{f_j(t)\}$ (to avoid the large outliers’ impact on the PSD). Specifically, we remove the large samples with the condition $x(t) > Median_t\{x(t)\}+5 \times IQR_t\{x(t)\}$.
\item After the calculation of the PSD, we take the 31 points-moving median of the raw PSD (see the section \ref{section_data_ana}) and use this value for the estimation (to ease week-peaks, sampling noise, and outliers of the PSD).  
\end{itemize}
}
\par
\textcolor{black}{Note that we sampled the $x_i$ in a logarithmic way to avoid that the contributions of data for the small $\{x_i \}$ are neglected substantially.}  \par 

\renewcommand{\theequation}{D\arabic{equation}}
\renewcommand{\thefigure}{D-\arabic{figure}}
\renewcommand{\thetable}{S-D\arabic{table}}
\setcounter{figure}{0}
\setcounter{equation}{0}
\section{The derivation of the relation between the random walk model Eq. \ref{eq_rw} and the fractional Langevin equation}
\label{app_arfima} 
 \textcolor{black}{The derivation of the relation between the random walk model Eq. \ref{eq_rw} and the fractional Langevin equation Eq. \ref{eq_frac_lan} is as follows.} 
On the condition $0 \leq \beta < 1$, the weight of the summation in Eq. \ref{eq_rw} is approximated by 
\begin{equation}
\frac{(s+d_\beta)^{-\beta}}{\Gamma(1-\beta)} \approx \frac{\Gamma(s+1-\beta)}{\Gamma(s+1) \Gamma(1-\beta)}.
\end{equation}
This approximation indicates that the model given by Eq. \ref{eq_rw} is approximated by $ARFIMA(0, 1-\beta,0)$ (i.e. the right-hand side of the following equation), 
\begin{equation}
r_j(t) \approx \sum_{s=0}^{\infty}\frac{\Gamma(s+1-\beta)}{\Gamma(s+1) \Gamma(1-\beta)} \cdot \eta_j(t-s). \label{r_forget}
\end{equation}
$ARFIMA$ is the abbreviation for the autoregressive fractionally integrated moving average model, which is a well-known time-series model to describe a time-series with long memory, in the field of statistics \cite{burnecki2014algorithms}.
$ARFIMA(0, 1-\beta,0)$ is written in short form as
\begin{equation}
(1-B)^{1-\beta}x(t)=\eta(t), \label{arfima}
\end{equation}
where $B$ is the backshift operator, that is $Bx(t)=x(t-1)$, and by using the Taylor series expansion, $(1-B)^{\alpha}$ is given by \cite{granger1980introduction}
\begin{equation}
(1-B)^{\alpha}=\sum^{\infty}_{k=0} \left( \begin{array}{c} \alpha \\ k \end{array}  \right) (-B)^k.
\end{equation}
The equation given by Eq. \ref{arfima} is the discrete version of the fractional Langevin equation given by Eq. \ref{eq_frac_lan} \cite{burnecki2014algorithms, magdziarz2007fractional}. \par

\renewcommand{\theequation}{E\arabic{equation}}
\renewcommand{\thefigure}{E-\arabic{figure}}
\renewcommand{\thetable}{S-E\arabic{table}}
\setcounter{figure}{0}
\setcounter{equation}{0}
\section{The normalised noncentral t-distribution}
\label{app_tdist}
The normalised noncentral t-distribution $Z^{(0)}(\nu,\mu)$ can be sampled by
\begin{equation}
Z^{(0)}(\nu,\mu) \sim \frac{Z(\nu,\mu)-m(\nu,\mu)}{\sqrt{v(\nu,\mu)}}, 
\end{equation}
$Z(\nu,\mu)$ is the noncentral t-distribution is can be sampled by, 
\begin{equation}
Z(\nu,\mu) \sim \frac{U+\mu}{\sqrt{V/\nu}}, 
\end{equation}
where $\nu$ is the degrees of freedom, $\mu$ is the noncentrality parameter,
$U$ is sampled from the standard normal distribution $U \sim N(0,1)$, $V$ is sampled from the chi square distribution with the degree of  freedom $\nu$, $V \sim \chi^2_{\nu}$ \cite{johnson1994continuous}, 
$m(\nu,\gamma)$ is the mean of the noncentral t-distribution, 
\begin{equation}
m(\nu,\mu)=\mu \sqrt{\frac{\nu}{2}} \frac{\Gamma((\nu-1)/2)}{\Gamma(\nu/2)} \quad (\nu>1)
\end{equation}
and $v(\nu,\mu)$ is the variance of the noncentral t-distribution,
\textcolor{black}{
\begin{equation}
v(\nu,\gamma)=\frac{\nu(1+\mu^2)}{\nu-2}-\frac{\mu^2\nu}{2}\left(\frac{\Gamma((\nu-1)/2)}{\Gamma(\nu/2)}\right)^2 \quad (\nu>2).
\end{equation}
}

\renewcommand{\theequation}{F\arabic{equation}}
\renewcommand{\thefigure}{F-\arabic{figure}}
\renewcommand{\thetable}{S-F\arabic{table}}
\setcounter{figure}{0}
\setcounter{equation}{0}
\section{Mean-squared displacement of $\{f_j(t)\}$}
\label{app_msd_f}
The $f_j(t)$ given by Eq. \ref{eq_rd} can be decomposed into two independent random variables, 
\begin{equation}
f_j(t)=\hat{c}_jr_j(t)+ q_j(t),
\end{equation}
where $q_j(t) \equiv f_j(t)-\hat{c}_j r_j(t)$ and the variance of $q_j(t)$ is written by \cite{RD_base}
\begin{equation}
\left<(q_j(t)|r_j(t))^2 \right>=\frac{1}{m(t)} \hat{c}_j r_j(t) +  \hat{\delta}_j^2 \hat{c}_j^2 r_j(t)^2. 
\end{equation}
\textcolor{black}{
Thus, the MSD is obtained as follows:
\begin{eqnarray}
&&\sigma_j(L)^2 \approx \nonumber \left<\sum_{t=1}^{T-L} \frac{(f(t+L)-f(t))^2}{T-L} \right>  \\
&&\approx \sum_{t=1}^{T-L}\frac{\left<(f(t+L)-f(t))^2 \right>}{T-L} = \nonumber \\
&&\sum_{t=1}^{T-L} \frac{\left<\{\hat{c}_jr_j(t+L)-\hat{c}_jr_j(t)+q_j(t+L)-q_j(t)\}^2 \right>}{T-L} \nonumber \\
&&\approx \sum_{t=1}^{T-L} \frac{ \hat{c}_j^2   \left<(r_j(t+L)-r_j(t))^2 \right>}{T-L}+ \nonumber \\ 
&&\sum_{t=1}^{T-L}\frac{\left<q_j(t+L)^2 \right>+\left<q_j(t)^2 \right>}{T-L} \nonumber \\
&&\approx  \hat{c}_j^2   \left<(r_j(t+L)-r_j(t))^2 \right> \nonumber \\ 
&&+\hat{c}_j  \sum_{t=1}^{T-L}\frac{r_j(t+L)/m(t+L)+r_j(t)/m(t)}{T-L} \nonumber \\ 
&&+\hat{c}_j^2 \hat{\delta}_j^2 \sum_{t=1}^{T-L}\frac{r_j(t+L)^2+r_j(t)^2}{T-L} \nonumber \\
&&\approx  \hat{c}_j^2   \left<(r_j(t+L)-r_j(t))^2 \right> \nonumber \\ 
&&+ 2  \hat{c}_j \sum^{T}_{t=1}\frac{1}{T}\frac{1}{m(t)}   +2 \hat{c}_j^2 \hat{\delta}_j^2   \label{eq_msd_f_app0}
\end{eqnarray}
where $\left<r_j(t+L)-r_j(t) \right>$ is given by Eq. \ref{diff_ans}. } \par
\textcolor{black}{From Eq. \ref{diff_ans1b}, we can obtain the following in the case where $\beta=0.5$:
\begin{eqnarray}
\sigma_j(L)^2 \approx  a_j \log(L)+b_j \quad  (L\gg1)
\label{eq_msd_f_app}
\end{eqnarray}
}
where $a_{j}=2  \hat{c}_j^2 \hat{\eta}_j^2 /\Gamma(1/2)^2$, $b_j= \hat{c}_j^2 \hat{\eta_j}^2 (-2\log(4)-2\psi^{(0)}(\Gamma(1/2)^{-2}))/\Gamma(1/2)^2 +2 \hat{c}_j^2 \hat{\delta}_j+2 \hat{c}_j$ and $\psi^{(0)}(x)$ is the digamma function. 
We use herein the approximation $1/(T-L)\sum_{t=1}^{T-L}r_j(t+L) \approx 1/(T-L)\sum_{t=1}^{T-L}r_j(t) \approx 1$, 
$1/(T-L)\sum_{t=1}^{T-L}r_j(t+L)^2 \approx 1/(T-L)\sum_{t=1}^{T-L}r_j(t)^2 \approx 1$, $1/(T-L) \sum_{t=1}^{T-L}1/m(t+L) \approx 1/(T-L) \sum_{t=1}^{T-L}1/m(t) \approx 1$ and Eq. \ref{diff_ans1b}. \par

\renewcommand{\theequation}{G\arabic{equation}}
\renewcommand{\thefigure}{G-\arabic{figure}}
\renewcommand{\thetable}{S-G\arabic{table}}
\setcounter{figure}{0}
\setcounter{equation}{0}
\section{The power spectrum density of $\{f_j(t)\}$}
\label{app_p_f}
The $f_j(t)$ given by Eq. \ref{eq_rd} and can be decomposed into two independent random variables, 
\begin{equation}
f_j(t)=\hat{c}_jr_j(t)+ q_j(t),
\end{equation}
where $q_j(t) \equiv f_j(t)-\hat{c}_j r_j(t)$ and the variance of $q_j(t)$ is written by \cite{RD_base}
\begin{equation}
\left<(q_j(t)|r_j(t))^2 \right>=\frac{1}{m(t)} \hat{c}_j r_j(t) +  \hat{\delta}_j^2 \hat{c}_j^2 r_j(t)^2. 
\end{equation}
\textcolor{black}{
Using this decomposition, the power spectrum density of $f_j(t)$ is written by:
 \begin{eqnarray}
&&P(\nu;f_j) \approx  \hat{c}_j^2 \cdot P_r(\nu)+P(\nu;q_j(t)) \\
&&\approx \hat{c}_j^2 \cdot P_r(\nu)+\sum_{t=1}^{T}\left<q_j(t)^2 \right>/T. \\ \label{spect3} 
&&\approx v_j \times (2 \sin(2 \pi \nu/2))^{-2(1-\beta)} +w_j,   \label{eq_app_psd_f0} 
\end{eqnarray}
where $v_{j}=\hat{c}_j^2  \hat{\eta}_j^2$ and $w_j=\hat{c}_j^2 \hat{\delta}_j^2 +\hat{c}_j$.
Here, we use the approximation $1/T\sum_{t=1}^{T}r_j(t) \approx 1$, $1/T\sum_{t=1}^{T}r_j(t)^2 \approx 1$ and $1/T\sum_{t=1}^{T}1/m(t) \approx 1$. } \par
\textcolor{black}{
In the case of $\beta=0.5$, we can obtain
\begin{eqnarray}
P(\nu;f_j) \approx v_j (2 \sin(2 \pi \nu/2))^{-1} +w_j.   \label{eq_app_psd_f}
\end{eqnarray}
}

\clearpage
\end{document}